%% file: LSST_DESC_SRD.tex
\documentclass[11pt]{report}

\include{inc/doc_settings} 

\include{inc/macros}       

\usepackage{float,longtable,threeparttablex,booktabs}

\begin{document}
\pagenumbering{gobble}

\pagestyle{empty}

\vspace*{0.2\textheight}

\begin{center}
{{\LARGE Rubin Observatory}\\ {\Huge\bfseries LSST DESC\\
\bigskip Science Requirements Document}}

\vspace*{0.2\textheight}

\input{commitID.tex}

\vspace*{0.1\textheight}

\begin{figure}[!h]
\centering\includegraphics[width=5cm,angle=0]{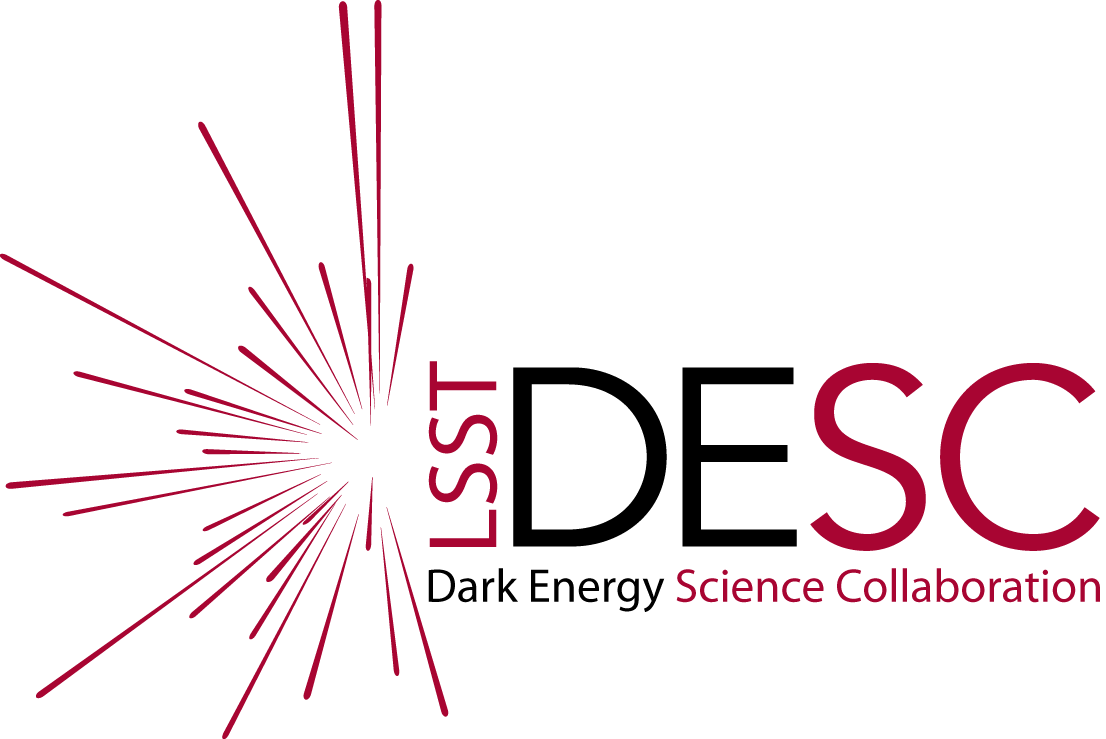}
\end{figure}

\end{center}

\clearpage
\include{inc/version}

\clearpage

\phantomsection\section*{Contributors}
\addcontentsline{toc}{section}{Contributors}

\input{inc/contrib.tex}

\clearpage
{\let\cleardoublepage\clearpage}

\vspace*{-1.0in}
\maketoc
\label{toc}

\clearpage

\pagestyle{fancy}
\fancyfoot{} 
\fancyfoot[R]{\thepage}  

\fancyhead[L]{}
\fancyhead[R]{LSST DESC Requirements}
\renewcommand{\footrulewidth}{1pt}

\setlength\parindent{0em}
\setlength{\parskip}{0.5em}



\renewcommand{\thepage}{\arabic{page}}
\setcounter{page}{1}

\renewcommand\thefigure{\arabic{figure}}
\setcounter{figure}{0}


\phantomsection\section*{Executive Summary and User Guide}
\addcontentsline{toc}{section}{Executive Summary and User Guide}

The Dark Energy Science Collaboration (DESC) was formed to design and implement dark energy analysis of the
data from the Vera C.\ Rubin Observatory's Legacy Survey of Space and Time (LSST) using five dark energy probes: weak and strong gravitational lensing, large-scale
structure, galaxy clusters, and supernovae.  Assuming the delivery of LSST data by 
Rubin Observatory according to the design specifications in the LSST Science Requirements
Document (LSST~SRD), the DESC will carry out further analyses with its own
infrastructure (software, simulations, computational resources, theory inputs, and re-analyses of
precursor datasets) to produce constraints on dark energy.

The first goal of this document is to
quantify the expected dark energy constraining power of all five DESC probes individually and
together, with conservative assumptions about analysis methodology and follow-up observational
resources (e.g., spectroscopy) based on our current
understanding and the expected evolution within the field in the coming years.  The second goal is
to define requirements on our analysis pipelines which, if met, will enable us to achieve our goal of carrying
out dark energy analyses consistent with the Dark Energy Task Force (DETF) definition of a Stage IV dark
energy experiment.
This is achieved through a forecasting process that includes the flow-down to detailed requirements
on multiple sources of systematic uncertainty.

We define two classes of systematic uncertainty: ``self-calibrated''
ones, for which we will build a physically-motivated model with nuisance parameters over which
we marginalize with priors that are either {\it uninformative} or mildly informative (where justified by other data); and ``calibratable'' ones, with nuisance parameters that may
not be physically meaningful and that relate to some error in the measurement process, for which
DESC simulations, theory, other software, or precursor datasets produce {\it informative} priors.  The
``total uncertainty'' consists of the statistical uncertainty, including the broadening of the posterior
due to marginalization over self-calibrated systematic uncertainties, combined with the calibratable
systematic uncertainty.  Our requirements are set such that these calibratable uncertainties will be a subdominant contributor to the total uncertainty.
As our understanding of systematic uncertainties changes, some may switch from calibratable to
self-calibrated.
We define detailed requirements through a process of error budgeting among different
calibratable systematic uncertainties, with forecasts used to check that meeting the detailed requirements will enable us to meet our high-level objectives.

Some of the key outcomes of this process are as follows.
\begin{itemize}
\item We have defined high-level objectives that the collaboration hopes to
  achieve in the next 15 years, including standards for control of systematic uncertainties.
\item We have defined a baseline analysis for each probe that is consistent with LSST being a stand-alone
  Stage IV dark energy experiment, with joint-probe marginalized
  uncertainties on dark energy equation-of-state parameters $(w_0, w_a)$ of $\sigma(w_0)=0.02$ and
  $\sigma(w_a)=0.14$ (combined $1\sigma$ statistical and systematic uncertainties), where $w(a) = w_0 + (1-a) w_a$.
\item We have defined a set of quantifiable requirements on each probe, including the flow-down to
  detailed requirements on the level of systematics control achieved by DESC infrastructure.  These can be compared with the current
  state-of-the-art and future plans in order to prioritize efforts in the coming years.  The detailed
  requirements in this first version of this document are a limited subset of those we expect to define in the end;
  here we focus on
  photometric redshift uncertainties, weak lensing shear, and
  photometry (through its impact on supernova light curves).  The high-level requirement that
  LSST be a stand-alone Stage IV dark energy experiment is expected to remain fixed, while the detailed
  requirements may change as our understanding of
  analysis methods improves.
\item We have defined a set of goals, which are quantifiable (like requirements) but are not
  prerequisites for collaboration success.
\item This exercise has highlighted the need for collaboration software for forecasting dark energy
  analyses self-consistently across all probes.  Aspects of the single-probe analyses and
  systematics models described in this document, whether they were
  implemented or not in this first DESC~SRD version, serve as guides for defining the
  capabilities of that collaboration software framework.
\end{itemize}

Future versions of this document will incorporate the following improvements: (a) evolution in our
software capabilities and analysis plans; (b) decisions by Rubin Observatory about survey strategy; (c) requirements on sufficiency of models for self-calibrated
systematic uncertainties; (d) requirements on calibratable systematic uncertainties beyond those in
this version of the DESC~SRD (particularly ones for which we currently lack a description of
their impact on the observables); and (e) self-consistent treatment of common systematic
uncertainties across probes.  Currently all objectives,
requirements, and goals relate to dark energy constraints; future DESC~SRD
versions may consider secondary science objectives such as constraints on neutrino mass.

\subsection*{How to Use This Document} When showing plots, forecasts, or requirements from this
document, it should be cited as ``the LSST DESC Science Requirements Document v1 (LSST~DESC 2018)''
in the text, and ``LSST~DESC~SRD~v1'' in figure legends.
(The ``DESC'' avoids ambiguity with the LSST~SRD developed by Rubin Observatory, and ``v1'' avoids confusion
with  later versions.)
On the LSST~DESC community Zenodo page\footnote{\url{https://zenodo.org/communities/lsst-desc}} we provide a
tarball with the following items: figures, all individual and
joint probe Fisher matrices from \autoref{fig:jointy1y10} along with the python script that produced
the plot, data vectors and covariances from the weak lensing, large-scale structure, and galaxy
clusters forecasts, MCMC chains, simulated strong lens and supernova catalogs, and the software for producing the supernova
requirements and forecasts. When using these data products, please cite the Zenodo DOI (for
which a BibTeX reference can be downloaded from the Zenodo page) in addition to the arXiv entry for
this document.  Care should be taken when
combining the Fisher matrices with those from other surveys, particularly to ensure common choices
of cosmological parameters and consistent choices of priors and that the Fisher matrices being added
are truly independent (which may not be case if the probed volume overlaps).
Finally, internal to the DESC, this
document will be used to inform analysis pipeline development, including the development of
performance metrics.

\clearpage

\section{Introduction}\label{sec:intro}

Understanding the nature of dark energy is one of the key objectives of the cosmological community
today.  The objective of the
LSST
Dark Energy Science Collaboration (DESC) is to prepare for and carry out dark energy analysis with
LSST \citep{2008arXiv0805.2366I,2009arXiv0912.0201L}.  Following acquisition of the LSST 
images and the processing with Rubin's LSST Science Pipelines, both carried out by Rubin
Observatory, the DESC will use its own ``user-generated'' software to analyze the LSST data and
produce cosmological parameter constraints.  In this document, the DESC
 Science
Requirements Document (DESC~SRD), we outline the DESC's scientific objectives, along with the performance
requirements that the DESC's software (including simulations and theoretical modeling capabilities)
must meet to ensure that the DESC meets those
scientific objectives.  Unlike requirements in a Science Requirements Document for a hardware project, the detailed requirements on software pipelines in the DESC~SRD may evolve with time, since they are sensitive to assumptions about the entire analysis pathway to cosmological parameters, about which our understanding will continually improve.

Rubin Observatory has its own science requirements document for LSST (the
LSST~SRD), which can be found on their
webpage\footnote{\url{https://docushare.lsstcorp.org/docushare/dsweb/Services/LPM-17}}.
The LSST~SRD outlines requirements on the hardware, observatory,
and the LSST Science Pipelines, all of which fall under the purview of Rubin Observatory.  In defining the performance requirements for DESC software, we
assume that 
Rubin Observatory is going to deliver survey data in accordance with the
``design specifications'' in the LSST~SRD (not the more pessimistic ``minimum
specifications'', or the more optimistic ``stretch goals''). We note
that the LSST observing strategy will continue to evolve as LSST
approaches first light, with the possibility of significant updates in cadence and
how depth is build up over time, while still satisfying the LSST~SRD requirements.  In the subsections below, we
highlight relevant LSST Project requirements; more
generally, our reliance on Rubin Observatory tools and requirements is
summarized in \appref{app:lsst}.

Following the convention for DOE projects, we quantify the constraining power of dark
energy measurements using the figure of merit (FoM) from the Dark Energy Task Force report (DETF;
\citealt{2006astro.ph..9591A}).  The definition of this quantity, and other relevant terminology for
the DESC~SRD, is in \autoref{sec:definitions}.  While the main text
summarizes
the calculations for the sake of brevity, detailed technical appendices describe exactly what was
calculated for each probe, with assumptions and systematics models described in a
manner designed to ensure
reproducibility of the results in this document.

In this document we make the reasonable assumption that already-funded
surveys will be carried out and that spectroscopic follow-up and other ancillary telescope resources
will continue to be available at similar rates as they are today.  We
do not assume the acquisition of substantial new ancillary datasets in order to mitigate
systematics.  See \appref{subsec:assump-followup} for a summary of assumptions about follow-up and
ancillary telescope resources for each DESC probe.

The outline of this document is as follows.  \autoref{sec:definitions} includes definitions for
terminology used throughout the DESC~SRD.
In \autoref{sec:objectives}, we outline the key
objectives of the LSST dark energy analysis, while in \autoref{sec:highlevelreq}
and \autoref{sec:detailedreq} we derive a set
of requirements on the DESC's analysis software, based on a flow-down from high-level (i.e.,
targeted constraining power on dark energy) to low-level details of the
tolerances for
residual systematic uncertainties.

Any changes to the DESC~SRD after the first official version (v1) is tagged will be proposed by the
Analysis Coordinator following consultation with the Working Groups, Rubin Observatory Liaisons and Management team,
and approved by the Spokesperson. In practice this will be achieved by a Pull Request to the master
branch of the DESC Requirements repository,
which is protected.  The Spokesperson will maintain a change log in the document, and tag the
repository as changes are merged.

\section{Definitions}\label{sec:definitions}

Below we define the terminology used throughout the document.

\begin{itemize}
\item {\em Objectives} (\autoref{sec:objectives}): The DESC's high-level objectives provide the
  scientific
  motivation for the LSST dark energy analysis\footnote{Some readers may notice that we have adopted similar terminology to the
    internal Dark Energy Survey (DES) science requirements document for objectives, requirements,
    and goals.  The choices made in this document were influenced by that document.}.  They provide the
  context for development of the science requirements and goals, but may not be directly
  testable themselves.
\item {\em Science requirements} (\autoref{sec:highlevelreq} and \autoref{sec:detailedreq}): Requirements are the testable criteria that must be satisfied in
  order for the collaboration to meet its objectives.
\item {\em Goals}: These are testable criteria that go beyond the science requirements. For
  example, these could be criteria that must be met in order to achieve secondary science
  objectives, such as constraining modified gravity theories or neutrino mass. They also could be
  criteria related to achieving an earlier, or more optimal, use of the data than is needed to meet
  our requirements. They are ``goals'' rather
  than ``science requirements'' because achieving them is not considered a prerequisite for collaboration
  success.
\item {\em Dark energy probes}: The DESC currently has five primary dark energy probes: galaxy clusters
  (CL), large-scale structure (LSS), strong lensing (SL), supernovae (SN), and weak lensing (WL).
  All details of the associated analyses are given in \appref{app:baselines}.  The general philosophy
  behind our calculations is that we aim for
  a state-of-the-art analysis with reasonable (neither overly aggressive nor overly conservative)
  assumptions about what data we will be able to successfully model to constrain dark energy. In
  some cases, the analysis choices were constrained by the capabilities of existing software, and
  hence will need to be updated to be more consistent with this philosophy in future DESC~SRD versions when
  improved software is available.  In
  brief, the baseline analysis for each probe is as follows:
\begin{itemize}
\item The baseline LSST CL analysis includes cluster counts and cluster-galaxy lensing.  It will be
  valuable to update the baseline analysis in future DESC~SRD versions to include cluster
  clustering, which can be beneficial in self-calibrating the mass-observable relation
  \citep[e.g.,][]{2004PhRvD..70d3504L}, when software with this capability is available.
\item The baseline LSST LSS analysis includes tomographic galaxy clustering to nonlinear scales, not
  just the baryon acoustic oscillation (BAO) feature studied in the DETF report.  Future DESC~SRD
  versions may define the baseline analysis in terms of a multi-tracer treatment
  \citep[e.g.,][]{2009PhRvL.102b1302S,2013MNRAS.432..318A}, which is beneficial in the cosmic
  variance-limited regime.
\item The baseline SL analysis includes time-delay quasars and compound lenses.  Future DESC~SRD
  versions should also include strongly lensed supernovae in the baseline analysis.
\item The baseline SN analysis includes WFD (Wide-Fast-Deep, the main LSST survey) and DDF (Deep Drilling Field) supernovae, with the assumption that a
  commissioning mini-survey will be used to build templates so that the photometric SN analysis can
  begin in year one of the survey.
\item Unlike in the DETF report, the baseline LSST WL analysis is a full tomographic ``3$\times$2pt'' analysis:
  shear-shear, galaxy-shear, and galaxy-galaxy correlations.  This analysis choice is consistent
  with the current state of the art in the field, but it means there is some statistical
  overlap between the LSS and the WL analysis.  For completeness we will also report on the
  constraints from shear-shear alone. When forecasting combined constraints across all
  probes, we include just the 3$\times$2pt analysis to avoid double-counting.
\end{itemize}
As our understanding of these analyses improves, the baseline analysis may need to be updated, resulting in
updates to the forecasts and the requirements.
\item {\em Dark Energy Task Force (DETF) figure of merit (FoM)}: given a dark energy equation of
  state model with
  $w(a) = w_0 + (1-a) w_a$, the DETF Report defines a FoM in terms of the Fisher matrix for
  $(w_0,w_a)$ marginalized over all other parameters as $\sqrt{|F|}$, corresponding to the area of
  the 68\% credible region.  See the DETF Report for
  details.\footnote{The text of the DETF Report defines the FoM in terms of the
    area of the 95\% contour.  However, all numbers tabulated in the
    report correspond to simply $\sqrt{|F|}$ without the additional factor needed to get the area of
    the 95\%
    credible region, and it has become common
    in the literature to refer to numbers calculated this way as the
    ``DETF FoM'', despite what the text of the report says.  The
    DESC~SRD follows this convention as well.}
\item {\em Overall uncertainty}: Following the DETF, we quantify
  overall uncertainty as the {\it width}\footnote{See
    \appref{app:requirements} for a discussion of how ``width''
    is determined.}
  of the posterior probability distribution in the $(w_0, w_a)$ plane,
  after marginalizing over nuisance parameters associated with
  systematic uncertainties.
\item {\em Error budget:}
    A target overall uncertainty on Dark Energy parameters sets the
    {\it error budget} for our LSST analysis.  The DESC~SRD describes how this error budget can be
    allocated: each probe will contribute to the
    overall  uncertainty by an amount that will depend on how much
    information  the LSST data contain, how much external information we
    can  provide, and how the probes' likelihood functions interact with
    each other.   Estimating the error budget for each probe, and for
    each measurement step within those probes, must be done iteratively,
    making forecasts of overall uncertainty given a set of assumptions,
    varying those assumptions, and repeating.  See the start of \autoref{sec:detailedreq} and
    \autoref{fig:errorbudget} for details.
\item {\em Statistical uncertainty}: We use the term ``statistical \
    uncertainty'' to describe the width of the
    posterior probability distribution in the $(w_0, w_a)$ plane when
    the nuisance parameters associated with systematic uncertainties
    are fixed at their fiducial values. We expect the statistical
    uncertainty for each LSST dark energy probe to be small compared to
    the additional posterior width introduced by marginalizing over
    systematic effects. This is what we mean by LSST cosmological
    parameter measurements being ``systematics limited.''
\item {\em Systematic biases or systematic errors}: These are known/quantified offsets in our measurements
  due to some observational or astrophysical issue.
  We use the noun ``systematic'' as an abbreviation for ``systematic bias'' and take ``error'' and ``bias'' to be synonymous.
  Their amplitude is not relevant for the
  DESC~SRD because the known part is  presumed to
  have been removed and does not impact our dark energy constraining power.  However, quantifying the uncertainty in
  these corrections is critical.
\item {\em Systematic uncertainties}:
  All sources of systematic uncertainty are treated, either explicitly or
  implicitly, by extending the model to include additional ``nuisance
  parameters'' that describe the effect. Marginalization over these
  nuisance parameters allows us to propagate the uncertainty, which is
  captured by the prior PDF for their values,
  through to the cosmological parameters.
  These systematic uncertainties  are hence associated with {\em residual}
  (uncorrected or post-correction) offsets, resulting from imperfect
  knowledge applied in the treatment of systematic biases.
  Two types of systematic effects are considered in the DESC~SRD,
  defined as follows:
\begin{itemize}
\item {\em Calibratable systematics}: We refer to systematic
  biases that can be estimated with some precision, or equivalently,
  modeled with nuisance parameters that have {\it informative priors}
  as ``calibratable.'' Such biases tend to be associated with some aspect of the measurement process, and their nuisance parameter
  priors can typically be
  derived by validating the relevant analysis algorithm
  against external data or sufficiently realistic
  simulations.  Generally the nuisance parameter values themselves are of no physical importance.
Selection bias may also be treated as calibratable, though in that case a meta-analysis may be
needed to place priors on its magnitude, since the bias is associated with sample definition rather than
per-object measurements.  In many cases the marginalization over calibratable
  systematic nuisance parameters can be done in advance of the
  cosmological inference, resulting in the apparent application of a
  ``correction'' and the corresponding introduction of some additional
  uncertainty. In the other cases, no well-defined model is available for the nuisance parameters or
  their priors, and we must estimate the potential impact and propagate this uncertainty.
  {\bf A key part of any
  dark energy analysis is demonstrating that {\it systematic uncertainties due to calibratable effects
  do not dominate},
  and hence we place requirements on
  calibratable systematic uncertainties in
  \autoref{sec:highlevelreq} and~\autoref{sec:detailedreq} below.}
  These can
  be thought of as requirements on the size of the informative priors that we can set on these effects.
Informative priors are important in the typical case that we do not have a sufficient model for
them. In principle, with a sufficiently descriptive model for a particular source of systematic
uncertainty, it could be allocated a larger fraction of our total error budget, moving it
into the self-calibrated category defined below.
\item {\em Self-calibrated systematics}:
  These are sources of systematic uncertainty that cannot be
  estimated in advance, but that can be ``self-calibrated'' by marginalizing over the nuisance parameters of a
  model for them at the same time that the cosmological parameters are constrained. They tend to be astrophysical in nature. Examples include the cluster mass vs.\ observable relation, galaxy bias, and
  galaxy intrinsic alignments.  The nuisance
  parameters associated with self-calibrated effects will generally have uninformative or mildly informative priors when considering the analysis of LSST
  data on their own, and often correspond to astrophysically-meaningful quantities.
  As mentioned in \autoref{sec:intro}, we do not place requirements on
  factors outside of the DESC's control, such as the acquisition of substantial ancillary datasets that
  would provide additional terms in the likelihood to constrain those nuisance parameters more
  tightly.
  {\bf When setting requirements on our control of calibratable systematic effects, our convention is
    to include the additional uncertainty
    caused by marginalizing over these self-calibrated effects
    together with the statistical uncertainty, referring to their combination
    as the {\em marginalized statistical uncertainty}.}  The marginalized statistical uncertainty
  differs from the {\em overall uncertainty} in that the latter also includes calibratable
  systematic uncertainties.
\end{itemize}
While we do not place requirements on self-calibrated
systematic uncertainties in this version of the DESC~SRD, one could in principle place requirements on them in
the future by
requiring {\em model sufficiency}.   Models for self-calibrated systematics must be sufficiently complex, flexible and extensive
so as to
span the range of realistic possibilities for the physical phenomena in question. If they are not,
then our overly-simplified modeling assumptions could result in a bias in
cosmological parameter estimates.  This bias is often referred to as `model bias', and some
meta-analysis may be required to estimate its magnitude.
Our current approach, however, is to assume that our models for
self-calibrated systematics (which are a topic of active R\&D within the DESC analysis working groups) are sufficient.
There is a subtlety associated with
which systematic uncertainties' nuisance parameters we marginalize over at different steps of the analysis.  When
setting requirements on calibratable systematic uncertainties, we
marginalize
{\em only over self-calibrated systematic uncertainties} in order to
check how the additional uncertainty caused by calibratable systematic uncertainties compares with
the marginalized statistical uncertainty.
When considering the final dark energy figure of merit, we marginalize over {\em both self-calibrated
and calibratable systematic uncertainties} to determine the {\em overall uncertainty}, just as we would in the real joint analysis.
\item {\em Cosmological parameters}: Due to practical considerations associated with the
  software framework used in defining the requirements, we consider a flat $w$CDM  cosmological model,
  which results in a seven-dimensional parameter space
  consisting
  of $(\Omega_m, \sigma_8, n_s, w_0, w_a, \Omega_b, h)$.  Future DESC~SRD versions may expand this
  parameter space, e.g., to include massive neutrinos and curvature.  Fiducial parameter values and priors are
  outlined in  \appref{subsec:assump-cosmo}.  For requirements that are placed using forecasts of the constraining power
  of a single probe, we carry out the likelihood analysis only with the parameters that that probe is able
  to constrain (e.g., SL and SN do not constrain $\sigma_8$).
\item {\em ``Year 1'' (Y1) and ``Year 10'' (Y10)} forecasts, requirements, and goals: Several of our
  requirements and goals are relevant at all times (not just at the end of the survey), so we
  provide forecasts for dark energy constraining power with the full survey and with approximately
  1/10 of the data.  We use ``Year 10'' (Y10) and ``Year 1'' (Y1) as shorthand terms for these
  datasets.  See \appref{subsec:assump-strategy} for details of how we define the Y1 and Y10 survey
  depths and areas.  Note that the time at which we receive a dataset corresponding to this Y1
  definition may differ significantly from a single calendar year after the survey starts plus the
  time for the Project to process and release that data.  The LSST~SRD has requirements on
  single-exposure and full-survey performance, but no specifications that collectively guarantee
  that the Y1 dataset
  as defined in this document will be delivered by a particular time.
\end{itemize}

\section{Objectives}\label{sec:objectives}
\setcounter{objectives}{1}

The DESC's primary scientific objectives are listed and described below.

\labelobj{obj:key}{LSST will be a key element of the cosmological community's Stage-IV dark energy program.}

The DETF report \citep{2006astro.ph..9591A} specifies that the ``overall Stage-IV program should
achieve, in combination, a factor of 10 improvement over Stage-II.''.  In principle, we need not
apply this criterion to LSST dark energy analysis on its own, since LSST is being carried
out in the context
of a broader Stage-IV dark energy program that includes, e.g., DESI.  We will nonetheless do so.

\labelobj{obj:complementary}{DESC will produce multiple (at least two) independent dark energy constraints with
  substantially different dependencies on the growth of structure and the cosmological expansion
  history.}

While one could in principle imagine optimizing dark energy
constraints by focusing exclusively on obtaining extremely precise
constraints from a single probe or class of probes (e.g., structure
growth only, with a focus on WL, CL, LSS), a key part of the Stage-IV
dark energy program will be demonstrating consistent results with
methods that probe dark energy in different ways and with distinct
sets of systematic uncertainties.

\labelobj{obj:notsysdom}{For the LSST dark energy constraints,
calibratable systematic uncertainty
should not be the dominant contribution to the overall uncertainty.
}

In practice, meeting this objective means ensuring that the calibratable
systematic uncertainty in the dark energy parameters, which is the most difficult type of
uncertainty to model accurately, does not exceed
the combination of the statistical uncertainty and the self-calibrated
systematic uncertainty (the ``marginalized statistical uncertainty'').
The latter can be estimated by conditioning on
fiducial values of the calibratable biases' nuisance parameters, and
marginalizing over the self-calibrated biases' nuisance parameters.

\section{High-level requirements}\label{sec:highlevelreq}

\setcounter{requirementh}{1}
\setcounter{goal}{1}

In this section, we derive the high-level science requirements from the objectives in
\autoref{sec:objectives}.  We start by quantifying requirements on the
overall
uncertainties, both jointly and from each probe.

\labelreqh{high:combinedfom}{DESC dark energy probes will achieve a combined FoM exceeding 500 ($\sim$10$\times$ Stage-II) with the full LSST Y10  dataset when including both statistical and systematic
uncertainties and using Stage III priors.}

This requirement is essentially a statement that the DESC dark energy analysis should meet the Stage
IV program requirements independent of other Stage IV experiments (which we refer to as being a `stand-alone Stage IV experiment'), when combining all probes and using the full ten-year
dataset.  The Stage II FoM in the DETF report (page 77) includes CL, SN, and WL analysis, corresponding to a
FoM of 54.  Stage IV surveys should {\em collectively} exceed this by approximately a factor of 10.

Proper incorporation of Stage III priors
for all probes is complicated, especially given the overlap between the LSST and DES
footprints.  For this
reason, we use only SDSS-III BOSS, Planck, and Stage III supernova survey priors, and an $H_0$ prior
(described in more detail in \appref{subsec:assump-cosmo}) rather than all Stage III
priors.  In future, using SDSS-IV eBOSS will be possible as well.

Note that, when imposing \ref{high:combinedfom}, the FoM includes the overall uncertainty:
pure statistical uncertainties and marginalization over {\em both}
self-calibrated and
calibratable biases (see \autoref{sec:definitions} for details of these
categories).  Indeed, \ref{high:combinedfom} is the first step in our systematic error budgeting process: If our
forecast FoM exceeds 500 without accounting for calibratable systematic uncertainties, we adjust the
amount of the error budget that goes into calibratable systematic uncertainties such that the final
FoM after including them is exactly 500.  This process is the first step in deriving detailed
requirements in \autoref{sec:detailedreq}.

Satisfying this
requirement will enable DESC to achieve its first objective, \ref{obj:key}.

\labelg{high:indivfom}{Each probe or combination of probes that is included as an independent term in
  the joint likelihood function for the full LSST Y10 dataset will achieve FoM $> 2 \times$ the corresponding
Stage-III probe
 when including both statistical and systematic uncertainties.
The relevant thresholds for the individual DESC probes\footnote{The origin of the
Stage-III figures of merit, which are $0.5\times$ the thresholds quoted here, is described
in \appref{subsec:assump-stage3}.  Since the completion of the DETF report, the landscape of measurement has changed significantly and
the actual obtained Stage III FoMs are in some cases well below those forecasted in the original
report.} are 12, 1.5, 1.3, 19,
and 40 for CL, LSS, SL, SN, and WL, respectively.}

In addition to the overall FoM requirements in~\ref{high:combinedfom}, our goal is to substantially
improve over the previous state of the art in each
individual probe analysis.  As for~\ref{high:combinedfom}, the FoM comparison implied by this goal
includes the overall uncertainty.   Another motivation behind this goal is to ensure that the DESC
meets its
objective~\ref{obj:complementary} of deriving dark energy constraints from multiple complementary
dark energy probes\footnote{Note that \protect\ref{high:indivfom} is phrased such that not all
  probes must meet it, only those probes that enter the final joint likelihood analysis.  However,
  this goal is only part of what is needed to meet~\protect\ref{obj:complementary};
  \protect\ref{high:twotypes} is also relevant to that objective.}. To test whether we will meet
\ref{high:indivfom} for any given dark energy probe, we must define baseline
analyses for LSST as well as a corresponding Stage-III FoM.
The baseline analyses for LSST are outlined in
\autoref{sec:definitions}, and all analysis choices and sources of systematic uncertainty are described in
\appref{app:baselines}.

In principle the factor of two in \ref{high:indivfom} is arbitrary.  However, it is empirically the
case that for some of our probes, the LSST Y10 forecasts indicate greater degeneracy breaking
between probes such as SN and WL than the Y1 forecasts.  By implication, the SN and WL
degeneracy-breaking power for Stage IV surveys should be greater than for Stage III surveys assuming
that the LSST Y1 and Stage III degeneracy directions may be similar.  In that case, the combined
probe Stage IV constraining power (a factor of three in overall FoM compared to Stage III; see \ref{high:combinedfom}) can
be achieved with an increase in FoM for individual probes that is less than a factor of three.

\labelreqh{high:sys}{Each probe or combination of probes that is included as an independent term in
 the likelihood function will achieve total calibratable systematic uncertainty
 that is less than the marginalized statistical uncertainty in the
($w_0,w_a$) plane.}

{\em This requirement, which is the only one of our requirements that can be applied to the Y1 analysis (or
any analysis before the completion of LSST),
 is a way of quantifying whether we have achieved our high-level objective~\ref{obj:notsysdom}.}  It is
important to note that by comparing against the marginalized statistical uncertainty, we are including self-calibrated systematic uncertainties (e.g., due to astrophysical effects
such as scatter in the cluster mass vs.\ observable relation, galaxy intrinsic alignments, galaxy
bias).
Hence we are not requiring that systematic uncertainty due to {\em
  any} non-statistical error be less than the purely statistical error.  We are only requiring that
residual uncertainty in {\em calibratable} systematics be less than the
uncertainty
after marginalizing over self-calibrated systematics.  The reason to frame this requirement in this
way is that realistically, some dark energy probes may have  astrophysical systematic
uncertainties that will always exceed the statistical
error for LSST.  This basic feature of those probes should not be considered a failure of the DESC's efforts to utilize those
probes.
Also note that the line between self-calibrated and calibratable systematics is potentially movable;
given a better model for calibratable uncertainties (and possibly a different approach to the analysis of
LSST data), they
could become self-calibrated.  In that case, they would enter \ref{high:sys} differently, since they
could acceptably become a dominant contributor to the overall uncertainty (leaving \ref{high:combinedfom}
and \ref{high:indivfom} as indirect constraints on how much additional uncertainty they can
contribute), modulo any requirements on model sufficiency which would be treated as calibratable uncertainty.

We have not specified precise tolerances (systematic uncertainty equals $X$ times marginalized statistical
uncertainty for some value of $X$) in \ref{high:sys} in recognition of the fact that many elements
of these forecasts will change as our understanding improves, so $X$ can be specified only
to one significant figure.  Changing $X$ from 1 would coherently shift all of our
requirements to be more/less stringent but is unlikely to strongly modify our understanding of which
systematics are more/less challenging to control.

While \ref{high:sys} applies at all times, its implications for the analysis of each
probe depend on time.  When the full survey dataset exists, \ref{high:combinedfom} constrains
how much statistical constraining power we can
lose to carry out a more conservative analysis that makes it easier to meet \ref{high:sys}.  Before
then, only \ref{high:sys} is relevant.  Hence, \autoref{sec:detailedreq} has Y10 detailed requirements
associated with \ref{high:sys}, along with Y1 goals.  The Y1 goals quantify the needed level of
systematics control to enable us to carry out our
desired baseline analysis with Y1 data, without sacrificing statistical precision due to
difficulties achieving the required control of systematic uncertainties.  However, if we cannot meet
those goals in Y1 (for example, due to unanticipated systematic uncertainties in the data that require
additional time to understand and mitigate), then meeting \ref{high:sys} is sufficient.

Finally, note that \ref{high:sys} may at times be directly satisfied due to the constraints
imposed by \ref{high:combinedfom}.  As mentioned in the description of \ref{high:combinedfom}, we
may decrease the allowable calibratable systematic uncertainty if needed to ensure that we meet our
high-level requirement of being a stand-alone Stage IV dark energy experiment.  In cases where that
occurs, as in this version of the DESC SRD, meeting \ref{high:combinedfom} automatically ensures
that \ref{high:sys} will be met.

\labelreqh{high:twotypes}{At least one probe of structure growth and one probe of the
  cosmological expansion history shall satisfy~\ref{high:indivfom} and~\ref{high:sys} for the full LSST Y10 dataset.}

This requirement ensures that we achieve our objective \ref{obj:complementary}.
\ref{high:twotypes} is motivated by the fact that a Stage IV dark energy experiment should ideally
provide not only constraints on the equation of state of dark energy, but also provide a stringent
test of gravity.  For the purpose of this requirement, we consider CL and WL as probes of structure growth
(though they carry a small amount of information about geometry), SN and SL as probes of the
expansion history, and LSS in both categories since it includes measurement of the baryon acoustic
oscillation feature in addition to smaller-scale clustering.
Deviations from General Relativity are best detected with two complementary probes.

\labelg{high:twotypesys3}{At least one probe of structure growth and one probe of the
  expansion history should satisfy~\ref{high:sys} for the full LSST Y3 dataset.}

We do not require that \ref{high:twotypes} is met in our early analyses, given the
level of technical challenge involved in carrying out an analysis that is not
dominated by calibratable systematics with a new dataset.  However, we would ideally like to be well
on our way to including multiple complementary dark energy measurements after several years -- hence
the definition of this goal~\ref{high:twotypesys3}.  Similarly to the definition of Y1
(\autoref{sec:definitions}), the definition of Y3 in \ref{high:twotypesys3} corresponds to the
science analyses after a time when roughly 3/10 of the WFD images over the full area have been
observed, processed, and released, rather than strictly the end of the third year of the survey.

\labelreqh{high:blinding}{DESC will use blind analysis techniques for all dark energy analyses to
avoid confirmation bias.}

Confirmation bias has been highlighted by the field as an important issue for cosmological
measurements \citep{2011arXiv1112.3108C}.  Carrying out blind analyses is becoming increasingly
common for probes of large-scale structure \citep{2017arXiv170801530D,2017MNRAS.465.1454H} and will
be even more important in the era of LSST.  Development of methods for blinding that will work for
LSST is a non-trivial task, and this procedural requirement is tantamount to saying that this work is
high-priority.  It will help us to work in a way that is consistent with our
objective~\ref{obj:notsysdom}; carrying out blinded analyses avoids
confirmation bias.  Currently, our inclusion of Stage III priors that originate from non-blinded
cosmological analyses in the derivation of our detailed requirements may appear to be in tension with \ref{high:blinding}. However, \ref{high:blinding} refers to the actual analyses carried out, and in practice we
would strive to use more up-to-date analyses from surveys that are not currently available (DESI,
Simons Observatory, CMB-S4, etc.), which will be both
more powerful than our current Stage III priors and will hopefully utilize blinded analysis methods
given the evolution of the cosmological community in this direction.

\section{Detailed requirements}\label{sec:detailedreq}

This section contains detailed requirements on
systematic uncertainties,
broken down by DESC dark energy probe.  These requirements are derived through a process of {\em
  error budgeting}.  Our total error budget for
calibratable systematic uncertainties that would
enable the DESC to meet its high-level requirements is allocated among calibratable systematic
effects in order to derive detailed
requirements on the treatment of each one.  We must make choices about how much of the error
budget to allocate to effects that are under our control; these allocations may change in future as we learn
more about various sources of systematic uncertainty.  Since the error budgeting process will be
affected by improvements in our understanding and analysis methods and by the inclusion of
requirements on model sufficiency, the detailed requirements (unlike the high-level ones) will
evolve in future versions of the DESC~SRD.

As a reminder of our overall methodology and a guide to the contents of this section, we note that
(as detailed in \autoref{sec:intro} and \autoref{sec:definitions}), we consider two categories of
systematic uncertainties for each probe: self-calibrated systematics (for which we typically have
uninformative priors on nuisance parameters) and calibratable ones (for which DESC simulations,
theory, other software, or precursor datasets produce informative priors).  While both types of
systematic uncertainties could be mitigated using new ancillary datasets, we only consider what can
be gained from LSST data, precursor and planned ancillary datasets, and follow-up at rates
consistent with what can be obtained now.

Consequently, we assume conservative priors for self-calibrated systematics,
and {\em only place requirements on calibratable systematic uncertainties.}
If new external data unexpectedly become available, it will be folded
into the joint likelihood, and will improve our ability to marginalize over the nuisance parameters
of both types of systematic uncertainties, reducing our overall error budget. Given a better model
for a given source of calibratable systematic uncertainty, it might be moved into the
self-calibrated category, which would change the way it is treated with respect to detailed
requirements below.  In particular, it would no longer have an associated detailed requirement, and
instead would increase the marginalized statistical uncertainty, which would have the additional
impact of increasing the tolerances for the remaining sources of calibratable systematic uncertainty
and hence loosening other requirements.  This tradeoff is acceptable as long as
\ref{high:combinedfom} can still be met.  We also note the need
for model sufficiency to reduce systematic biases, but do not place requirements on model
sufficiency in this DESC~SRD version.

In the subsections below, we place requirements on multiple sources of
calibratable systematic uncertainty.
In general, our approach (as defined by the need to jointly satisfy \ref{high:combinedfom} and
\ref{high:sys}) is to  compute the marginalized systematic uncertainty by
conditioning on fiducial values of the calibratable systematic effects'
nuisance parameters, and then  allocating
some fraction of this marginalized statistical uncertainty across all
sources of calibratable systematic uncertainty.  The fraction $f_\text{sys}$ that is allocated, i.e., the ratio of
calibratable systematic uncertainty to marginalized statistical uncertainty, is determined by
\ref{high:combinedfom} and \ref{high:sys} as follows. Schematically, if we want our overall FoM to be 500, and our FoM with
marginalized statistical uncertainty is FoM$_\text{stat}$, then \ref{high:combinedfom} implies that $f_\text{sys}$ is determined as
\begin{equation}\label{eq:fsys}
\frac{\text{FoM}_\text{stat}}{500} = 1 + f_\text{sys}^2
\end{equation}
because the FoM scales like an inverse variance.  Clearly if $\text{FoM}_\text{stat}$ exceeds 1000,
then $f_\text{sys}>1$, which would cause a violation of \ref{high:sys}, and hence we cap
$f_\text{sys}$ at precisely 1 to jointly meet \ref{high:combinedfom} and \ref{high:sys}.  If
$\text{FoM}_\text{stat}$ is only slightly above 500, we would have little room for
systematic uncertainty, and the requirements would be extremely tight.  In practice, the
$f_\text{sys}$ determination is a bit more subtle than \autoref{eq:fsys} implies, for two reasons.  First, the
requirement that the overall FoM be 500 includes Stage III priors, which do not get degraded by
the LSST systematic uncertainty.  Accounting for this involves degrading the Fisher matrices for the
DESC probes by the above factor and combining with our Stage III priors, optimizing $f_\text{sys}$
until \autoref{eq:fsys} is met\footnote{There is yet another subtlety, which is that degrading
  the individual Fisher matrices is not quite the right thing to do; the systematic uncertainties
  may have a different direction in the 7-dimensional cosmological parameter space.  We defer
  consideration of this effect to future versions of the DESC~SRD.}.
Second, the above discussion presumes that all probes will have the
same value of $f_\text{sys}$.  While this may be a reasonable default, preliminary calculations with
this assumption  resulted in unachievably stringent photometric calibration requirements for the
supernova science case.  As a result, we gave a slightly larger fraction of the systematic error
budget to supernovae, and lower fractions for all other probes: $f_\text{sys}^{(\text{SN})}=0.7$,
and $f_\text{sys}^{(\text{non-SN})}=0.62$, again optimizing using the appropriate
generalization of \autoref{eq:fsys}.

Once we determined the overall systematic uncertainty fraction for each probe, we then considered
all sources of calibratable systematic uncertainty, and divided them up based on quadrature
summation to the probe-specific $f_\text{sys}$ value.  This process results in a set of Y10
requirements, as described in \autoref{sec:highlevelreq} below \ref{high:sys}.  The Y10 error budgeting
process described here is illustrated in \autoref{fig:errorbudget}. For Y1 goals, the
process is slightly different, since \ref{high:combinedfom} does not apply, only \ref{high:sys}.
Hence we use $f_\text{sys}^{(\text{Y1})}=1$ to set the overall size of the total calibratable systematic
error budget for all probes in Y1, while keeping the same breakdown between
different sources of systematic uncertainty for a given probe as for Y10.

The mathematical implications of the adopted $f_\text{sys}$ values are described in
\appref{app:requirements}. We define $N_\text{class}$ classes of
calibratable systematic uncertainty for
each probe, with our current understanding of the tall poles in each analysis being used to define
major/minor classes that should get a larger/smaller fraction of that error budget.  For example, if
$N_\text{class}=2$ then the more major one might get $0.8f_\text{sys}$ and the more minor one
$0.6f_\text{sys}$ (note that 0.8 and 0.6 add in quadrature to 1).  Each class is thus given a fraction $f_\text{class}f_\text{sys}$.  Within each
class, there might be $N_\text{sub}$ sources of uncertainty that contribute; these each get
$f_\text{class}f_\text{sys}/\sqrt{N_\text{sub}}$ of the error budget.
A crucial assumption here is that there is no covariance between systematic offsets. For example, it
is imaginable that the error in mean photometric redshifts could correlate with the error in the
redshift scatter determination. Given that the sign of cross-probe correlations can be either
positive or negative (i.e.\ making results better or worse compared to quadrature addition), we
proceed with this assumption and will, if necessary, modify our parametrization in the future so
that individual contributions will be roughly uncorrelated, or properly account for correlations as
needed.
In general, we include in the tally of
$N_\text{class}$ and $N_\text{sub}$ all sources of
calibratable systematic uncertainty outlined in
\appref{app:baselines} for a given probe, even those for which we do not yet have the infrastructure to set
requirements now.  This means that in future DESC~SRD versions we will not have to revise the
fraction of the error budget given to sources of calibratable systematic uncertainty for which requirements
already exist when we add requirements on new sources of systematic uncertainty.  This statement is
only true to the extent that the different contributors to each class of systematic uncertainty are
independent of all others (within that class or otherwise).

We emphasize here that there are several layers of subjective choices in the error budgeting beyond what is
deterministically specified by our high-level requirements.  These include whether to give each
probe the same calibratable systematic error budget (specified as a fixed fraction of its statistical error
budget), and how to divide up the error budget amongst the different sources of calibratable systematic
uncertainty.  This feature of our error budgeting provides flexibility, should some of our detailed
requirements prove difficult to meet even given a reasonable amount of additional resources (which would be the first
avenue to meeting challenging requirements).  In short, the paths to dealing with tight
requirements on individual sources of systematic uncertainty are (a) devote additional resources to
the problem, (b) re-budget within different sources of uncertainty for a given probe to give more
room for this source of systematic uncertainty, (c) re-budget
the calibratable systematic error budget across probes, and finally (d) re-think where our constraining power is
coming from across all probes, potentially changing analysis methods in ways that enable
requirements to be loosened.

Several of our technical appendices summarize information and methodology that went into the
detailed requirements enumerated below.
The full list of self-calibrated and calibratable systematic uncertainties that should be considered
for each probe is given in \appref{app:baselines}, including both the subset that we can
currently model and/or place requirements on, the current parametrization, and future improvements.
A synthesis of the calibratable effects on which we place requirements across probes is in
\autoref{subset:synth:thisversion}.  As DESC software pipelines evolve, future DESC~SRD
versions will naturally be able to describe requirements on additional effects.  Finally, the
details of how requirements were defined are described in \appref{app:requirements}.
Several plots that illustrate key aspects of the results in this section are
in \appref{app:plots}.

\subsection{Large-scale structure}\label{subsec:lssreq}
\renewcommand{\probe}{LSS}
\setcounter{requirementd}{1}

Here we derive requirements for the galaxy clustering measurements (\appref{app:lss}), which
carry information about structure growth and the expansion history of the Universe.

The baseline galaxy clustering analysis used here involves tomographic clustering (auto-power
spectra only) across a wide
range of spatial scales.  This baseline analysis does not include explicit
measurement of the baryon acoustic oscillations (BAO) peak, but BAO information is implicitly
included (albeit suboptimally) by extending the power spectrum measurements to $\ell_\text{min}=20$.
In addition to
the statistical uncertainties,
our cosmological parameter
constraints incorporate additional uncertainty due to marginalization over a model for galaxy bias with
one nuisance parameter per tomographic bin.  Following \ref{high:indivfom}, our target FoM for this
analysis after Y10 is 1.5; the forecast FoM with statistical and self-calibrated systematic
uncertainties after Y1 and Y10, with informative priors on the non-($w_0,w_a$) subset of the space
(see \appref{app:requirements}), is 13 and 14, respectively.  The similarity of these
two numbers results from the current design of the baseline analysis for LSS being suboptimal in a way that
prevents it from benefiting from the increase in constraining power of the survey as time proceeds;
this should be improved in future versions of the DESC~SRD.
If we achieve
\ref{high:sys}, then
inclusion of calibratable systematic uncertainties should multiply these numbers by a
factor of 0.72 in Y10, which comes from the $1/(1+(f_\text{sys}^{(\text{non-SN})})^2)=1/(1+0.62^2)$ factor motivated in the introduction to
\autoref{sec:detailedreq}.  In practice the FoM reduction is not as severe as that, since it does not
apply to the Stage III priors.

We define two classes of calibratable systematic uncertainty for LSS measurements, as described in
\appref{app:lss}: redshift and number density uncertainties.  The total calibratable systematic
uncertainty for LSS split into 0.8 and 0.6 for the two categories, respectively.  These numbers are
chosen such that the quadrature sum is 1, but the redshift uncertainties (which are expected to be
more challenging to quantify and remove) are given a greater share of the error budget.  In this
version of the DESC~SRD, we do not place any detailed requirements on number density uncertainties, and only
place requirements on two out of six contributors (see \autoref{fig:lsssys-cal}) to the redshift uncertainties: the uncertainty
associated with the mean redshift $\langle z\rangle$ in each tomographic bin, and the uncertainty in
the width of the redshift distribution in the tomographic bin (presumed to be identical for each
bin, modulo a standard $1+z$ factor).  Hence there are two {\probe} requirements below,  and both
effects are allowed to contribute a fraction equal to $0.8/\sqrt{6}\sim
0.3$ of the total calibratable systematic uncertainty.  Here the $\sqrt{6}$ indicates that
eventually we will place requirements on a total of six sources of redshift uncertainty, allocating
the total redshift uncertainty budget to each one equally.  Finally, as noted previously, the total
calibratable systematic uncertainty is allowed to be a factor of
$f_\text{sys}^{(\text{non-SN})}=0.62$ times the marginalized statistical uncertainty in Y10.

\labelreqd{reqd:lssmeanz}{the mean redshift of each tomographic bin}{$0.005(1+z)$}{$0.003(1+z)$}

The above requirement was determined by coherently shifting the mean redshift of all
tomographic bins by the same amount, resulting in tighter requirements than when
considering shifts in individual bins, but looser requirements than if we had included a pattern
specifically chosen to mimic the impact of dark energy on the tomographic galaxy power spectra.
While tighter than what is routinely achieved by existing surveys
\citep[e.g.,][]{2017arXiv171002517D}, which are currently limited by systematic uncertainties that
will require additional work to overcome, these requirements are well above the $0.0004(1+z)$ accuracy that should be achievable through
cross-correlation analyses with a DESI-like survey covering the full LSST footprint
\citep{2015APh....63...81N}.  With the
expected 4000 square degrees of overlap between LSST and DESI, combined with the more-dilute 4MOST
galaxy and quasar samples covering the remainder of the imaging area, the expected accuracy will be
worse than this by a factor of $\sqrt{2}$ or less for Y10, still well within the requirements.

\labelreqd{reqd:lssscatterz}{the photometric redshift scatter $\sigma_z$}{$0.1(1+z)$}{$0.03(1+z)$}

The above requirement was determined by computing a data vector in which we coherently broadened the photometric redshift scatter while
computing model predictions with the original baseline photometric redshift scatter.

\subsection{Weak lensing (3$\times$2-point)}\label{req:wl}
\renewcommand{\probe}{WL}
\setcounter{requirementd}{1}

Here we derive requirements for the weak lensing ($+$LSS, i.e., 3$\times$2-point) measurements (\appref{app:wllss}), which
carry information primarily about structure growth, with a small contribution from the expansion
history.

The baseline weak lensing analysis in this
version of the DESC~SRD involves tomographic shear-shear, galaxy-shear, and galaxy-galaxy power
spectra across a wide range of spatial scales.  Cross-bin correlations are included for shear-shear
and shear-galaxy power spectra, while only auto-power spectra are included for galaxy-galaxy.
In addition to pure statistical errors, our
cosmological parameter constraints incorporate additional uncertainty due to marginalization over a
model for galaxy bias with one nuisance parameter per tomographic bin, and
due to
intrinsic alignments with
four nuisance parameters overall. Following \ref{high:indivfom}, our target FoM for this analysis
after Y10 is 40; the forecast FoM with statistical and self-calibrated systematic uncertainties
after Y1 and Y10 is 37 and 87, respectively.  If we achieve \ref{high:sys}, then inclusion of
calibratable systematic uncertainties will multiply these numbers by a factor of $\sim$0.72 (see
\autoref{subsec:lssreq} for details) for Y10, which still enables us to meet \ref{high:indivfom}.
 Note that if we consider just the shear-shear
contribution to the dark energy constraining power, the forecast FoM with statistical and
self-calibrated systematic uncertainties after Y1 and Y10 is 19 and 52, respectively. The reason to
consider the shear-shear aspect of the analysis separately is that in practice we begin by
separately analyzing shear-shear versus galaxy-galaxy correlations to ensure consistent results.

There are four classes of calibratable systematic uncertainty for this analysis, as described in
\appref{app:wllss}: redshift, number density, multiplicative shear, and additive shear
uncertainties.  We allocate 0.7, 0.2, 0.7, and 0.2 of the total calibratable systematic error budget
to these categories, respectively.  These numbers are chosen such that the quadrature sum is 1;
the two categories that are given less
room in the error budget are more easily diagnosable directly through null tests on the data.
Finally, as in \autoref{subsec:lssreq}, the total
calibratable systematic uncertainty is allowed to be a factor of
$f_\text{sys}^{(\text{non-SN})}=0.62$ times the marginalized statistical uncertainty in Y10.

We place requirements on two out of seven sources of systematic uncertainty associated with
redshifts (see \autoref{fig:wlsys-cal}): the uncertainty associated with the mean source redshifts $\langle z\rangle$ in
each bin and the uncertainty in the photometric redshift scatter (presumed to be identical in each
bin, modulo a standard $1+z$ factor). These are each allowed to contribute a fraction equal to
$0.7/\sqrt{7}\sim 0.25$ times the total calibratable systematic uncertainty. We also place
requirements on our overall knowledge of multiplicative shear calibration (0.7 of the total uncertainty), as well
as derived requirements on (a) our knowledge of PSF model size errors, and (b) stellar contamination
in the source galaxy sample.  Hence there are five {\probe} requirements below.  Requirements on control of additive shear systematic biases are deferred to future
DESC~SRD versions.  In general, the sensitivity to additive shear biases depends on their
scale dependence and how well it mimics changes in scale dependence due to changes in cosmological
parameters; hence more meaningful requirements will be placed after we have
templates for the scale dependence of the relevant systematics effects.

\labelreqd{reqd:wllssmeanzs}{the mean redshift of each source tomographic bin}{$0.002(1+z)$}{$0.001(1+z)$}

The above requirement was determined by coherently shifting the mean redshift of all source
tomographic bins by the same fraction for the shear-shear analysis.  Currently the analysis setup
for 3$\times$2-point does not allow separate consideration of biases in the lens and source
populations, so we rely on \ref{reqd:lssmeanz}~for lens sample requirements on knowledge of ensemble
mean redshifts and \ref{reqd:wllssmeanzs} for source sample requirements on knowledge of ensemble
mean redshifts (considered entirely separately).

Because of the relatively larger constraining power in this measurement, this requirement is
stricter than \ref{reqd:lssmeanz}.  The magnitude of this requirement on
the systematic uncertainty in the mean redshifts is comparable to those
forecast by \citet{2006ApJ...636...21M}, who use different default analysis assumptions but noted
that the requirements on knowledge of redshift distribution parameters are especially tight when
forecasting requirements with $w_a \ne 0$, i.e., in the $(w_0,w_a)$ space rather than assuming a
constant dark energy equation of state.  Nonetheless, per discussion following \ref{reqd:lssmeanz}, the
magnitude of the Y10 requirement in \ref{reqd:wllssmeanzs} should be within reach for cross-correlation-based calibration
methods alone.

\labelreqd{reqd:wllssscatterz}{the source photometric redshift scatter $\sigma_z$}{$0.006(1+z)$}{$0.003(1+z)$}

The above requirement was determined by computing a data vector in which we coherently broadened the photometric redshift scatter while
computing model predictions with the original baseline photometric redshift scatter.
This was done specifically for shear-shear, since the analysis setup does not currently enable us to
separately vary the lens and source photo-$z$ scatter values for the 3$\times$2-point analysis. This requirement is substantially more stringent than the requirements for a
clustering-only analysis \ref{reqd:lssscatterz}, reflecting the greater statistical power in the
weak lensing analysis.

\labelreqd{reqd:wllssshearm}{the redshift-dependent shear calibration}{$0.013$}{$0.003$}

The assumption behind this requirement is that the DESC will carry out its cosmological weak lensing
analysis using shear catalogs provided by Rubin Observatory, but will use its own software to
quantify and remove
any redshift-dependent calibration biases in the ensemble shear signals, and to place bounds on the residuals.  This requirement is therefore on our knowledge of the shear calibration: how well can we
constrain {\em the sum of all effects that cause uncertainty in the redshift-dependent shear
  calibration}, i.e., the residual calibration bias after subtracting off known effects?  (For
a listing of all effects implicitly included, see \appref{subsubsec:wl-sysuncert}.)
This requirement was placed based on the 3$\times$2-point analysis, though shear-shear requirements
are only slightly larger.

The canonical requirement on residual shear calibration that is often quoted in the literature for Stage-IV
surveys, $\Delta m \lesssim 0.002$, comes from \citet{2013MNRAS.429..661M}.  The Euclid forecasts in that work
naturally differ from these forecasts in basic survey parameters (Euclid vs.\ LSST) and use of
shear-shear only, but also in basic methodology: they use Gaussian rather than non-Gaussian
covariances; they do not marginalize over intrinsic alignments; and they define shear
calibration requirements with $r\approx 0.15$ (\autoref{eq:requirement} on page \pageref{eq:requirement}) rather than $0.4=0.7f_\text{sys}^{(\text{non-SN})}$ as
we have done here. Use of shear-shear alone may make their requirements marginally less stringent than ours, while
all the other differences should make them more stringent.  In short, it may be surprising that the
requirements for Euclid and for LSST Y10 agree so well.  In the context of the field, the best
state-of-the-art methods can already achieve uncertainty on $\Delta m=5\times 10^{-3}$ in the simplest
scenario, {\em
  without} accounting for all sources of systematic uncertainty that we are including in this requirement (e.g.,
blending effects tend to lead to larger shear calibration uncertainties than this).  Hence meeting this
requirement requires some improvement on the current state of the art to tackle specific
contributors to uncertainties on shear calibration
that are less well understood such as blending -- but does not require an order of magnitude
improvement and hence is likely to eventually be achievable with variants of the existing state of
the art.

\labelreqd{reqd:wllsspsfsize}{the PSF model size defined using the trace of the second moment matrix}{0.4\%}{0.1\%}

It is well known \citep[e.g.,][]{2004MNRAS.353..529H} that biases in the PSF model size can cause a
coherent multiplicative bias in the weak lensing shear signals.  While the LSST~SRD places explicit
requirements on how well the PSF model {\em shapes} are known, it places no requirement on PSF
model size (except the indirect and non-quantitative constraint that most algorithms that can
accurately infer the PSF model shape also estimate the PSF model size fairly accurately).
Fortunately, there are well-established null tests that can uncover the presence of PSF model size
residuals in the real data \citep[e.g.,][]{2016MNRAS.460.2245J,2018PASJ...70S..25M}.  DESC pipelines
will use those null tests along with our analysis of image simulations to constrain the magnitude of
PSF model size residuals; the above requirement is on our knowledge of those residuals.
Many physical effects can cause PSF model size errors; this version of the DESC~SRD does not
drill down to place separate requirements on each of those effects.  While the exact magnitude of
the shear calibration bias induced by a PSF model size error depends on the size of the galaxy
population compared to the PSF, to within a factor of $\sim 2$ it is typically the case that the
shear calibration bias is set by the size of the typical fractional PSF model size error (i.e., $\delta
T_\text{PSF}/T_\text{PSF}$, where $T_\text{PSF}$ is the trace of the moment matrix of the PSF and
hence is related to the area covered by the PSF).

This requirement was derived without additional forecasts; rather, it comes from
\ref{reqd:wllssshearm}, along with the aforementioned formalism for estimating how PSF model size
residuals propagate directly into shear calibration biases.  Since there are many effects that can
contribute to shear calibration bias (of order ten) we allocate $1/\sqrt{10}$ of the shear
calibration bias error budget to PSF model size uncertainty.

\labelreqd{reqd:wllssstars}{the stellar contamination of the source sample}{0.4\%}{0.1\%}

Inclusion of stars in the {\probe} source sample can, if unrecognized, cause a dilution of the
estimated shear signal that is directly related to the fraction of the sample that is stars\footnote{Or rather,
  the total {\em weighted} stellar contamination fraction for whatever weighting scheme is used to
  infer the ensemble weak lensing shear.},
because the stars contribute zero shear
signal.  Hence our overall requirement on shear calibration \ref{reqd:wllssshearm} can be translated
directly into a requirement on how well we have quantified the redshift-dependent contamination of the source sample by
stars.  Similarly to \ref{reqd:wllsspsfsize}, we allocate $1/\sqrt{10}$ of the
shear calibration bias uncertainty to stellar contamination.

\subsection{Galaxy clusters}\label{subsec:cl-req}
\renewcommand{\probe}{CL}
\setcounter{requirementd}{1}

Here we derive requirements for the galaxy clusters analysis (\appref{app:cl}), which carries information about
structure growth.

The baseline galaxy clusters analysis in this
version of the DESC~SRD involves tomographic cluster counts and stacked cluster WL
profiles in the 1-halo regime.  In addition to pure statistical errors, our cosmological parameter
constraints incorporate marginalization over a relatively flexible parametrization of the cluster
mass-observable relation (MOR).  Following \ref{high:indivfom}, our target FoM for this analysis
after Y10 is 12; the forecast FoM with statistical and self-calibrated systematic uncertainties
after Y1 and Y10 is 11 and 22, respectively.  If we achieve \ref{high:sys}, then as described in
\autoref{subsec:lssreq}, inclusion of
calibratable systematic uncertainties will multipy these numbers by a factor of $\sim$0.72 (see
\autoref{subsec:lssreq} for details) in Y10, which
enables us to meet \ref{high:indivfom}.

There are four classes of calibratable systematic uncertainty for this analysis, as described in
\appref{app:cl}: redshift, number density, multiplicative shear, and additive shear
uncertainties.  We allocate 0.7, 0.2, 0.7, and 0.2 of the total calibratable systematic uncertainty
to these categories, respectively.  These numbers are chosen such that the quadrature sum is 1; the
two categories that are given less
room in the error budget are more easily diagnosable directly through null tests on the data.
We place requirements on two out of seven sources of systematic uncertainty associated with
redshifts (see \autoref{fig:clsys-cal}): the uncertainty associated with the mean source redshifts $\langle z\rangle$ in
each bin, and the uncertainty in the source redshift bin width (presumed to be identical in each
bin, modulo a standard $1+z$ factor). These are each allowed to contribute a fraction equal to
$0.7/\sqrt{7}\sim 0.25$ times the total calibratable systematic uncertainty. We also place
requirements on our overall knowledge of shear calibration (0.7 of the total calibratable systematic
uncertainty).  Hence there are three {\probe} requirements below. Note that as in \autoref{subsec:lssreq}, the total
calibratable systematic uncertainty, for which we have just described its detailed allocation
between effects, is allowed to be a factor of
$f_\text{sys}^{(\text{non-SN})}=0.62$ times the marginalized statistical uncertainty in Y10.

\labelreqd{reqd:clmeanz}{the mean redshift of each source tomographic bin}{$0.008(1+z)$}{$0.001(1+z)$}

Like \ref{reqd:wllssmeanzs}, the above requirement was determined by coherently shifting the mean
redshift of all source
tomographic bins by the same amount.
This requirement is comparable to the corresponding requirement for WL for Y10,
\ref{reqd:wllssmeanzs}, despite differences in cosmological constraining power.

\labelreqd{reqd:clscatterz}{the source photometric redshift scatter}{$0.02(1+z)$}{$0.005(1+z)$}

Like \ref{reqd:wllssscatterz}, the above requirement was determined by computing a data vector in which
we coherently broadened the photometric redshift scatter for all source tomographic bins while
computing model predictions with the original baseline photometric redshift scatter.

\labelreqd{reqd:clshearm}{the redshift-dependent shear calibration}{0.06}{0.008}

As for WL, the assumption behind this requirement is that the DESC will carry out its cosmological
weak lensing analysis using shear catalogs provided by Rubin Observatory, but will use its own
software to remove any redshift-dependent calibration biases in the ensemble shear signals and to place bounds on any
residual calibration biases.  This requirement is therefore on our knowledge of the
redshift-dependent shear
calibration.

\ref{reqd:clshearm} is weaker than the corresponding shear calibration requirement for WL,
\ref{reqd:wllssshearm}.  Thus any associated requirements defined in \autoref{req:wl} will also be more stringent
than similarly derived requirements for CL, so we do not proceed to define requirements related to
knowledge of PSF model size and stellar contamination in the source sample for CL analysis.

\subsection{Supernovae}\label{subsec:sn}
\renewcommand{\probe}{SN}
\setcounter{requirementd}{1}

Here we derive requirements for the supernova analysis (\appref{app:sn}), which carries
information about the expansion rate of the Universe.  The detailed requirements presented in this
subsection are directly connected to several requirements in the LSST Project~SRD, as will be explicitly
noted below.  Rubin Observatory is responsible for many aspects of photometric calibration,
combining information from the in-dome hardware, other system diagnostics,
auxiliary telescope data, and the raw science images.
As in the rest of this document, we assume that the basic photometric dataset provided by the LSST Facility
will meet the requirements of the LSST Project~SRD.
Where the detailed DESC requirements derived in this subsection are more stringent than those in the LSST Project~SRD,
the implication is that DESC will need to provide additional resources and expertise, and deploy them in
close collaboration with LSST Facility staff, in order to achieve a more precise photometric calibration.
Depending on the factors that limit the photometric calibration, this may not be achievable in
practice; the LSST SRD requirements were used to set hardware requirements and inform hardware
design, resulting in fundamental limitations in some aspects of the system.  In that case, what is
needed in practice is for the DESC analysis methods to improve (e.g., by updating modeling methods
such that any of the DESC probes becomes more constraining, leaving more room in the error budget
for the systematic uncertainty associated with photometric calibration).
After presenting all the detailed requirements in
this section, we
will briefly summarize which of them relate to aspects of photometric calibration that are
carried out by the Project, and outline the strategy for meeting them.

The baseline supernova analysis in
this version of the DESC~SRD includes supernova samples derived from both the WFD and the DDF, with
conservative estimates of supernova numbers based on simulations that incorporate the
\texttt{minion\_1016}\footnote{\url{https://www.lsst.org/scientists/simulations/opsim/opsim-v335-benchmark-surveys}} cadence strategy and with plausibly achievable numbers of host spectroscopic
redshifts. Currently we neglect the cosmological constraining power of those supernovae for which
host spectroscopic redshifts cannot be obtained, relying on them purely for building templates and
constraining models for astrophysical systematic uncertainties.  The forecasts include
marginalization over several self-calibrated systematic uncertainties associated with
standardization of the color-luminosity law (including redshift dependence), intrinsic scatter, and
host mass-SN luminosity correlations.  Following \ref{high:indivfom}, our target FoM for supernova analysis after Y10 is 19; the forecast FoM with statistical and self-calibrated systematic
uncertainties after Y1 and Y10 is 44 and 211, respectively. As in the previous subsections,
 these include informative Stage III priors on the non-($w_0,w_a$) subset of the parameter space,
 which is particularly important due to the $\Omega_m$ vs.\ $w_a$ degeneracy for the supernova
 constraints and hence substantially increase the FoM. If we achieve \ref{high:combinedfom} and \ref{high:sys}, then
inclusion of calibratable systematic uncertainties will multipy these numbers by a factor of $\sim 1/(1+(f_\text{sys}^{(\text{SN})})^2) \approx 0.67$ (see
\autoref{subsec:lssreq} for details) in Y10.  This which would still enable us to meet \ref{high:indivfom}.

There are two classes of calibratable systematic uncertainty for {\probe} as described in
\appref{app:sn} and shown in \autoref{fig:snsys-cal}: flux measurement calibration and identification uncertainties. While in
general one would include redshift error, we ignore this as we assume that each supernova has an
identified host with spectroscopically determined redshift. We allocate 0.95 and 0.3 of the
calibratable systematic error budget to these classes (with the constraint that their quadrature sum
is 1 and that calibration takes up the largest fraction of the budget because many factors
contribute to it).

We then distribute the systematic uncertainty associated with photometric calibration such that the
largest fraction of the error budget is given to the source of systematic uncertainty that will most
affect cosmology: the zero point uncertainty in each band. We allocate 0.69 of the total systematic
uncertainty to zero point uncertainties, and a further 0.39 to the filter mean wavelength
uncertainties. In order to account for the fact that the systematic zero point or mean wavelength
uncertainties may differ in each band, we draw zero point offsets or mean wavelength offsets from a
4-dimensional normal distribution with a standard deviation set by the magnitude of the systematic uncertainty in
either wavelength or zero point.  There are only 4 bands because all cosmological constraining power
comes from $griz$ only\footnote{The initial forecasts were carried out with $ugrizy$, but comparison
of results with $griz$-only calculations indicated that $u$ and $y$ provide negligible cosmological
information and hence are neglected for the rest of this work.}.  The fact that we can use $griz$-only is beneficial because $u$-
and $y$-band come with additional calibration challenges.
The bias in the observable quantity $\mu$, $\Delta\mu$, is computed from the vector sum of these per-band
biases. We then Monte Carlo over this space to determine the covariance in $(w_0,w_a)$ space due to the
systematic uncertainties in zero point and mean wavelength. Given the quadratic relationship between
the magnitude of the zero point/wavelength offset and the systematic covariance, one can use the
`allowed' systematic error fraction to set requirements on the zero point or mean wavelength
uncertainty.  In our case, we found that naively allowing equal contributions to the systematic
uncertainty from all calibration uncertainties using this process resulted in unachievably tight
requirements on the zero point and mean wavelength uncertainty, so we set minimum `floor'
values (see Y10 requirements below), and that is what determined the numbers 0.69 and 0.39 given
earlier in this paragraph.

This leaves a remaining allowed systematic uncertainty of $\sqrt{1-0.3^2 - 0.69^2 - 0.39^2}/\sqrt{5}
= 0.24$ for each of the other five sources of calibration uncertainty.  While two of these are
currently unmodelled (nonlinearity and wavelength-dependent flux calibration), the constraint on the
allowed uncertainty above can be translated to three of the remaining source of photometric
calibration error as they affect the light curve quality: wavelength-dependent flux calibration; SN light curve modeling; and Milky Way
extinction corrections.
Finally, we emphasize that in addition to the error budgeting within the supernova systematic
uncertainties, there is the factor of $f_\text{sys}^{(\text{SN})}=0.7$ described above in order to
satisfy our high-level requirements.  Hence we impose an additional scaling of 0.7 to all
requirements in Y10. See \appref{app:plots} for a plot illustrating
how these requirements were set, where the relevant numbers come from, and where the systematics
trend lines cross the $r=0.34$ line for Y1 and $r=0.24=0.34\times0.7$ line for Y10 (with $r$ defined as in \autoref{eq:requirement}).

The first two requirements below, \ref{reqd:snfilterzp} and~\ref{reqd:snfiltermean}, depend on
observations of standard stars by LSST.

\labelreqdcomment{reqd:snfilterzp}{the $griz$-filter zero points}{5~mmag}{1~mmag}{As the $griz$ requirements represent an ambitious improvement versus the LSST~SRD (5
  mmag in $griz$), an alternative way to meet this requirement is to improve our analysis methods
  for all probes until the LSST~SRD requirement is sufficient.}

By ``zero point
uncertainty'', we mean the difference between the synthetic brightness prediction obtained by
integrating the spectra of calibrated standard stars (e.g., HST `Calspec' standards) through the LSST
passbands, and the observed LSST magnitudes. Relative zero-point and astrometric corrections are
computed for every visit. Sufficient data are kept to reconstruct the normalized system response
function (see Eq.~5, LSST~SRD) at every position in the focal plane at the time of each
visit as required by Section~3.3.4 of the LSST~SRD.
\ref{reqd:snfilterzp} puts strong constraints on (1) the accuracy
of the primary flux reference, and (2) the metrology chain, i.e., the chain of flux measurements
that links the objects on one image to observations of the primary flux reference.

Table~16 of the LSST~SRD gives design specifications of 5~mmag for filter zero points except for
$u$-band.  Improvement beyond that level in
$griz$ in later years is primarily a question of resources rather than intrinsic hardware
limitations (unlike for $y$-band, which we have not used for SN).
Doing so will require the DESC to further constrain the
residuals through some other method, such as linking the calibration of the sources to GAIA
observations. We expect that using the GAIA Bp/Rp catalog as an external anchor, the uniformity of
the LSST measurements may be controlled at the per-mil level. Crucially, the zero point calibration should be valid over a broad color range
(e.g., $0.5<g-i<3$).  Note that the $griz$ design specifications are
comparable to our Y1 $griz$ goal.

Given the repeat observations required to build up a light curve over time, and the need to have a
calibrated dataset across the sky, the LSST~SRD requirements in Table~14 (specifications for
photometric repeatability) and Table~15 (specifications for spatial uniformity of filter zero
points) have impact on the DESC SN science case. These are given as 5, 15 mmag for PA1 and PA2 for
the repeatability, and 5, 10~mmag for PA3 and PA4 respectively for spatial uniformity.
Our ability to calibrate the LSST photometric system for the supernovae depends on the number of standard stars used for calibration, as any systematic uncertainty related to spatial uniformity reduces as $\sqrt{N_\text{standards}}$. Observing multiple standard stars over the field of view is therefore central to achieving our calibration goals while staying within LSST~SRD requirements.

Regarding photometric repeatability, we expect that our strategies to deal with zero point fluctuations will be beneficial in improving repeatability.  Our simulations did not include spatial variation in
zero point fluctuations and hence we cannot comment on the quantitative benefit of the LSST~SRD
requirements on spatial uniformity in this version of the DESC~SRD.

\labelreqd{reqd:snfiltermean}{the $griz$-filter mean wavelength}{6~\AA{}}{1~\AA{}}

The very large size of the LSST SNIa sample sets very strong
requirements on the calibration systematics, including these filter
mean wavelength uncertainties. The wavelength uncertainty of 1\AA\ for Y10 is a strict requirement,
but this currently assumes that the errors are not mitigated through light curve modelling. Characterizing
filter mean wavelengths at the level of 1--2~\AA\ is well within reach of current metrology techniques.
In future simulation-based investigations, it will be valuable to explore the joint marginalization
over this systematic uncertainty along with others, rather than considering it completely
independently as we have done now.

Unlike \ref{reqd:snfilterzp}, \ref{reqd:snfiltermean} has no direct analog in the LSST~SRD.

\labelreqd{reqd:snglobalcal}{the wavelength-dependent flux calibration}{a slope of 5~mmag per
  5500~\AA\ in wavelength}{a slope of 4.4~mmag per 5500~\AA\ in wavelength}

This source of systematic uncertainty relates to our knowledge of the wavelength-dependent flux
calibration of the external photometric system to which we tie LSST's calibration (e.g., the HST
photometric system). This calibration extends over the entire wavelength range considered. In this analysis we restrict ourselves to considering only the $griz$ bands, hence we require this slope on the external calibration over $\simeq 5500$~\AA.

The LSST~SRD has an overall 10~mmag calibration requirement. The
supernova dark energy science case is not sensitive to an overall calibration offset, as this is
degenerate with the instrinsic magnitude of the supernova population (or alternatively the Hubble
constant).  However, the calibration of the LSST photometric system to the standards \citep[e.g., the
  HST standards][]{2014arXiv1403.6861B} cannot vary in a wavelength-dependent way by more than
2.2~mmag per 7000~\AA\ for the Y10 survey.
This requirement is linked to \ref{reqd:snfilterzp} in that the zero point calibration allows the
LSST SNe to map onto the low-$z$ sample, while \ref{reqd:snglobalcal} relates to our ability to
relate the LSST supernovae (and the low-$z$ sample itself) to the standard stars \citep[see
  e.g.][]{2015ApJ...815..117S}.  However, unlike \ref{reqd:snfilterzp}, \ref{reqd:snglobalcal} has
no direct analog in the LSST~SRD.

Our ability to meet this requirement depends on how well the standard stars are modeled.
The extent to which the wavelength-dependent flux calibration is required also depends on how
low-redshift samples are included in any analysis. In this simulation, an independent low-$z$ sample was included, and the SALT model uncertainty and HST calibration uncertainty were varied for this low-$z$ sample. Care will have to be taken to ensure calibration between high-quality low-$z$ surveys and the LSST sample. This is an active area of research, and future studies will investigate the optimal combinations of current and future low-redshift data to anchor the LSST Hubble diagram.

\labelreqd{reqd:snlightcurve}{the light curve modeling}{22\% of current SALT2 model errors}{3\% of
  current SALT2 model errors}

This requirement is on our uncertainty in the SALT2 light curve model due to calibration
uncertainties in and statistical limitations of the training sample.
Improving the error on the SALT2 model will be critical especially for the Y10 LSST SN analysis;
there is a clear path to dealing with this systematic uncertainty by using some of the
LSST sample to retrain the SALT2 models. The training sample we will obtain from LSST observations is expected to be 10 to 50 times larger than the current SALT2 training sample. This
will permit us to significantly decrease the statistical uncertainty
affecting the light curve model. Futhermore, this will allow us to capture
more of the SN variability than currently captured by SALT2.  Finally, the calibration requirements
in \ref{reqd:snfilterzp}--\ref{reqd:snglobalcal} will be important in enabling an improved
calibration of the SALT2 model.

\labelreqd{reqd:snmw}{Milky Way extinction corrections}{100\% of current systematic Galactic extinction uncertainties}{30\% of current systematic Galactic extinction uncertainties}

The reason why this source of systematic uncertainty is so important is that the WFD and DDF
locations have different extinction and their supernovae have different redshift distributions.
For context, the Milky Way extinction model is generally determined using data and methods external
to LSST, and the current uncertainty in the normalization of the global Milky Way extinction is $\pm
5$\% \citep{2014ApJ...789...15S}. The Y1 sample does not require any improvement on the current Milky Way extinction model. However, the Y10 sample will require the model of systematic uncertainty related to Galactic extinction to be 30\% of its current value, \citet{2014ApJ...789...15S} from \autoref{tab:snsys-cal}.

Work that will enable the above requirements to be met occurs in several different contexts.  The first
two requirements in this subsection, \ref{reqd:snfilterzp} (zero points) and \ref{reqd:snfiltermean}
(filter mean wavelengths), relate to
aspects of photometric calibration that are driven by Rubin Observatory.
These detailed requirements are more stringent than their counterparts in the LSST~SRD;
as indicated at the start of this subsection,
by providing additional DESC resources and expertise, and working closely with the LSST Facility, we
aspire to achieve, together, a more precise photometric calibration than the Facility is
{\it required} to produce on its own.
Moreover, the
DESC is actively pursuing
research methods that might result in eventual loosening of the detailed requirements on photometric
calibration.  Some of these are methods that directly relate to photometric calibration; e.g., methods of marginalizing over astrophysical systematics that may be able to
reliably absorb certain photometric calibration uncertainties.  This
is particularly the case for requirements \ref{reqd:snglobalcal}, \ref{reqd:snlightcurve}, and
\ref{reqd:snmw}; \ref{reqd:snglobalcal} partly depends on the photometric calibration provided by
Rubin Observatory and partly on what external datasets that DESC chooses to include in its analysis, while
\ref{reqd:snlightcurve} and \ref{reqd:snmw} are most likely to be met (or loosened) by further
development of modeling methods on the DESC side.  However, even \ref{reqd:snfiltermean}, which is
ostensibly dependent on photometric calibration provided by Rubin Observatory, has the potential to be
loosened in the future given that filter mean wavelength
uncertainties may be absorbed in the light curve model retraining that will be performed for the LSST
supernova sample, and may additionally be absorbed by our techniques for marginalizing over
astrophysical systematic uncertainties.  Progress
on this high-priority work will be reflected in future versions of the DESC~SRD.  However, we
reiterate that the high-level requirement of being a Stage IV dark energy survey ties together all
of the probes, such that even methodological improvements in e.g.\ weak lensing that result in
increased constraining power could enable a relaxation of the photometric calibration requirements
for SN.

Finally, as a consistency check, we note that an independently conducted study carried out within
the DESC Photometric Corrections working group\footnote{F.~Hazenberg, M.~Betoule, S.~Bongard,
  L.~Le~Guillou, N.~Regnault, P.~Gris, {\em et al.}, 2018, ``Impact of the calibration on the
  performances of the LSST SN survey'', DESC
internal note} with
somewhat different methodology, including the retraining of the SALT2 models, came to similar
conclusions as \ref{reqd:snfilterzp} and \ref{reqd:snfiltermean}.

\subsection{Strong lensing}
\renewcommand{\probe}{SL}
\setcounter{requirementd}{1}

This section describes the strong lensing analysis, which yields information about
the expansion rate of the Universe.

As described in \appref{app:sl} in more detail, the baseline strong lensing analysis in this
version of the DESC~SRD includes time delay and compound lens systems, with sample sizes defined
based on conservative assumptions about follow-up resources.  The forecasts include marginalization
over several self-calibrated systematic uncertainties; this marginalization is implicitly done, via
increased uncertainties in the per-lens distance measurements, rather than explicitly via
marginalization over models for those uncertainties. Following \ref{high:indivfom}, our target FoM
for strong lensing after Y10 is 1.3; the forecast FoM with statistical and self-calibrated systematic
uncertainties after Y1 and Y10 is 2.0 and 9.4, respectively.   As in the previous subsections,
 these include informative Stage III priors on the non-($w_0,w_a$) subset of the parameter space. If
 we achieve \ref{high:sys}, then inclusion of calibratable systematic uncertainties will multipy these numbers by a factor of $\sim 0.72$ (see
\autoref{subsec:lssreq} for details) in Y10,  which still enables us to meet \ref{high:indivfom}.

In this version of the DESC~SRD, we do not place requirements on calibratable systematic
uncertainties for strong lensing, because developing models for how those systematic uncertainties
affect the observable quantities is work that will begin during DC2.

\subsection{Combined probes and other requirements}\label{subsec:combprobes}

\renewcommand{\probe}{J}
\setcounter{requirementd}{1}

Now that we have described the detailed requirements and the baseline forecasts for
individual probes, we revisit our requirement on the joint probe constraining power of
LSST. First, we describe how the joint forecast was carried out.  In principle, we would like to
undertake
a full joint forecast with software that is capable of describing the likelihood analyses for all
five probes, and combine all probes at the level of likelihoods.  However, at present three
different software tools are used to produce the forecasts (see \appref{app:software}).
Currently our method for combining them is through Fisher matrix approximations of their posterior
probability distributions for cosmological parameters (ignoring any non-Gaussian distributions).
Future DESC~SRD versions should work at the likelihood level as additional DESC software products
become available.
Following \ref{high:combinedfom}, our target FoM for all probes including Stage III priors and
factoring in all sources of systematic uncertainty is
500.  The forecast joint FoMs with statistical and self-calibrated systematic uncertainties after Y1
and Y10 are 156 and 711, respectively, after including Stage III priors.  See \appref{app:plots} for a plot illustrating the
joint constraining power in the $(w_0,w_a)$ plane. The error budgeting process described in the
preamble of \autoref{sec:detailedreq}, aimed at jointly satisfying \ref{high:combinedfom}
and~\ref{high:sys}, ensures that calibratable systematic uncertainties will lower the Y10
number to something very close to 500 (actually 505 in practice).

In addition, here we include requirements that are not probe-specific.  These relate to our high-level
blinding requirement \ref{high:blinding}.

\labelreqdjoint{det:blindingmethod}{(Y1)}{Blinding methods will involve failsafes to avoid
  accidental unblinding (e.g., redundancy of blinding both summary statistics and cosmological
  parameter plots, use of public key encryption).}

\labelreqdjoint{det:blindingj}{(Y3)}{DESC dark energy analyses will employ blind analysis techniques
  that self-consistently work for individual and joint probe analyses.}

Requirements~\ref{det:blindingmethod} and~\ref{det:blindingj} apply not only to the analyses listed
explicitly (Y1 and Y3, respectively) but to all later analyses.

\section{Conclusion and outlook}

The baseline analysis for all five probes defined in this document represents the first
DESC-wide forecasting exercise, including adoption of common analysis methodology and key
sources of systematic uncertainty across all probes.  In this section we briefly summarize the
key findings from this exercise.

First, the estimated DETF FoM values for each of our baseline analyses and for all probes together,
based on current forecasts,
are summarized in \autoref{tab:fom_summary}.

\begin{table}[!hp]
\begin{center}
\begin{tabular}{|p{0.6in}|p{1.3in}|p{0.7in}|p{0.7in}|p{0.7in}|} \hline
Analysis & Priors & Y1 FoM (ceiling) & Y10 FoM (ceiling) & Target \\
\hline
LSS & Stage III (not $w_0$, $w_a$) & 10 (13) & 10 (14) & 1.5 \\
LSS & None & 6.7 (8.4) & 6.6 (9.1) & 1.5 \\
WL$+$LSS & Stage III (not $w_0$, $w_a$) & 31 (37) & 66 (87) & 40 \\
WL$+$LSS & None & 22 (27) & 49 (68) & 40 \\
CL & Stage III (not $w_0$, $w_a$) & 9 (11)  & 17 (22) & 12\\
CL & None & 6.5 (8.2) & 12 (17) & 12 \\
SN & Stage III (not $w_0$, $w_a$) & 36 (44) & 157 (211) & 19 \\
SN & None & 10 (12) & 32 (48) & 19 \\
SL & Stage III (not $w_0$, $w_a$) & 1.6 (2.0) & 6.9 (9.4) & 1.3 \\
SL & None & 1.3 (1.7) & 4.4 (6.1) & 1.3 \\
\hline
All & Stage III & 142 (156) & 505 (711) & 500 \\
All & None & 108 (135) & 461 (666) & - \\
\hline
\end{tabular}
\caption{Summary of forecast DETF FoMs after Y1 and Y10 for each probe and their combination,
  synthesizing target and forecast numbers from across \autoref{sec:detailedreq}. The FoM values correspond to our
  calculated baselines from current forecasts including all sources of uncertainty, while ``ceiling'' values in parenthesis
  indicate those without any contribution from
  calibratable systematic uncertainties, which in practice should not be reachable with current
  analysis methodology.  The ``Target''
  column corresponds to the Y10 targets defined by \ref{high:combinedfom} and \ref{high:indivfom},
  to be compared with the first number in the previous column.  For individual probes,
  our goal (\ref{high:indivfom}) is that the Y10 forecast FoM should meet or exceed the target
  value, while for combined probes, it is a requirement (\ref{high:combinedfom}).  Note that since
  our methodology is to make a forecast with reasonable assumptions of what is achievable in all
  probes (including statistical and astrophysical systematic uncertainties), and then degrade that
  forecast by tuning the size of the calibratable systematic error budget to meet
  \ref{high:combinedfom}, in general the Y10 forecast and target will match precisely as described
  in \autoref{subsec:combprobes} and the introduction to \autoref{sec:detailedreq}.  This is a
  feature of our flowdown from our high-level science requirements rather than a feature of the baseline
  analysis for each probe that goes into the forecasts.  Note that only the individual-probe FoM
  values without priors should be compared with the individual probe contours in
  \autoref{fig:jointy1y10}.  Finally, the individual probe results include Stage III priors on
  non-dark energy parameters (to stabilize the Fisher matrix calculations), while the ``All''
  results include full Stage III priors because they are included in the definition of overall
  constraining power in \ref{high:combinedfom}.  The Stage III priors on their own give a FoM of 23.
 \label{tab:fom_summary}}
\end{center}
\end{table}

Second, as noted in \autoref{sec:intro}, the overarching purpose of the exercise carried out here
is to place requirements on the DESC's analysis pipelines to enable an analysis of LSST data
corresponding to a stand-alone Stage IV dark energy experiment.  Here we briefly discuss the connection
between the requirements placed in this version of the DESC~SRD and the relevant analysis
pipelines:
\begin{itemize}
\item We placed requirements on our knowledge of mean redshifts and redshift bin widths for
  tomographic LSS, WL, and CL analyses.  These essentially amount to requirements on the performance
  of the \texttt{PZCalibrate} pipeline that is being developed by the PZ working group during the
  DC2 era.  The purpose of that pipeline is to use cross-correlation analysis and spectroscopic
  training data to provide calibrated $N(z)$ (not just mean and width, but full $N(z)$ including the
  impact of photometric redshift outliers) for tomographic samples defined using photometric
  redshifts.  The requirements for WL and CL analyses, such as \ref{reqd:wllssmeanzs}, represent a substantial improvement beyond the
  current state of the art.  Current analyses are limited by the
  availability of spectroscopic surveys within the footprint of the imaging survey, so the
  4000~deg$^2$ overlap with DESI and the more dilute samples expected from 4MOST in the rest of the
  area will provide some of the improvement in control of redshift uncertainties for LSST.  The rest
  of the improvement will rely on algorithmic improvements, and rigorous development and testing of
  \texttt{PZCalibrate}.  Additional spectroscopic redshifts for training and calibration, particularly at the faint
  end (near and beyond the limit of the C3R2 survey; \citealt{2017ApJ...841..111M}) would provide additional margin on
  this source of systematic uncertainty. However, this approach has several serious challenges:
  (a) the necessary amount of telescope time on 8m-class telescopes
  \citep{2015APh....63...81N} is very large, (b) spectroscopic incompleteness is still an issue and
  difficult to assess at the necessary tolerances, and
  (c) direct calibration is very sensitive to small numbers of incorrect redshifts.
\item We placed requirements on our knowledge of the redshift-dependent ensemble shear calibration
  (\ref{reqd:wllssshearm}).  This corresponds to a requirement on \texttt{shearMeasurementPipe},
  the pipeline being developed by the WL working group in order to quantify the calibration of the
  LSST Science Pipelines shear estimator for the selected tomographic shear
  samples at the required level of accuracy.  As mentioned in the text associated with~\ref{reqd:wllssshearm}, state-of-the-art shear estimation
  methods can already achieve uncertainty on shear calibration that is comparable to this requirement, but {\em
    without} accounting for all effects that are included in this requirement (e.g., blending).  The
  implications for weak lensing pipeline development are that it is important to integrate at least
  one of these state-of-the-art shear estimation methods into the pipeline in the near term to enable
  work that must be done on less well-understood effects such as blending.
\item We also placed requirements on our knowledge of specific contributors to shear calibration
  bias, such as PSF model size errors and stellar contamination.  These will be quantified with the
  WL null testing pipeline, \texttt{WLNullTest}.
\item The requirements on photometric calibration for the SN science case correspond to
  requirements on (a) the software provided by the Photometric Corrections working group to quantify
  aspects of photometric calibration that go beyond what is provided by Rubin Observatory (for
  \ref{reqd:snfilterzp} and \ref{reqd:snfiltermean}), (b) the SN working group light curve
  modeling software (\ref{reqd:snlightcurve}), and (c) the SN working group likelihood
  analysis software, since our stringent requirements suggest it will be important to investigate
  avenues for jointly (and efficiently) marginalizing over observational and astrophysical
  systematic uncertainties, converting some of our calibratable systematic uncertainties to
  self-calibrated ones through development of appropriate models (for \ref{reqd:snglobalcal} and
  \ref{reqd:snmw}).  In some cases, meeting the detailed requirements on photometric calibration may
  require investment of DESC resources on work with the Facility to
  achieve more accurate photometric calibration than the design specifications in the LSST~SRD.
\end{itemize}

Finally, the baseline analyses defined in this version of the DESC~SRD do not necessarily correspond
to each working groups' aspirations; limitations were imposed both by the capabilities in existing
software and the fact that further R\&D is needed into several key questions about the analysis
process.  Here we briefly summarize anticipated updates in future DESC~SRD versions (with further
details available in \appref{app:software}, \ref{app:assumptions}, and
\ref{app:baselines}).  Such updates will inevitably lead to improved forecasts and hence revised values in
\autoref{tab:fom_summary}:
\begin{itemize}
\item The LSS analysis may be defined with different samples (e.g., including multitracer analysis),
  a longer redshift baseline, inclusion of cross-power spectra between redshift bins, modified
  $\ell$ binning to better resolve the BAO feature, and other updates to make it more optimal.  In addition,
  marginalization over nonlinear (rather than just linear) bias must be included, which will also
  enable the $k_\text{max}$ value to be shifted to smaller scales, potentially providing more
  cosmological information.
\item The WL$+$LSS analysis may be defined with a different sample than the LSS-alone analysis for
  lenses, and may benefit from the LSS improvements described above.  The powerfully constraining 3$\times$2-point
  analysis is particularly sensitive to what scales can be used; improvements in theoretical
  modeling of nonlinear bias has the potential to produce substantial gains.
\item The galaxy clusters baseline analysis will be updated to include further
  realistically-achievable priors on the MOR, and to include large-scale cluster clustering and
  cluster lensing, which can improve the self-calibration of the MOR.
\item For all probes of structure growth, more optimal choices of tomographic binning schemes will
  be explored.  Aside from possible gains in statistical constraining power, different choices may
  enable more optimal self-calibration of redshift-dependent effects, and/or changes in the
  requirements in control of redshift-related systematic biases.
\item The SN analysis will include the impact of photometric redshift uncertainties for the
  photometric SN sample, probabilistic SN inference, and SN type misclassification.
\item The SL analysis should include lensed supernovae.
\item We will place requirements on model sufficiency for self-calibrated systematic uncertainties.
\item The cosmological parameter space should be widened to include massive neutrinos.
\item In this DESC~SRD version we only placed requirements on a subset of calibratable
  systematic uncertainties, occasionally due to limitations in existing software but in other cases
  due to insufficient knowledge in the field as to how to parametrize the effects of interest in
  terms of how they affect our observable quantities.  The DESC's DC2 and other non-simulation-based
  work happening during the DC2 era should deepen our
  understanding of how to describe these sources of systematic uncertainty, resulting in both
  requirements and more capable analysis software.  Notable areas in which substantial
  development of systematics models is needed include the impact of photometric redshift errors on
  photometric SN analysis; more flexible photometric redshift
  uncertainties for structure growth probes; the impact of residual blending systematics on number densities,
  redshifts, and shear; the impact of survey inhomogeneity on the galaxy density field; a description of how many types of observational systematics impact the
  strong lensing observables.
\item The interaction between models for different types of systematic uncertainties, and their
  potential for very different behavior in the 7-dimensional cosmological parameter space, will be more
  thoroughly considered.  In addition, we will take care to adopt common models of specific sources
  of systematic uncertainty across probes wherever possible.
\item Future versions of this document will use DESC software for describing cosmological
  observables and their covariances, so as to enable collaboration-wide development of forecasts and
  requirements within a common software framework that meets DESC coding guidelines.  It will also
  incorporate lessons learned about the dependence of the forecasts on observing strategy produced
  by the DESC's Observing Strategy Task Force during the second half of 2018, along with any
  subsequent updates in the LSST baseline survey definition (see baseline used for the DESC~SRD in
  \appref{subsec:assump-strategy}).
\item More concrete statements should be made about blinding as our understanding of blinding
  techniques develops.
\item Requirements should be placed on the accuracy of modeling of cosmological quantities such as
  power spectra, mass functions, etc.\
\end{itemize}

Some of the above improvements will be the subject of R\&D in the coming years, the results of
which will be incorporated into our baseline analyses as our understanding evolves.
Further details of planned updates are given in the aforementioned appendices.  All changes will be
subject to the change control process outlined at the end of \autoref{sec:intro}.

\clearpage

\phantomsection\section*{Acknowledgments}
\addcontentsline{toc}{section}{Acknowledgments}

\input{desc-tex/ack/standard.tex}

We are grateful for the support of the University of Chicago Research Computing Center for
assistance with the calculations carried out in this work, and also acknowledge the use of the
Glamdring cluster in the Department of Physics at the University of Oxford.
Part of the research was carried out at the Jet Propulsion Laboratory, California Institute of
Technology, under a contract with the National Aeronautics and Space Administration and is supported
by NASA ROSES ATP 16-ATP16-0084 grant.
We thank Chris Hirata for providing input on various approaches to placing requirements.
\clearpage

\bibliographystyle{inc/apj-mod}
\phantomsection\bibliography{inc/references}

\clearpage
\appendix
\renewcommand\thefigure{\thesection\arabic{figure}}
\renewcommand\thetable{\thesection\arabic{table}}
\renewcommand{\thesection}{\Alph{section}}
\renewcommand{\thesubsection}{\Alph{section}\arabic{subsection}}

\phantomsection\section*{Appendices}
\addcontentsline{toc}{section}{Appendices}

\section{Connections to Rubin Observatory tools and documents}\label{app:lsst}

\input{inc/lsst.tex}

\section{Software}\label{app:software}

\input{inc/software.tex}

\section{Assumptions}\label{app:assumptions}

\input{inc/assumptions.tex}

\section{Baseline analyses}\label{app:baselines}

Each subsection within this appendix outlines the baseline analysis for a single
probe or probe combination.

\subsection{Large-scale structure}\label{app:lss}

\input{inc/lss.tex}

\subsection{Weak lensing (3$\times$2-point)}\label{app:wllss}

\input{inc/wl.tex}

\subsection{Galaxy clusters}\label{app:cl}

\input{inc/cl.tex}

\subsection{Supernovae}\label{app:sn}

\input{inc/sn.tex}

\subsection{Strong lensing}\label{app:sl}

\input{inc/sl.tex}

\section{Synthesis of systematic uncertainties across all probes}

\input{inc/synthesis.tex}

\section{Defining number densities}\label{app:dndmag}

\input{inc/dndmag.tex}

\section{Forecasting-related plots}\label{app:plots}

\begin{figure}
  \begin{center}
    \includegraphics[width=\textwidth]{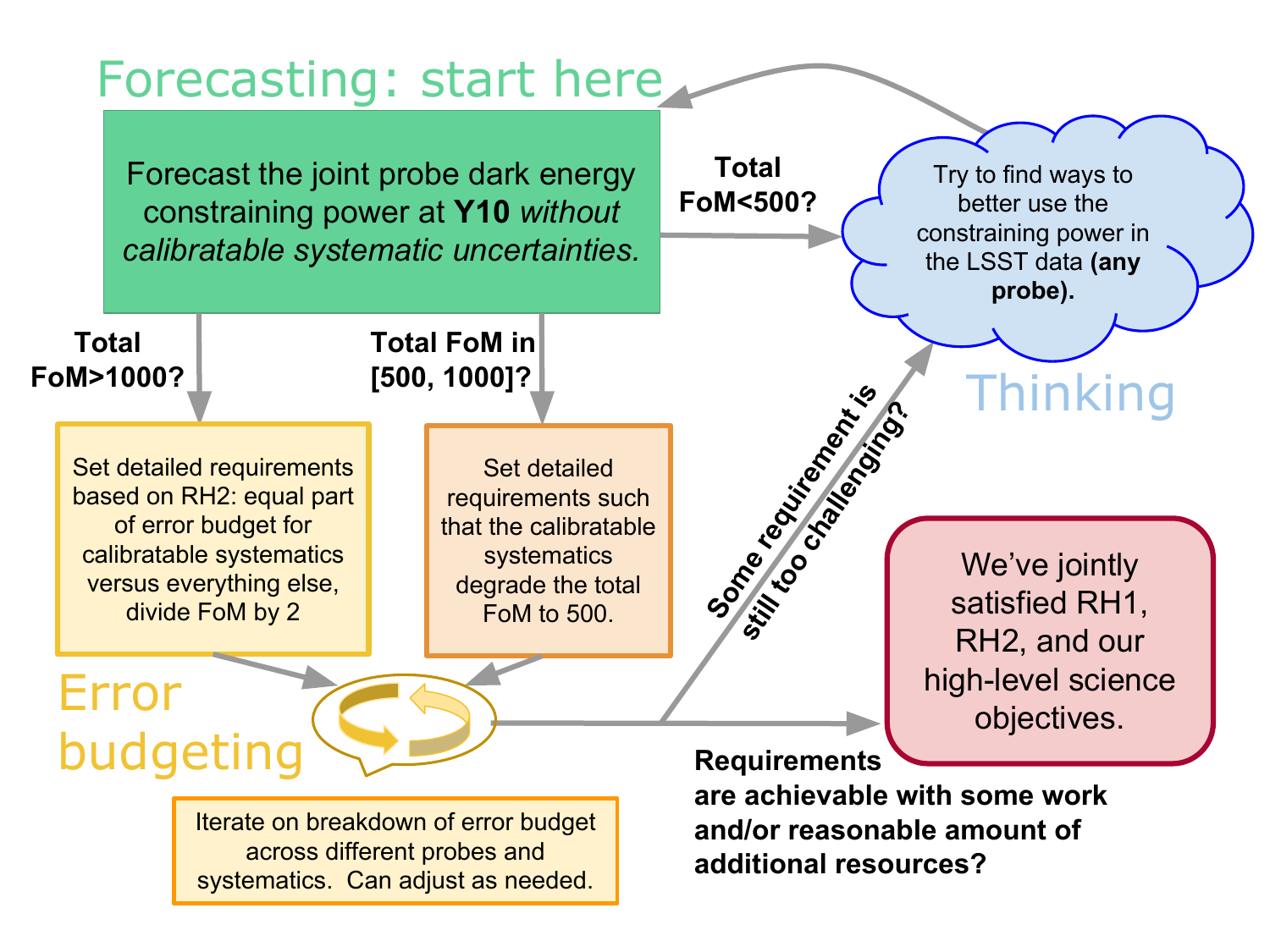}
  \end{center}
  \caption{A schematic illustrating the error budgeting process through which the Y10 detailed
    requirements in \autoref{sec:detailedreq} are derived.
  \label{fig:errorbudget}}
\end{figure}

\begin{figure}
  \begin{center}
    \includegraphics[width=0.47\textwidth]{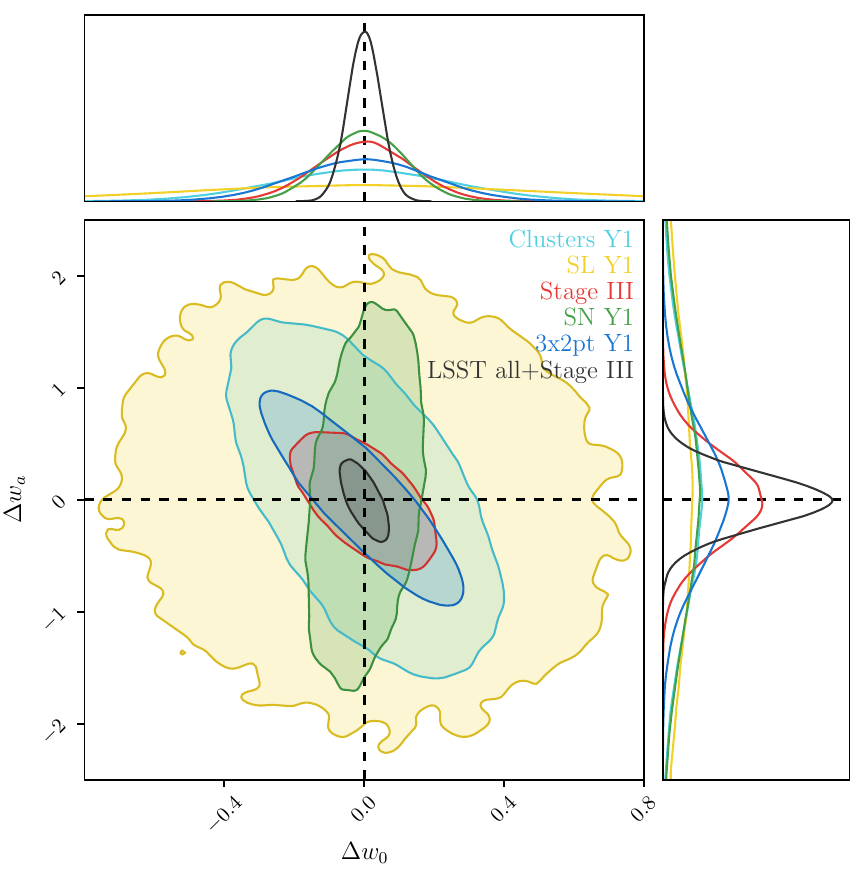}
    \includegraphics[width=0.47\textwidth]{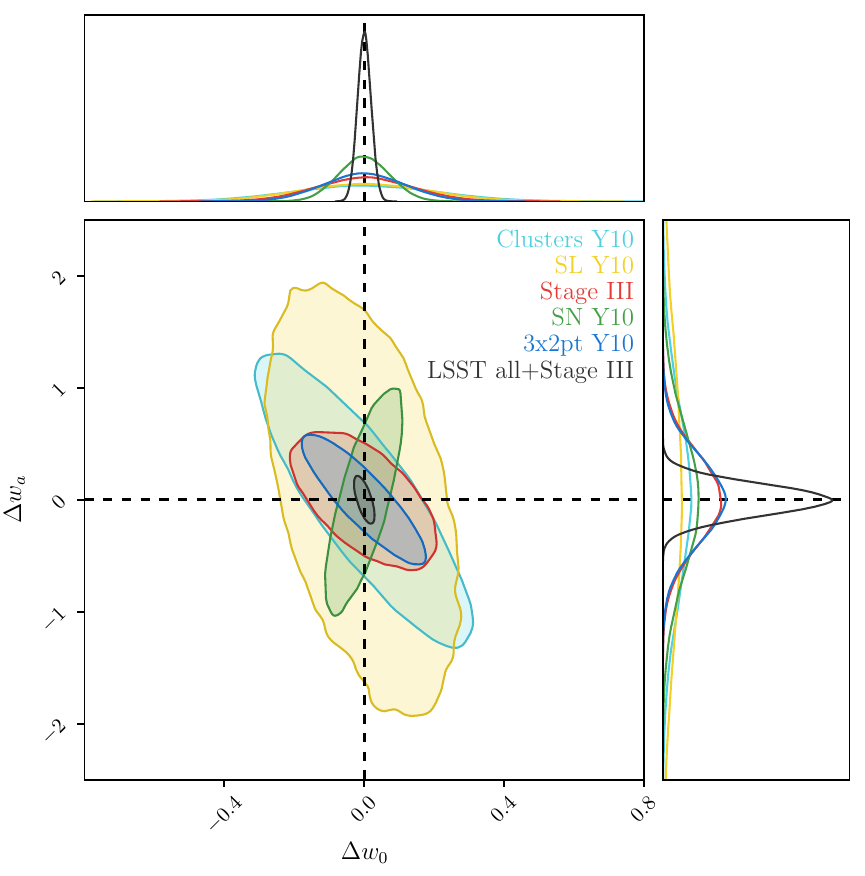}
    \includegraphics[width=0.52\textwidth]{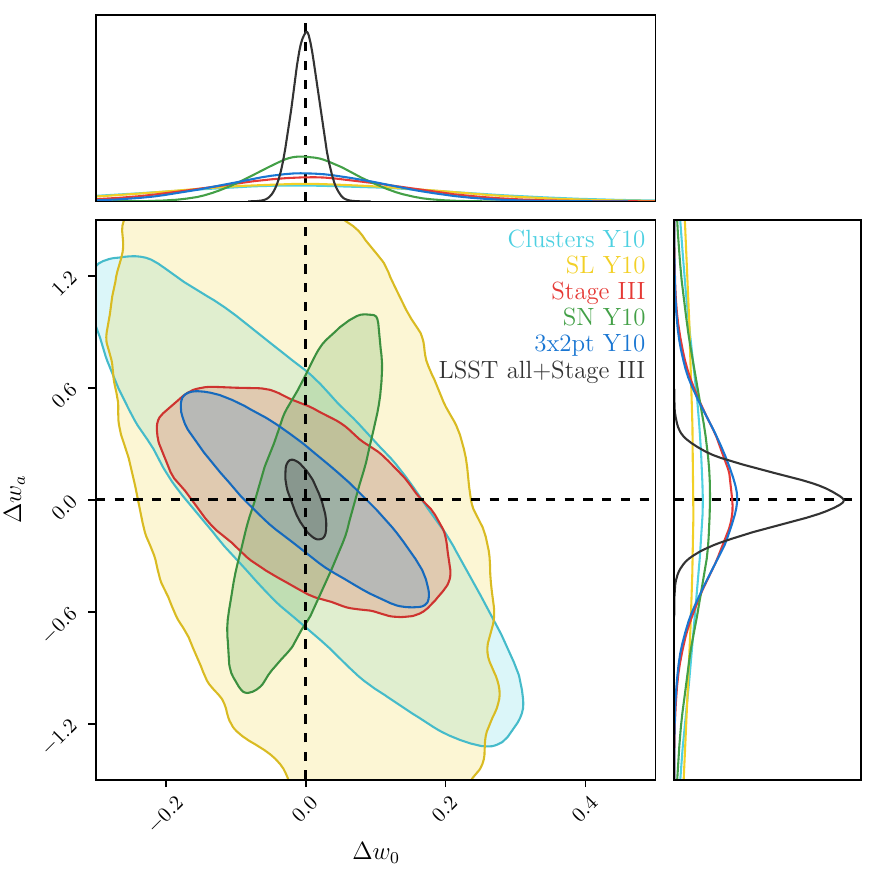}
  \end{center}
  \caption{The forecast dark energy constraints at Y1 (top left) and Y10 (top right; bottom) from each probe
    individually and the joint forecast including Stage III priors.  For consistency, the same
    axes are used on the Y1 and the top Y10 plot, while the bottom Y10 plot is zoomed in further.  Note that the supernova
    contours appear to be tilted clockwise with respect to typical forecasts shown in the literature, because most papers include a
    Stage III prior when generating the contour for SN.   68\% confidence intervals are shown in
    all cases; the plotted quantities $\Delta w_0$ and $\Delta w_a$ are the difference between $w_0$ and $w_a$ and their fiducial
    values of -1 and 0. The contours in this figure for individual probes do not
    include Stage III priors, so they should only be compared with the individual probe
    FoM values in \autoref{tab:fom_summary} that have no Stage III prior included.
  \label{fig:jointy1y10}}
\end{figure}

\begin{figure}
  \begin{center}
    \includegraphics[width=0.7\textwidth]{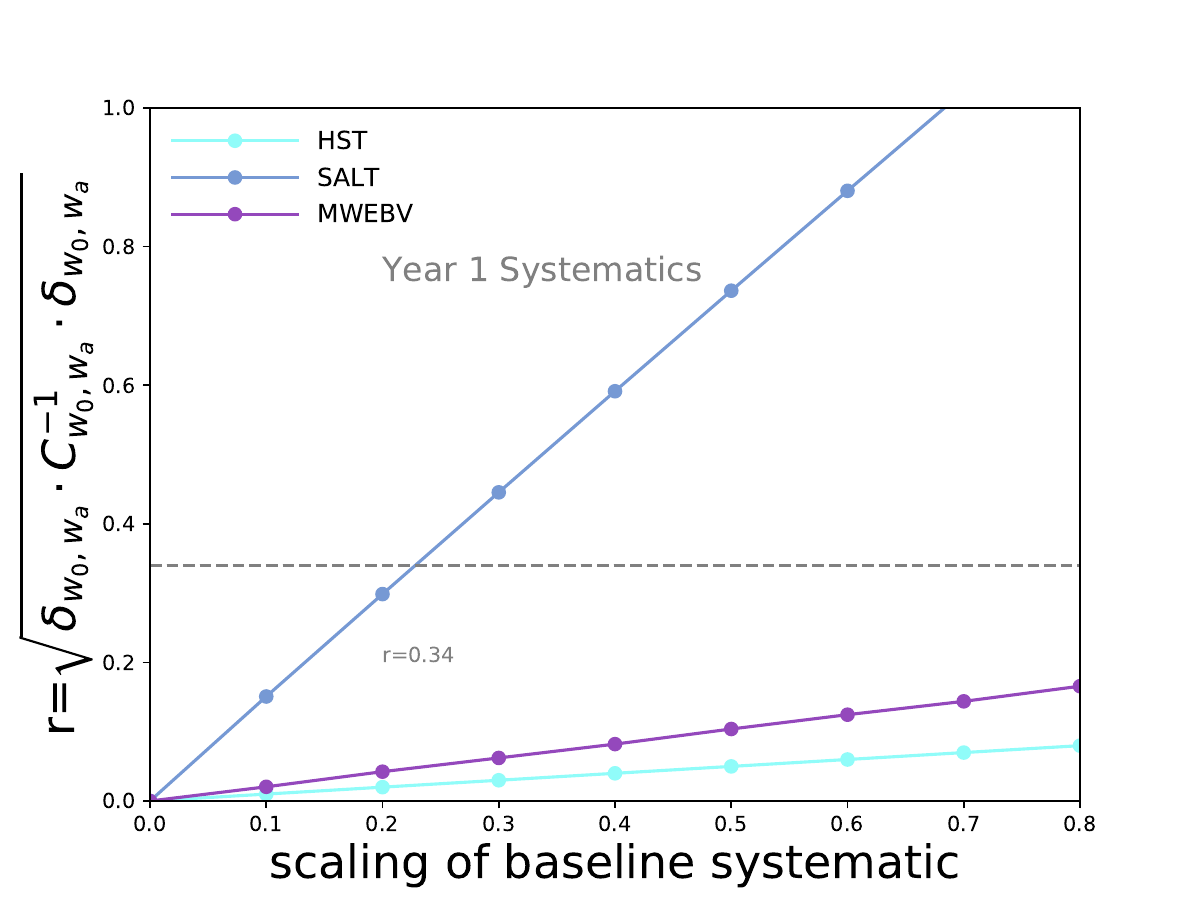}
    \includegraphics[width=0.7\textwidth]{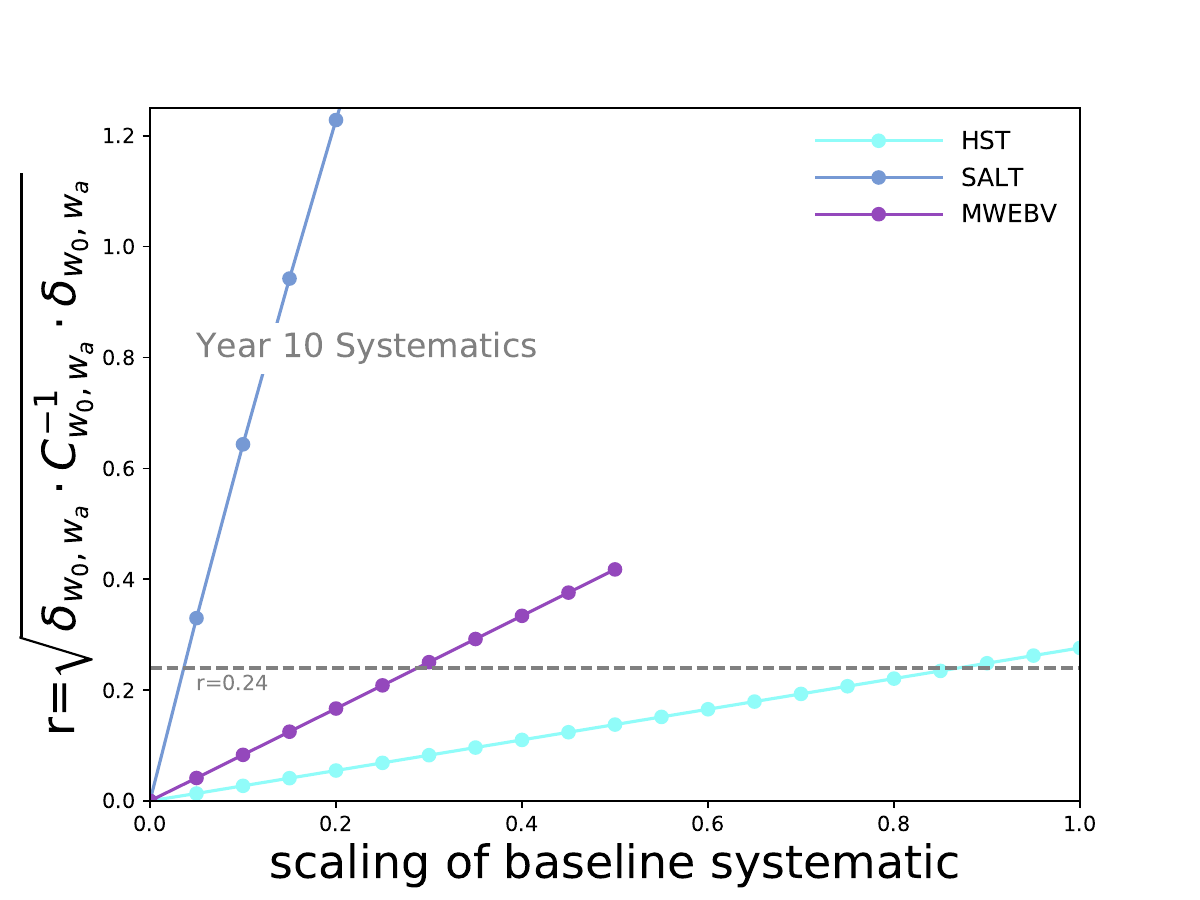}
  \end{center}
  \caption{Figures illustrating how requirements are set for supernova, for Y1 (top) and Y10
    (bottom), for the systematics that are given the tightest restrictions.  The horizontal axis gives the scaling of the systematic uncertainty with respect to
    the baseline value given in \autoref{tab:snsys-cal}. The vertical axis shows the bias in
    cosmological parameters with respect to the statistical plus self-calibrated systematic
    uncertainties, with the horizontal lines indicating the point at which a given individual
    source of calibratable systematic uncertainty would take up its allocated fraction of that error budget ($r=0.24$ for Y10 and $r=0.34$ for Y1, where $r$ is defined in
    \autoref{eq:requirement}).
  \label{fig:snreq}}
\end{figure}

In this section, we collect a subset of representative plots from the forecasts:
\begin{itemize}
\item \autoref{fig:errorbudget} illustrates the main steps involved in the error budgeting process
  described at the start of \autoref{sec:detailedreq}.
\item \autoref{fig:jointy1y10} shows the $(w_0, w_a)$ constraints from all five probes individually,
  and the joint forecast including Stage III priors as well\footnote{\autoref{fig:jointy1y10} was
    produced using \href{https://samreay.github.io/ChainConsumer/}{ChainConsumer}
    \citep{Hinton2016}, with advice and support provided by the ChainConsumer team.}.
\item \autoref{fig:snreq} shows how requirements were set for the calibratable systematic
  uncertainties for the supernova analysis.
\end{itemize}

\end{document}

%% file: inc/doc_settings.tex
\usepackage{datetime}
\usepackage{fancyhdr}
\usepackage[outermarks]{titlesec}
\usepackage[utf8x]{inputenc}
\usepackage[T1]{fontenc}
\usepackage{amsmath}
\usepackage{amssymb}
\usepackage{epsfig}
\usepackage{graphics}
\usepackage{graphicx}
\usepackage[usenames]{color}
\usepackage{helvet}
\usepackage{times}
\usepackage{natbib}
\usepackage{import}
\usepackage{tabularx}
\usepackage{xspace}
\usepackage{scrextend}
\usepackage{mathrsfs}
\usepackage[normalem]{ulem}
\usepackage{scrwfile} 
\usepackage{xcolor}
\usepackage{colortbl}
\usepackage{pdflscape}
\usepackage{afterpage}

\usepackage{pdftexcmds}

\usepackage[linktocpage=false]{hyperref}
\hypersetup{
    colorlinks=true,
    citecolor=DESCred,
    filecolor=DESCred,
    linkcolor=DESCred,
    urlcolor=DESCred,
}
\usepackage{hypcap}

\def\bfseries{\fontseries \bfdefault \selectfont \boldmath} 
\renewcommand{\thesection}{\arabic{section}}
    \titleformat{\section}
      {\normalfont\large\bfseries}{\thesection}{1em}{}
    \titleformat{\subsection}
      {\normalfont\normalsize\bfseries}{\thesubsection}{1em}{}
    \titleformat{\subsubsection}
      {\normalfont\normalsize\bfseries}{\thesubsubsection}{1em}{}
    \titleformat{\paragraph}
      {\normalfont\normalsize\bfseries}{\theparagraph}{1em}{}
    \titleformat{\subparagraph}
      {\normalfont\normalsize\bfseries}{\thesubparagraph}{1em}{}

    \titlespacing\section{0pt}{12pt plus 4pt minus 2pt}{0pt plus 2pt minus 2pt}
\titlespacing\subsection{0pt}{12pt plus 4pt minus 2pt}{0pt plus 2pt minus 2pt}
\titlespacing\subsubsection{0pt}{12pt plus 4pt minus 2pt}{0pt plus 2pt minus 2pt}

\def\vs{\vspace{0.5cm}}

%% file: inc/macros.tex
\usepackage{xifthen}
\usepackage{xspace}

\let\counterwithin\relax
\usepackage{chngcntr}
\counterwithin{table}{section}
\counterwithin{figure}{section}

\usepackage{multirow}
\usepackage{tocloft}

\usepackage{datetime}

\makeatletter

\newcommand{\appref}[1]{\hyperref[#1]{Appendix~\ref{#1}}}

\newcommand{\relabel}[3]{\@bsphack
    \protected@write\@auxout{}%
        {\string\newlabel{#2}{{#1}{\thepage}{\relax}{#3}{}}}%
    \@esphack}

\newcommand{\listofrecommendations}{
  \@starttoc{tor}
}

\newcommand{\listofjusttheserecommendations}[1]{
  \vskip\baselineskip
  {\bf \nameref{sec:\headerstring} Recommendations:}
  \phantomsection\label{tab:\headerstring:recommendations}
  \medskip
  \hrule
  \@starttoc{\headerstring.tor}
  \hspace{-0.5em}
  \medskip
  \hrule
  \vskip\baselineskip
  \vskip\baselineskip
}

\newcounter{recommendation}

\newcommand{\headerstring}{}

\newcommand{\maketoc}{
   \renewcommand{\contentsname}{\hrule{\large\flushleft\sffamily\bfseries Contents\vspace{-1.2cm}}}
   \setlength{\parskip}{0.1cm}
   \phantomsection\tableofcontents 
   \flushbottom
   \vs\hrule
}

\newcommand{\recommendationnum}{%
    \headerstring:\therecommendation
}

\newcommand{\recommendationstring}[2]{\hyperref[tor:\recommendationnum]{\textbf{Recommendation \recommendationnum: We should #2}}}

\newcommand{\footernavigationbar}
{
    \it\footnotesize Go to:
    \hyperref[toc]{The Table Of Contents} 
    \ifthenelse{\equal{\headerstring}{}}
   	    {}
        {
            $\bullet$ \hyperref[sec:\headerstring]{the start of this section}
  			$\bullet$ \hyperref[sec:TOR]{the list~of~recommendations}
        }
}

\def\EndofDCOneDA{06/17}
\def\EndofDCOne{\EndofDCOneDA}

\def\EndofDCTwoDA{12/18}
\def\EndofDCTwo{\EndofDCTwoDA}

\def\EndofDCThreeDA{03/20}
\def\EndofDCThree{\EndofDCThreeDA}

\newcommand{\FY}[2]{%
\newcount\year
\ifthenelse{\equal{#2}{Q1}}%
{%
  \year=\numexpr#1-1\relax
  {11/\the\year}%
}{%
  \year=#1%
  \ifthenelse{\equal{#2}{Q2}}{02/\the\year}{%
  \ifthenelse{\equal{#2}{Q3}}{05/\the\year}{%
  {08/\the\year}%
}}}%
}

\newcommand{\deadline}[1]{%
\ifthenelse{\equal{#1}{??/??}}%
    {\def\era{\thisrecommendationsera}}%
    {\def\era{#1}}%
\ifthenelse{\equal{\era}{DC1}}{\def\duedate{\EndofDCOne}}{%
\ifthenelse{\equal{\era}{DC2}}{\def\duedate{\EndofDCTwo}}{%
\ifthenelse{\equal{\era}{DC3}}{\def\duedate{\EndofDCThree}}{%
\ifthenelse{\equal{#1}{??/??}}{\def\duedate{\era}}{%
\def\duedate{{\bf{\era}}}%
}}}}
\hyperref[fig-DC]{\duedate}
}

\bibpunct[, ]{(}{)}{;}{a}{}{,}

\newboolean{appendix}
\setboolean{appendix}{false}

\renewenvironment{thebibliography}[1]{%
  \begin{oldthebibliography}{#1}%
    \addcontentsline{toc}{section}{References}
}%
{%
  \end{oldthebibliography}%
}

\makeatother

\newcommand{\etal}{et~al.~}

\newcommand{\dndmag}{d$N$/dmag}
\newcommand{\neff}{\ensuremath{n_\text{eff}}}

\definecolor{gray}{rgb}{0.5,0.5,0.5}
\definecolor{green}{rgb}{0.0,0.5,0.0}
\definecolor{DESCred}{rgb}{0.63,0.00,0.20} 
\definecolor{dkblue}{rgb}{0, 0.30, 0.99 }

\newcommand{\lookup}[1]{\hyperref[#1]{#1}}

\newcounter{requirementh}
\newcounter{requirementd}
\newcounter{objectives}
\newcounter{goal}
\newcommand{\probe}{\@empty}

\newcommand{\requirementhname}{%
    RH{\therequirementh}%
}
\newcommand{\requirementdname}{%
    \probe{\therequirementd}%
}
\newcommand{\objectivename}{%
    O{\theobjectives}%
}
\newcommand{\goalname}{%
    G{\thegoal}%
}

\makeatletter
\newcommand{\customlabel}[2]{%
   \protected@write \@auxout {}{\string \newlabel {#1}{{#2}{\thepage}{#2}{#1}{}} }%
   \hypertarget{#1}{\textbf{#2}:}
}
\makeatother

\makeatletter
\newcommand{\labelreqh}[2]{%
   \protected@write \@auxout {}{\string \newlabel {#1}{{\requirementhname}{\thepage}{\requirementhname}{#1}{}} }%
   \hypertarget{#1}{\textbf{\textcolor{cyan}{High-level requirement \requirementhname}: #2}}\stepcounter{requirementh}}
\makeatother

\makeatletter
\newcommand{\labelreqdjoint}[3]{%
   \protected@write \@auxout {}{\string \newlabel {#1}{{\requirementdname}{\thepage}{\requirementdname}{#1}{}} }%
   \hypertarget{#1}{\textbf{\textcolor{cyan}{Detailed joint probes requirement {\requirementdname} {#2}}: {#3}}}\stepcounter{requirementd}}
\makeatother

\makeatletter
\newcommand{\labelreqd}[4]{%
   \protected@write \@auxout {}{\string \newlabel {#1}{{\requirementdname}{\thepage}{\requirementdname}{#1}{}} }%
   \hypertarget{#1}{\textbf{\textcolor{cyan}{Detailed requirement {\requirementdname} (Y10)}: Systematic uncertainty in {#2} shall not exceed {#4} in the Y10 DESC {\probe} analysis.\\ \textcolor{cyan}{Goal {\requirementdname} (Y1)}: Systematic uncertainty in {#2} should not exceed {#3} in the Y1 DESC {\probe} analysis.}}\stepcounter{requirementd}}
\makeatother

\makeatletter
\newcommand{\labelreqdcomment}[5]{%
   \protected@write \@auxout {}{\string \newlabel {#1}{{\requirementdname}{\thepage}{\requirementdname}{#1}{}} }%
   \hypertarget{#1}{\textbf{\textcolor{cyan}{Detailed requirement {\requirementdname} (Y10)}: Systematic uncertainty in {#2} shall not exceed {#4} in the Y10 DESC {\probe} analysis. #5\\ \textcolor{cyan}{Goal {\requirementdname} (Y1)}: Systematic uncertainty in {#2} should not exceed {#3} in the Y1 DESC {\probe} analysis.}}\stepcounter{requirementd}}
\makeatother

\makeatletter
\newcommand{\labelobj}[2]{%
   \protected@write \@auxout {}{\string \newlabel {#1}{{\objectivename}{\thepage}{\objectivename}{#1}{}} }%
   \hypertarget{#1}{\textbf{\textcolor{cyan}{Objective \objectivename}: #2}}\stepcounter{objectives}}
\makeatother

\makeatletter
\newcommand{\labelg}[2]{%
   \protected@write \@auxout {}{\string \newlabel {#1}{{\goalname}{\thepage}{\goalname}{#1}{}} }%
   \hypertarget{#1}{\textbf{\textcolor{cyan}{Goal \goalname}: #2}}\stepcounter{goal}}
\makeatother

%% file: commitID.tex
 {\Large\bfseries Version~1.0.2}

\medskip
Date: \today

%% file: inc/version.tex
\newpage
\section*{Change Record}

\begin{table}[!thp]
\begin{center}

\renewcommand{\arraystretch}{1.2}
\begin{tabular}{|p{1.5cm}|p{2cm}|p{6.cm}|p{5cm}|}
\hline
{\bf Version}  &{\bf Date}  &{\bf Description}  & {\bf Owner name}
\\ \hline
  \href{https://github.com/LSSTDESC/Requirements/releases/tag/v0.9}{v0.9}
  & 03/27/2018
  & Initial pre-release, for collaboration feedback prior to the 2018 operations review.
  & Rachel Mandelbaum
\\ \hline
  \href{https://github.com/LSSTDESC/Requirements/releases/tag/v0.99}{v0.99}
  & 04/26/2018
  & Pre-release, for the May 2018 LSST DESC DOE OHEP operations review.
  & Phil Marshall
\\ \hline
  \href{https://github.com/LSSTDESC/Requirements/releases/tag/v1}{v1}
  & 09/05/2018
  & Initial public release. Used internally by the LSST DESC for prioritization of software and dataset development effort, data challenge design, and derivation of performance metrics.
  & Phil Marshall
\\ \hline
  \href{https://github.com/LSSTDESC/Requirements/releases/tag/v1.0.1}{v1.0.1}
  & 05/01/2019
  & Fixed a minor issue with an incorrect file in the DESC SRD v1 release, clarified a few details of the signal calculations for the WL and LSS analyses. No changes in the analysis, forecasts, requirements, etc, but data products are now easier to use.
  & Phil Marshall
\\ \hline
  \href{https://github.com/LSSTDESC/Requirements/releases/tag/v1.0.2}{v1.0.2}
  & 09/06/2021
  & Updated to reflect observatory renaming, clarified SN forecasting process and degeneracy
  direction.  No change in goals, requirements, or other results.
  & Rachel Mandelbaum
\\ \hline
  \href{}{}
  &
  &
  &
\\ \hline
\end{tabular}
\renewcommand{\arraystretch}{1.0}
\end{center}
\end{table}

%% file: inc/contrib.tex
\noindent Contributors to the DESC~SRD effort are listed in the table below, with leading contributions
shown in bold and affiliations indicated in the notes below the table.

\begin{ThreePartTable}
\begin{TableNotes}
\footnotesize
\item [1] School of Physics and Astronomy, Cardiff University
\item [2] Department of Physics, University of Oxford
\item [3] Department of Physics and Astronomy, Rutgers University
\item [4] Oskar Klein Centre, Department of Physics, Stockholm University
\item [5] SNSF Ambizione, Laboratory of Astrophysics, \'Ecole Polytechnique F\'ed\'erale de Lausanne (EPFL)
\item [6] Center for Cosmology and Astroparticle Physics, Ohio State University
\item [7] Department of Physics, Stanford University
\item [8] Kavli Institute for Particle Astrophysics and Cosmology (KIPAC), Stanford University
\item [9] Institute of Cosmology and Gravitation, University of Portsmouth
\item [10] Department of Physics, Brown University
\item [11] SLAC National Accelerator Laboratory
\item [12] Steward Observatory/Department of Astronomy, University of Arizona
\item [13] Jet Propulsion Laboratory, California Institute of Technology
\item [14] Fermi National Accelerator Laboratory
\item [15] Kavli Institute for Cosmological Physics, University of
Chicago
\item [16] California Institute of Technology
\item [17] Lawrence Berkeley National Laboratory
\item [18] Department of Astronomy, University of California, Berkeley
\item [19] Department of Astronomy and Astrophysics, University of Toronto
\item [20] Dunlap Institute for Astronomy and Astrophysics, University of Toronto

\item [21] Department of Physics, Lancaster University
\item [22] Department of Astronomy, University of Washington
\item [23] Rubin Observatory Project
\item [24] Department of Physics and Astronomy, University of California, Irvine
\item [25] Mullard Space Science Laboratory, University College London
\item [26] McWilliams Center for Cosmology, Department of Physics, Carnegie
Mellon University
\item [27] Nagoya University
\item [28] Kavli Institute for the Physics and Mathematics of the Universe (Kavli IPMU, WPI)
\item [29] Department of Physics and Astronomy and PITT PACC, University of Pittsburgh
\item [30] Physics Department, Brookhaven National Laboratory
\item [31] Department of Physics and Astronomy, University of Southampton
\item [32] Department of Physics, The Ohio State University
\end{TableNotes}
\begin{longtable}{|p{4cm}|p{11cm}|}
\endfirsthead
\endhead
\endfoot
\insertTableNotes  
\endlastfoot
\hline
Name & Contribution \\ \hline
David Alonso$^{1,2}$ & Forecasting guidance, LSS analysis definition, WL forecast cross-check \\
Humna Awan$^3$ & Input on survey definition based on OpSim v3 \verb minion_1016 \\
Rahul Biswas$^4$ & SN analysis definition, systematics treatment \\ %
Jonathan Blazek$^{5,6}$ & Forecasting guidance, review of complete draft \\
Patricia Burchat$^{7,8}$ & Initial organization, WL systematic uncertainties list\\
Elisa Chisari$^2$ & Forecasting guidance, review of complete draft \\
{\bf Tom Collett}$^9$ & {\bf SL forecaster}, SL analysis definition\\
Ian Dell'Antonio$^{10}$ & CL analysis definition \\ %
Seth Digel$^{8,11}$ & Review of complete draft \\ %
{\bf Tim Eifler}$^{12,13}$ & {\bf Lead forecaster}, WL+LSS+CL analysis and systematics definition, joint probe covariance computation\\
Josh Frieman$^{14,15}$ & Review of complete draft \\ %
{\bf Eric Gawiser}$^3$ & Project management, {\bf scientific guidance}, text editing \\
Daniel Goldstein$^{16,17,18}$ & SL analysis definition \\
{\bf Ren\'{e}e Hlo\v{z}ek}$^{19,20}$ & {\bf SN forecaster}, SN analysis definition, systematics treatment\\
Isobel Hook$^{21}$ & Consultation on 4MOST capabilities for SN science case \\ %
\v{Z}eljko Ivezi\'{c}$^{22}$ & Review of complete draft, feedback on connection to LSST~SRD\\ %
Steven Kahn$^{7,8,11,23}$ & Review of document and feedback on connection to LSST~SRD \\ %
Sowmya Kamath$^{7,8}$ & WL systematic uncertainties list\\ %
David Kirkby$^{24}$ & Initial organization, source sample characterization, review of draft \\
Tom Kitching$^{25}$ & WL analysis definition \\
Elisabeth Krause$^{12}$ & Forecasting guidance, Fisher software, CosmoLike
infrastructure \\
Pierre-Fran\c{c}ois Leget$^{7,8}$ & WL systematic uncertainties list\\ %
{\bf Rachel Mandelbaum}$^{26}$ & {\bf Lead author, scientific oversight}, project management \\
Phil Marshall$^{8,11}$ & Initial organizational work, scientific guidance, text editing \\ %
Josh Meyers$^{7,8}$ & WL systematic uncertainties list\\ %
Hironao Miyatake$^{13,27,28}$ & Cluster mass-observable relation software, CL forecasting guidance\\
Jeff Newman$^{29}$ & Input on WL, LSS, and SN sample definition, review of complete draft \\ %
Bob Nichol$^{9}$ & Consultation on 4MOST capabilities for SN science case \\
Eli Rykoff$^{8,11}$ & WL, LSS, and CL area definition including dust, depth constraints\\ %
F. Javier Sanchez$^{24}$ & WL source sample characterization \\ %
{\bf Daniel Scolnic}$^{15}$ & {\bf SN analysis definition and forecasting} \\
An\v{z}e Slosar$^{30}$ & LSS analysis definition\\
Mark Sullivan$^{31}$ & Consultation on 4MOST capabilities for SN science case \\ %
Michael Troxel$^{6,32}$ & Review of complete draft \\
\hline
\end{longtable}
\end{ThreePartTable}

%% file: desc-tex/ack/standard.tex
The DESC acknowledges ongoing support from the Institut National de Physique Nucl\'eaire et de Physique des Particules in France; the Science \& Technology Facilities Council in the United Kingdom; and the Department of Energy, the National Science Foundation, and the LSST Corporation in the United States.  DESC uses resources of the IN2P3 Computing Center (CC-IN2P3--Lyon/Villeurbanne - France) funded by the Centre National de la Recherche Scientifique; the National Energy Research Scientific Computing Center, a DOE Office of Science User Facility supported by the Office of Science of the U.S.\ Department of Energy under Contract No.\ DE-AC02-05CH11231; STFC DiRAC HPC Facilities, funded by UK BIS National E-infrastructure capital grants; and the UK particle physics grid, supported by the GridPP Collaboration.  This work was performed in part under DOE Contract DE-AC02-76SF00515.

%% file: inc/lsst.tex
In this Appendix, we briefly summarize how this document depends on Rubin Observatory tools and
requirements.

First, our assumptions about the cadence (affecting the time-domain science cases), the
coadded depth as a function of time in each band, and the area reaching some criteria for
homogeneous coverage as outlined in \appref{subsec:assump-strategy} are entirely based on Rubin Observatory tools, specifically the operations simulator (OpSim\footnote{\url{https://github.com/lsst/sims\_operations}}) \verb minion_1016 ~run.  The LSST
observing strategy has not been finalized, and hence these assumptions may need to be revisited.
The DESC is working to quantify the impact of LSST observing strategy on the dark energy science
cases so as to communicate with Rubin Observatory on this important topic through the mechanism of
contributing to the community white paper on the LSST Observing
Strategy\footnote{\url{https://github.com/LSSTScienceCollaborations/ObservingStrategy}}.

We also rely on the LSST catalog simulator
(CatSim\footnote{\url{https://www.lsst.org/scientists/simulations/catsim}}) to estimate the weak lensing source number
density and redshift distribution (\appref{app:dndmag}) using simulated LSST images that have
parameters based on Rubin Observatory Project inputs such as filter throughputs, and anticipated survey image
characteristics such as typical PSF FWHM, sky brightness, and so on from Table 2 of
\citet{2008arXiv0805.2366I}.

Finally, there are many relevant requirements in Rubin Observatory's LSST~SRD\footnote{\url{https://docushare.lsstcorp.org/docushare/dsweb/Services/LPM-17}}.  Below we
briefly comment on the LSST~SRD requirements and their relevance to enabling our science cases.  All
appendices, tables, and equations listed below without links are in the LSST~SRD, while
those with direct links are in this document.
\begin{itemize}
\item Basic aspects of the instrument in Appendix A, Table 1 (filter complement), Table 11 (pixel
  size specification), Table 22 (area coverage), Table 23 (median number of visits per filter),
  Table 24 (coadded depth), Table 25 (distribution of visits over time) were implicitly encoded in
  our forecasts through our reliance on OpSim to define depths and cadence.
\item Tables 5 (single image depth) and 6 (variation of single image depth with bandpass)
  are primarily relevant in enabling transient science.  Our
  assumptions about the number of supernovae with a given light curve quality described in
  \appref{app:sn} depend heavily on this specification; the excellent single-image depth is an
  important enabler of our dark energy constraints from supernovae, including
  both the statistical and systematic uncertainties (e.g., associated with light curve modeling).
\item There are several requirements associated with image quality, along with assumptions that are
  not framed as requirements.  First, regarding assumptions, Appendix D of the LSST~SRD shows the distribution of
  atmospheric seeing at the site.  The LSST~SRD then places requirements on image quality through
  Table 9, which is effectively a requirement that the total PSF size for a given atmospheric seeing
  should have no more than 15\% contribution due to the LSST system.  Our weak lensing cosmology
  constraints are enabled by the excellent image quality that is implied by the expected atmospheric seeing
  and the requirement on the overall PSF size in Table 9, given that the constraining power of weak
  lensing improves when the image quality is better.  Image quality has a more difficult-to-quantify
  impact on all probes through its impact on blending systematics.  At fixed depth, blending
  effects become worse as the PSF size increases, with impacts on galaxy photometry and shear
  estimation that are not yet well-quantified.
\item At a lower level, we also are sensitive to Tables 12 and 13 in the LSST~SRD, which
  cover the spatial profile of the PSF (not too much power in the wings) and the PSF ellipticity
  distribution, respectively.  If the latter is imperfectly removed in software when estimating
  shear, it can generate additive systematics in the shear-shear correlation functions.  However,
  state-of-the-art shear estimation methods are quite effective at removing the PSF anisotropy from
  weak lensing shear estimates, and there are null tests to effectively diagnose this issue, so
  these are lower level effects than the image quality assumptions in the bullet point above.
\item Tables 14-17 in the LSST~SRD cover various aspects of photometric calibration.  The
  connection between these requirements and the DESC supernova science case is explored in detail in
  \autoref{subsec:sn}.  In this version of the DESC~SRD, we have not quantified to what extent the
  Rubin Observatory requirements on photometric calibration are important for meeting our goals with
  respect to control of photometric redshift uncertainties.
\item Section 3.3.5 covers requirements on the astrometry, which enters our dark energy observables
  in ways that we have not explicitly quantified in this version of the DESC~SRD.
\item Table 27 provides requirements on the PSF model ellipticity residuals (i.e., difference
  between PSF model ellipticity and the true PSF ellipticity).  Significant PSF model ellipticity
  residuals would mean we are effectively removing the wrong PSF anisotropy from galaxy shear
  estimates, which generates additive biases in the weak lensing shear-shear correlation functions
  that must be quantified and removed through null tests.  Hence, the requirements in Table 27 in the LSST~SRD
  reduce the burden on the DESC in diagnosing such effects.
\end{itemize}

%% file: inc/software.tex
\subsection{Software packages}

Here we briefly describe the software used for forecasting and setting requirements in this version
of the DESC~SRD.

For weak lensing, galaxy clustering, and galaxy cluster analysis, we use
CosmoLike\footnote{\url{https://github.com/CosmoLike}, \url{http://www.cosmolike.info/}}
\citep{2017MNRAS.470.2100K}.  Use of the same software is important, as these three probes of
large-scale structure are correlated with each other and hence must be treated self-consistently
especially in joint probe forecasts.  CosmoLike can model all cross-correlations among probes, with
analytical non-Gaussian covariances, and a variety of self-calibrated and calibratable systematic
uncertainties.  Indeed, the choice of which systematic uncertainties to include (and in what form)
in this version of the DESC~SRD was largely driven by the existing capabilities of CosmoLike,
though in a few highlighted cases in \appref{app:baselines} our approach represents a
departure from \cite{2017MNRAS.470.2100K}.  All CosmoLike forecasts in the DESC~SRD are carried out
through a Fisher forecasting approach.  The limitations of this approach include the fact that it
assumes a multivariate Gaussian likelihood and the fact that the results are numerically somewhat sensitive to
the step size of the derivatives (resulting in potentially 5--10\% variations in figures of merit).
The assumption of a multivariate Gaussian likelihood is generally more problematic for
geometric probes than it is for probes of structure growth \citep{2012JCAP...09..009W}.
Having many poorly-constrained directions in parameter space, even in dimensions that are being
marginalized over, can be particularly problematic for convergence of the FoM calculated from the
Fisher matrices. Our default
priors on cosmological parameter space (\appref{subsec:assump-cosmo}) are relatively broad.  To
achieve more stable results, the individual probe calculations from CosmoLike used Stage III priors on the five cosmological parameters that are marginalized over, i.e., everything
but $w_0$ and $w_a$, as described in \appref{subsec:assump-cosmo}.

As a comparison point with CosmoLike, we used
GoFish\footnote{\url{https://github.com/damonge/GoFish}}, a completely independent code base that
can carry out Fisher forecasting.  Our comparison between CosmoLike and GoFish\footnote{See the end
of \url{https://github.com/LSSTDESC/Requirements/issues/6} for details.} involved forecasting
constraints in the $(w_0, w_a)$ plane for the 3$\times$2pt analysis without any systematic
uncertainties, but
very similar baseline data vectors. CosmoLike used non-Gaussian covariance matrices,
took a conservative approach in not including LSS cross-bin correlations, and both codes used slightly
different redshift bins and bandpowers. This comparison showed that the parameter constraints from CosmoLike
were indeed weaker than those found with GoFish by roughly 30--40\% for $w_0$ and 10--20\% for $w_a$. This
can be expected given the differences in both baseline analysis and approaches to covariances.
As a more quantitative validation of CosmoLike, both codes generated forecasts for exactly the same
baseline data vector (the shear-shear power spectrum for the Y1 survey parameters) in the absence of
systematics. To isolate the impact of the non-Gaussian covariance, GoFish was modified to take in the
covariance estimated by CosmoLike as input. In this case, the values for the DETF FoM found by both
codes were the same within $1\%$, well within the expected numerical fluctuations due to the
instability associated to the numerical derivatives. This test case was also used to quantify the impact
of the non-Gaussian terms in the covariance matrix. For this GoFish was re-run using the same data vector
as well as its internally-computed Gaussian covariance. The effect of the non-Gaussian terms was found to
be minimal ($\sim$2--5\% of the FoM, in agreement with expectations given the large sky fraction assumed
for LSST).

For the shear-shear forecasts, an MCMC vs. Fisher matrix comparison was carried out using CosmoLike,
including marginalization over self-calibrated systematic uncertainties.  The results between the
two methods were found to agree to within our aforementioned $\sim$10\% tolerance.

The software frameworks used for supernova and strong lensing forecasts are dedicated probe-specific
frameworks that are described in Appendices~\ref{app:sn} and~\ref{app:sl}, respectively.

The joint forecasts were carried out in CosmoLike, by combining Fisher matrices from the three
different code bases described above.  As noted in \autoref{subsec:combprobes}, this implies that
we are treating the SL, SN, and the combination of WL$+$LSS$+$CL as carrying completely independent
information, allowing us to combine Gaussian approximations to their posterior probability
distributions rather than doing a full joint likelihood analysis.
This should be an excellent assumption for the foreseeable future. 

Future versions of the DESC~SRD will use DESC software for describing cosmological observables and
their covariances, TJPCosmo and TJPCov, so as to enable the forecasts to use the same models for
systematic uncertainties and how they impact dark energy observables as DESC working groups are
defining for their likelihood analysis of real LSST data.  This will allow us to carry out
cross-checks on the software more easily than can be done now, while also ensuring that all
calculations use software that has undergone collaboration-wide review.

\subsection{How requirements are set}\label{app:requirements}

In practice, the detailed requirements in \autoref{sec:detailedreq} were placed by (a)
generating contaminated data vectors with some systematic bias, (b) carrying out cosmological parameter
constraints via Fisher forecasts while ignoring the fact that the initial data vectors had a
systematic bias in some calibratable effect (while marginalizing over self-calibrated systematic
uncertainties).  This process was carried out until the best-fitting parameters in the $(w_0, w_a)$
plane reached the edge of the $1\sigma$ ellipse defined by the marginalized statistical
uncertainties. The question of whether a particular systematic reached the edge is determined as
follows: given a choice of baseline systematic parameter $s_0$ (such as a bias in shear calibration); a 2D vector of biases in $w_0$ and $w_a$ determined by comparing the best-fitting and
fiducial cosmological parameters, called $\mathbf{b}$; and a covariance
matrix for $w_0$ and $w_a$ after marginalization over nuisance parameters and the other cosmological
parameters\footnote{To be more specific, given a covariance $\mathbf{C}$ with rows for all
  cosmological parameters and nuisance parameters, we take the submatrix corresponding to the
  $(w_0,w_a)$ dimensions.  This is implicitly marginalized over all other dimensions.  In contrast,
  taking the subset of the Fisher matrix ($\mathbf{C}^{-1}$) would correspond to all other
  dimensions being fixed, resulting in overly small uncertainties and too-conservative
  requirements.}, $\mathbf{C}$, the linear distance between the best-fitting point and the edge of the ellipse is
\begin{equation}\label{eq:requirement}
r = \sqrt{\mathbf{b}\cdot\mathbf{C}^{-1}\cdot\mathbf{b}}.
\end{equation}
The $1\sigma$ (systematic and statistical uncertainty are equivalent) requirement
value is $s_0/r$.  In other words, if our calculation results in $r=1$, it means that our baseline
systematic parameter $s_0$ has
gotten us precisely to the boundary of the error ellipse defined by the covariance matrix.  If this
is the only systematic bias under consideration, and if $f_\text{sys}$ defined by \autoref{eq:fsys}
is 1, then our requirement would be that the residual
systematic bias must be less than or equal to $s_0$.
Since individual sources of systematic uncertainty are not allowed to take up the
entire error budget, this number is then further rescaled by whatever fraction of the error budget
is to be allocated to this effect.  In doing so, we use the quadrature summation, i.e., if there are
two sources of uncertainty that are to be given equal fractions of the error budget,
we set the requirement such that $r=1/\sqrt{2}$ for each of them;
this assumes independence of the systematic errors under consideration, such that variances of the
systematic error distribution add\footnote{It does not assume Gaussianity of these distributions,
  just that they can be considered independent, such that their error distributions are convolved.
  In that case, the variances add even if the distributions are non-Gaussian.}.

We implicitly assume a linear scaling of $\mathbf{b}$ with the
value of $s_0$, which was confirmed to be a good approximation for the sources of systematic
uncertainty considered in the supernova analysis (\appref{app:plots}) for small
values of systematic contamination, but is unlikely to be
true when considering very large variations in $s_0$.  For this reason, convergence tests were
carried out for a few limited cases where $r$ differed by orders of magnitude from $1$, by modifying
the value of $s_0$ and re-estimating the $\mathbf{b}$ and the requirement.

In the case that $w_0$ and $w_a$ confidence ellipses are aligned with the axes in that 2D space,
\autoref{eq:requirement} reduces to
\begin{equation}
r = \sqrt{\left( \frac{b(w_0)}{\sigma(w_0)}\right)^2 + \left(\frac{b(w_a)}{\sigma(w_a)} \right)^2}
\end{equation}
which intuitively makes sense: the biases for $w_0$ and $w_a$ are rescaled by their uncertainties,
and we measure the hypotenuse of a right triangle in that rescaled space.

The choice to aim for $r=1$ in \ref{high:sys} was motivated, as previously mentioned, by the intent to not allow
offsets in the $(w_0,w_a)$ plane in any direction to be outside the $1\sigma$ error ellipse.  One
could interpret the quantity $\mathbf{b}\cdot\mathbf{C}^{-1}\cdot\mathbf{b}$ as a $\Delta\chi^2$ for
the biased measurement; in our 2D parameter space, it may seem natural to set requirements at
$r^2=2.3$, the $\Delta\chi^2$ for $1\sigma$ offsets in the case of a Gaussian likelihood.
What would effectively happen in this case is
that there could be some specific subsets of directions in our 2D space for which the systematic
bias would go outside the $1\sigma$ ellipse.  Since we are likely to care about such offsets in the
era of LSST, when there are other Stage-IV experiments that will further reduce the
interesting area in this parameter space, we use the stricter cut at $r=1$.  Situations in which one
might cut based on the $\Delta\chi^2$ for a multivariate Gaussian with the dimensionality of the
space in which $\mathbf{b}\cdot\mathbf{C}^{-1}\cdot\mathbf{b}$ is calculated are generally those in which the tests for biases are being done in a high-dimensional space
that has many dimensions that are not physically interesting, unlike here.

 When estimating biases in cosmological parameter space due to use of a contaminated
data vector as described above, with marginalization over models for self-calibrated systematic uncertainties, it is possible that
the biases can be absorbed into the
self-calibrated systematic model, if it is sufficiently flexible (as are some of our adopted
models).  This has the potential to result in very loose requirements on calibratable systematic uncertainties.  In a
realistic measurement, we would have done much more careful tests for interactions between models
for various systematic uncertainties, and would be explicitly controlling for such effects rather
than allowing for very large calibratable systematic uncertainties while counting on self-calibrated
systematics models to absorb residual biases.  Understanding the interactions between various
types of systematic uncertainties, their parametrizations, and priors on nuisance parameters is an
active area of research within DESC, and future DESC~SRD versions will consider this issue more
carefully.

Finally, we note that \autoref{eq:requirement} is agnostic of the space in which the analysis
is carried out.  That is, one could in principle carry it out using the observables ($C_\ell$, etc.)
and their covariance, or in cosmological parameter space.  As described above, we have opted to do
this in cosmological parameter space.  There are advantages and disadvantages to this choice.  The
main advantage is that if you have a systematic error that causes a large shift in the observables
$\hat{X}$ that looks very different from $\partial\hat{X}/\partial w_0$ or $\partial\hat{X}/\partial
w_a$ (the response of the
observables to a change in the dark energy equation of state), working in $(w_0,w_a)$ space produces
appropriately weaker requirements on that source of systematic uncertainty compared to those that
actually do mimic the change in observables due to dark energy.  This may in principle be the right
thing to do, relying on the fact that systematic biases that look nothing like changes in dark
energy models would be
flagged and corrected due to what they do elsewhere in cosmological parameter space (e.g., produce
results for other parameters that are inconsistent with Stage III priors, or yield a terrible overall
$\chi^2$).  However, it does mean that our requirements are somewhat more sensitive than they might
otherwise be to our parametrization and whether it results in changes in observables $\Delta\hat{X}$
that mimic those induced by dark energy; see for example the relevant discussion on shear
calibration bias in \appref{subsubsec:wl-sysuncert}.

Several of these issues with systematics parametrizations and priors
have also been raised previously in e.g.\ \citet{2013MNRAS.429..661M} and
\citet{2017MNRAS.470.2100K}.

\subsection{Ensuring reproducibility}

To ensure reproducibility of the calculations in this version of the DESC~SRD, we take the following
steps.

First, for calculations in \appref{app:dndmag}, all analysis scripts are in the DESC's
Requirements repository\footnote{\url{https://github.com/LSSTDESC/Requirements}}.  The relevant
version of each script is associated with the tag of the repository for that document version.

For CosmoLike calculations, there are two private repositories that are relevant.  The first is the
`cosmolike\_core' repository, which has the bulk of the CosmoLike infrastructure.  The second
(DESC\_SRD\footnote{\url{https://github.com/CosmoLike/DESC_SRD}}) is a dedicated repository containing scripts, configuration files, and other analysis
routines specific to this document.  Both repositories have a v1 tag containing scripts used to
produce v1 of this document; moreover, plotting routines and associated data products are made
available in the tarball released with v1 of this document on Zenodo.

For strong lensing forecasts, the associated scripts are stored in the `forecasting/SL' directory in
the Requirements repository and in the tarball released with v1 of the document.  Instructions on
how to rerun the forecasts are given in the file
README.md in that directory.  Similarly, the software for supernova forecasts and requirements can
be found in the `forecasting/SN' directory of the repository and tarball.

%% file: inc/assumptions.tex
\subsection{The LSST observing strategy}\label{subsec:assump-strategy}

We use
the OpSim\footnote{\url{https://github.com/lsst/sims\_operations}} v3 \verb minion_1016 ~run
\footnote{\url{https://www.lsst.org/scientists/simulations/opsim/opsim-v335-benchmark-surveys}}
to define the Y1 and Y10 survey depths and area, and to simulate the sample of supernovae and
estimate the number with well-sampled light curves.  Our Y1 definition
involves taking the first 10\% of the WFD
observations\footnote{For details of why this is not simply the first 10\% of nights, see extensive
discussion in \url{https://github.com/LSSTDESC/Requirements/issues/9}.  In short, the OpSim run
has a spurious preference for the Deep Drilling Fields in the first year that leads to the wide survey area being
under-observed.  Since this is not expected to be part of the actual survey
strategy, we used an ad-hoc correction for this.}, which gives median (across the survey) $5\sigma$
point-source detection depths of 24.07, 25.60, 25.81, 25.13, 24.13, 23.39 in $ugrizy$ after 1 year,
and 25.30, 26.84, 27.04, 26.35, 25.22, 24.47 after 10 years. These median
depths were determined after discarding areas that have a depth shallower than $i<24.5$ and 26 for
Y1 and Y10, respectively. (These were chosen to be 0.4 and 0.7 magnitudes deeper than the LSS galaxy
samples for Y1 and Y10.) The resulting areas after this homogenization process are
12.3k deg$^2$ (Y1) and 14.3k deg$^2$ (Y10).  Note that the depth cut removes areas of high
extinction near the Galactic plane, because the depths used for the above cuts are after accounting
for dust extinction.  We have explicitly confirmed that regions near a Galactic latitude of zero and
those with $E(B-V) \gtrsim 0.2$ are
eliminated by our depth cut (more specifically, 3.8\% and 0.7\% of the area exceeds $E(B-V)=0.2$ in
Y1 and Y10, or 0.7\% and 0\% for $E(B-V)=0.3$).

\subsection{The cosmological parameter space}\label{subsec:assump-cosmo}

For forecasting, we assume a $w_0w_a$CDM cosmology.

Our
  forecasts for the static probes use the Fisher matrix formalism, with weak priors necessary to regularize the Fisher matrix when individual probes exhibit significantly non-Gaussian likelihoods. Following
  \citet{2017MNRAS.470.2100K} Table 1, we fit for the following seven parameters, with
  fiducial parameter values and $\sigma$ of the Gaussian prior listed in parenthesis:
%
$\Omega_m$ $(0.3156; 0.2)$; $\sigma_8$ (0.831; 0.14); $n_s$ (0.9645; 0.08); $w_0$ (-1.0; 0.8); $w_a$
  (0.0; 2.0); $\Omega_b$ (0.0492; 0.006); $h$ (0.6727; 0.063).

\begin{figure}
\begin{center}
\includegraphics[width=0.65\textwidth]{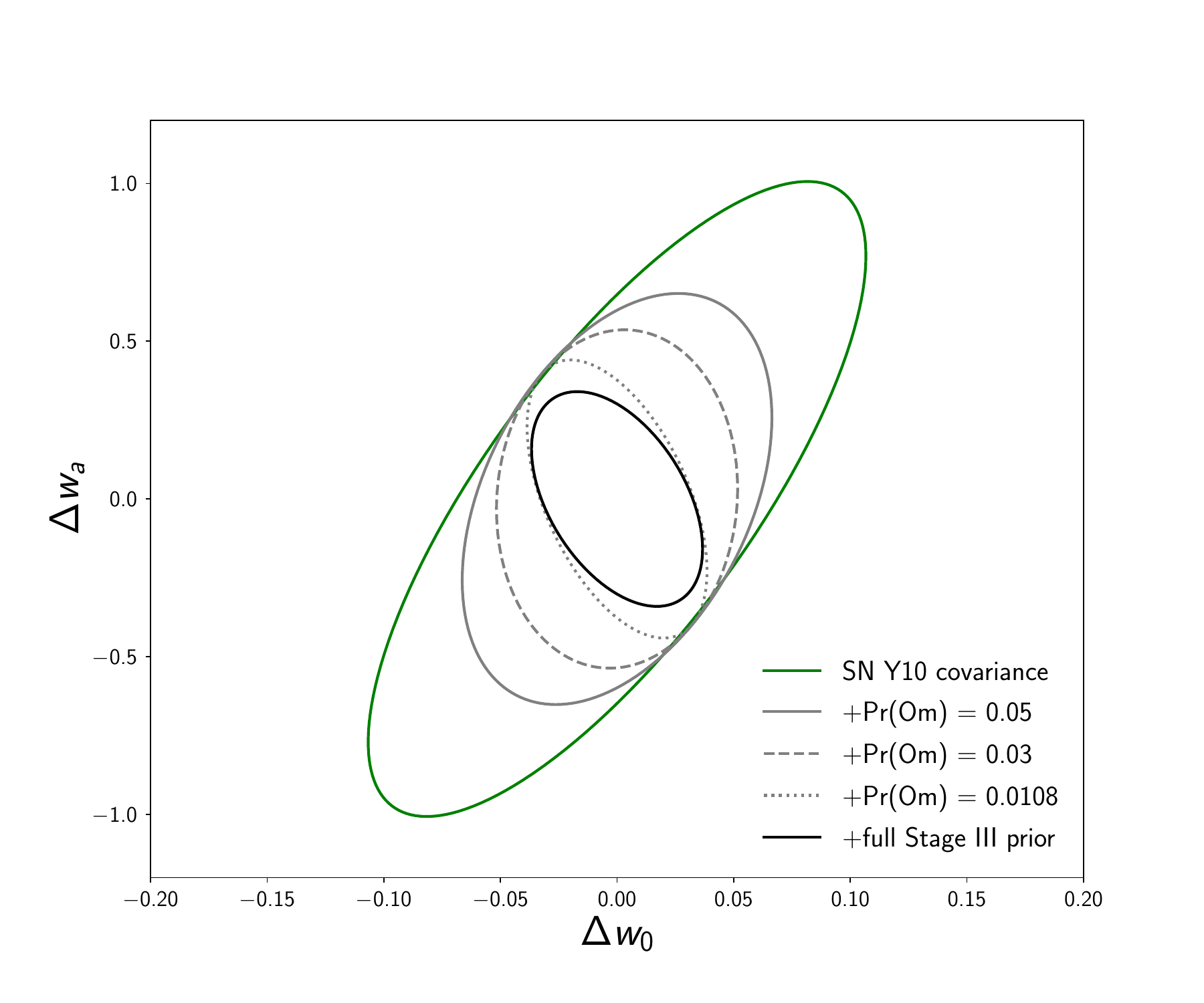}
\caption{$95\%$ confidence ellipses for the supernova science case as a function of prior on $\Omega_m$. For low-matter universes, the dark energy equation of state lies along the positive $w_0-w_a$ degeracy direction, however a tight prior around the fiducial matter density focuses and rotates the dark energy equation of state parameters along the degeneracy direction for acceleration. \label{fig:sndegen}}
\end{center}
\end{figure}

  The supernova forecasts begin with MCMC, and involve only the four of the aforementioned cosmological
  parameters to which supernova measurements are sensitive: $\Omega_m$, $w_0$, $w_a$, and $h$.  We also vary the supernova intrinsic magnitude $\mathcal{M}$ (which is completely degenerate with the Hubble constant) using a flat prior over the prior range $-20 <\mathcal{M} < -18$. 
Beyond the weak priors mentioned above, the supernova forecasts have one additional prior imposed within CosmoSIS: $w(a)<0$ for all $a$. We note that the direction of the degeneracy  ellipse in the $w_0-w_a$ plane depends sensitively on the width of the assumed priors on these or other parameters. In particular, we find that for wide priors on $\Omega_m$, the degeneracy lies along the $w_a \approx -(w_0+1)$ line, but rotates towards a line of $w_a \approx +(w_0+1)$ as the $\Omega_m$ priors are tightened (assuming uninformative priors on intrinsic magnitude parameter), as it changes the pivot point in redshift space where the dark energy constraints are tightest. This effect is seen for generic dark energy models and is not specific to the dark energy parameterization considered here \citep{PhysRevD.75.083503}. This effect has been studied previously for upcoming surveys \citep[see e.g.][]{2001A&A...380....6G}. The exact slope of the line is also dependent on the redshift distribution of the sample, the particular systematics included and the prior on the intrinsic magnitude; a more exhaustive study is forthcoming in a separate paper. We show the behavior for changing the $\Omega_m$ prior in Figure~\ref{fig:sndegen}. The supernova parameters ($\mathcal{M},H_0, \Omega_m, w_0,w_a$) are sampled via MCMC and the resultant chains are processed to produce a covariance matrix. That covariance matrix is then sampled with the other three cosmological parameters not relevant to the SN likelihood to produce the full 7-parameter chains. 

We
  show parameter constraints in the ($X,Y$) plane for any pairs of parameters as $(\Delta X,
  \Delta Y)$ after subtracting off the fiducial values, to de-emphasize our choice of fiducial
  values (which do have some impact on the forecasts).  Note that we are fixing the neutrino mass
  $m_\nu=0$ for the forecasts in the baseline analysis even though neutrinos have mass.  This is for
  the sake of expediency: fixing the neutrino mass to a nonzero value should have little impact on
  the $(w_0, w_a)$ constraints while inflating the run-time by a factor of $\sim 2$.  For future
  DESC~SRD versions, it may be worth
  revisiting this choice.  Also, this forecast does not allow for curvature, unlike the forecasts in
  the DETF report.  In principle, this means that our comparison of FoMs against ones from that
  report is overly optimistic.  However, this is only important for supernovae, not the other probes
  and not the joint forecast\footnote{To quote from the DETF report: ``Setting the spatial curvature
    of the Universe to zero greatly strengthens the dark energy constraints from supernovae, but has
    a modest impact on the other techniques once a dark-energy parameterization is selected. When
    techniques are combined, setting the spatial curvature of the Universe to zero makes little
    difference to constraints on parameterized dark energy, because the curvature is one of the
    parameters well determined by a multi-technique approach.''}.  Since our primary concern is the
  joint forecast, we do not attempt to account for this over-optimism in the comparison of the
  supernova forecasts against those from Stage III.

When including Stage III priors, the ones that we use correspond to the scenario described in section~6.3 and shown
with dark blue contours in figure~28
of~\citet{2016A&A...594A..13P}, the ``TT+TE+EE+lowP+lensing+ext'' chain with $w_0$ and
$w_a$ free.  This prior includes Planck
polarization and lensing (neglecting all cross-terms with LSST probes) and the following external
data (``ext''): BOSS, JLA, and $H_0$, with the exact analysis used for these three probes described
in subsections~5.2--5.4 of \citet{2016A&A...594A..13P}. Without a prior on the matter density,
 the supernova posterior is considerably non-Gaussian and so the contours produced by computing the covariance of the supernova chains are broader than the Fisher matrix contours from the SN likelihood. The addition of the Stage III prior significantly Gaussianizes the contours.

\subsection{Stage III dark energy surveys}\label{subsec:assump-stage3}

We must assume some Stage-III DETF FOM for each probe in order to quantify whether we meet \ref{high:indivfom}.
The derivation of the Stage-III numbers is described below.
\begin{itemize}
\item WL: Both scenarios presented in the DETF report for Stage-III WL are too optimistic in certain
  key respects, using an area comparable to the full DES\footnote{\url{https://www.darkenergysurvey.org/}} survey area but
  a much higher effective source number density (15/arcmin$^2$) and redshift ($\langle z\rangle\sim
  1$). These can be compared with the DES Y1 shear catalog paper \citep{2017arXiv170801533Z}, which reports number densities from two catalogs,
  the larger of which has 6.5/arcmin$^2$ and a mean redshift $\langle z\rangle\sim 0.6$, constructed using
  state-of-the art methodology that is unlikely to evolve in ways that bring DES substantially
  closer to the DETF configuration.  The increased depth in later years, with roughly doubled
  exposure time, should yield tens of percent higher number densities, but not a factor of 2.5.  In
  addition the DETF assumes 20\% lower shape noise (lower $\sigma_\gamma^2$) than is found in DES in
  practice.  For all of these reasons, the signal is lower and the noise is higher in DES
  than in the DETF forecast,
  meaning that the real Stage III numbers will be substantially more shape noise-dominated
  than even the DETF pessimistic scenario.  In the absence of a more realistic Stage III forecast,
  we use the DETF pessimistic FoM (20) as our benchmark for LSST WL.  The fact that we will compare
  it with an LSST forecast 3x2-point analysis partially accounts for the fact that even the
  pessimistic forecast may be too optimistic in terms of statistical precision.
\item LSS: The DETF report only considers BAO, not a full LSS analysis to smaller scales.  We
  nonetheless use a geometric mean of their Stage-III pessimistic and optimistic photometric BAO
  FoMs, which yields a value of 0.76.
\item CL: The DETF considers number counts and cluster lensing.  Their optimistic cluster lensing
  forecasts assume too much prior knowledge of the mass-observable relation compared to our
  assumptions here, so we use the
  DETF pessimistic FoM of 6.
\item SL: The DETF does not include strong lensing time delays among the probes they consider, so we
  must estimate the Stage III FoM some other way.  Based on reasonable
  assumptions about what Stage III strong lensing dark energy analysis will include\footnote{See discussion in \url{https://github.com/LSSTDESC/Requirements/issues/17}, which
    resulted in a Stage-III configuration slightly more conservative than that described in \cite{2016A&ARv..24...11T}
    for time delay lenses, but with the inclusion of the compound lenses not discussed there.},
  we assume 3 compound lenses each with 1.7\% distance measurements, and 15 time delay lenses each with
  7\% distance measurements.  Given this Stage III survey definition and our adopted priors, the Stage-III FoM is 0.65.
\item SN: The geometric mean of the optimistic and pessimistic scenarios in the DETF report gives a FoM of 9.4.
\end{itemize}

\subsection{Follow-up observations and ancillary data}\label{subsec:assump-followup}

As described in \autoref{sec:intro}, some of our science cases assume that already-funded surveys
will be carried out and that spectroscopic follow-up and other ancillary telescope resources will
continue to be available at similar rates as they are today.  Below the assumptions for specific
science cases are summarized, with references to where further discussion can be found.
\begin{itemize}
\item WL: The WL baseline analysis outlined in \appref{app:wllss} does not rely directly on
  ancillary telescope resources.  It is expected that spectroscopic samples will be needed to meet
  the requirements on
  knowledge of photometric redshift bias and scatter presented in \autoref{req:wl}.  However, as
  described in \autoref{subsec:lssreq}, the overlaps of the LSST footprint with DESI and 4MOST should enable us to meet
  those requirements.
\item LSS: The LSS baseline analysis outlined in \appref{app:lss} does not rely directly on
  ancillary telescope resources, though the same considerations about spectroscopic data to meet the
  requirements on knowledge of photometric redshift errors mentioned for WL apply to LSS.
\item CL: The CL baseline analysis outlined in \appref{app:cl} makes conservative
  assumptions about available multiwavelength (X-ray and SZ) data to help place priors on the
  mass-observable relationship, as described in \appref{subsubsec:cluncert}.
\item SL: The SL baseline analysis outlined in \appref{app:sl} assumes readily achievable amounts of
  follow-up both to constrain the lens model parameters over which we must marginalize and to
  provide lens and source spectroscopic redshifts.  The modest sample size is set in part by the
  need for follow-up, with the follow-up needs described in detail there.
\item SN: The SN baseline analysis outlined in \appref{app:sn} assumes (a) the presence of a
  low-redshift external supernova sample (based on an ongoing survey), and (b) host spectroscopic
  redshifts, with the usable sample size strictly limited by reasonable assumptions about host spectroscopy using
  4MOST for the WFD supernova sample and PFS or DESI for the DDF sample.  See that appendix for more
  details.
\end{itemize}

%% file: inc/lss.tex
The default LSS (galaxy clustering only) analysis is a tomographic one with galaxy two-point
correlations.  BAO information is implicitly included in the tomographic analysis of the 2-point
correlations.  For this DESC~SRD version, the baseline LSS analysis is simply the galaxy-galaxy part
of the WL$+$LSS analysis described in the next subsection.  That is, the WL$+$LSS analysis has a
``lens'' sample and a ``source'' sample for which auto- and cross-correlations are measured.  The
sample described in this section is the lens sample for WL$+$LSS (while the source sample is
described in \appref{app:wllss}).

\subsubsection{Analysis choices}

Here we describe the essential points of the LSS analysis setup in this version of the DESC~SRD:

\begin{itemize}
\item We assume 10 tomographic bins spaced by $0.1$ in photo-$z$ between
  $0.2\le z\le 1.2$ for Y10, and 5 bins spaced by $0.2$ in photo-$z$ in that same redshift range for
  Y1.
\item The data vector consists of the angular power spectra, $C_\ell$. For clustering, unlike for shear, we only use the auto-power spectra (which carry the vast majority
  of the cosmological information), not cross-spectra
  between different redshift bins.
\item Since the astrophysical issues determining what scales to use are tied to physical scale, we
  use $k_\text{max}\sim 0.3~h/$Mpc.  This cut is based on work in progress indicating that this is
  the scale where nonlinear bias results in $\sim$10\% deviations from the linear bias for $z\sim
  0.5$ galaxies. Following \cite{2014JCAP...05..023F}, this should be
  taken as an effective maximum wavenumber. Statistical precision of
  LSST will be considerably better than 10\%, but it is assumed that
  these differences
  will be accounted for by fitting for beyond-linear bias parameters
  using scales slightly larger than $k_\text{max}$.
\item In order to enable combined probe analysis, we define a common set of $\ell$ bins for all
  large-scale structure analyses (LSS, WL, CL).  There are $20$ logarithmically-spaced $\ell$ bins,
  covering $20\le \ell \le 15000$ (where this value is adopted to accommodate the galaxy cluster
  lensing profiles in the 1-halo regime).  These limits are the bin edges, not centers.
 For the LSS analysis, an $\ell_\text{max}$ is chosen
  for each redshift bin based on its $k_\text{max}$ and redshift distribution as follows:
\begin{equation}
\ell_\text{max} = k_\text{max} \chi(\langle z\rangle) - 0.5.
\end{equation}
All $\ell$ bins
  above that $\ell_\text{max}$ value in our standardized $\ell$ binning are discarded.
\item The nonlinear matter power spectrum and covariances are calculated
  following \cite{2017MNRAS.470.2100K}, using the \citet{2012ApJ...761..152T} prescription for the
  nonlinear power spectrum and the transfer function from \citet{1999ApJ...511....5E}. 
\item For the Y10 lens sample, we use the gold
  sample\footnote{See chapter 3 of the LSST science book, \citet{2009arXiv0912.0201L}, \url{http://www.lsst.org/sites/default/files/docs/sciencebook/SB\_3.pdf}
    (Equation~3.8).} which has a
  limiting magnitude of $i_\text{lim}=25.3$.  The overall normalization of the number density is
  estimated
  based on the HSC Deep survey (see \appref{subsec:overalln} for details), 48~arcmin$^{-2}$,
  and the intention was to use other parameters from the LSST Science book\footnote{This $\sigma_z$ value is optimistic; it is close to the LSST design
  specifications without inclusion of effects such as template uncertainty or the impact of
  incomplete spectroscopic training samples.  However, we defer a more realistic model (including a
  more realistic distribution, not just the width, and including stronger redshift evolution) to future DESC~SRD versions; in any case, the
  statistical power of the LSS analysis is largely limited by the need to marginalize over galaxy
  bias rather than by the photo-$z$ scatter.}: $b(z) = 0.95/G(z)$, where $G(z)$ is the normalized
    growth factor with $G(z=0)=1$; and
  $\sigma_z=0.03(1+z)$.   In practice, after tagging v1 we found that a different set of bias values
    was used\footnote{For Y1, the per-bin bias values that were actually used were $[1.562362,
    1.732963, 1.913252, 2.100644, 2.293210]$.  For Y10, they were $[1.376695,1.451179,1.528404,1.607983,1.689579,1.772899,1.857700,1.943754,2.030887,2.118943]$.} for the data vectors and covariances, but for a cosmic variance-dominated
    sample and when marginalizing over a separate galaxy bias in each bin, this should not affect the forecasts. We use a parametric redshift distribution,
\begin{equation}\label{eq:nz}
\frac{dN}{dz} \propto z^2 \exp{[-(z/z_0)^\alpha]}
\end{equation}
with $(0.28,0.90)$ for Y10.  See
  \appref{subsec:dndzphot} an explanation of the origin of the best-fitting values given here
  and a comparison with real
  and simulated data.
\item For the Y1 lens sample, we use the equivalent of the gold sample, cut off 1 magnitude
  brighter than the (shallower) median survey depth, giving $i_\text{lim}=24.1$ (see
  \appref{subsec:assump-strategy} for an explanation of how the limiting magnitudes were determined).
Again we define the number density following empirical results
  from the HSC survey (\appref{subsec:overalln}), finding a value of 18~arcmin$^{-2}$.  Similarly the parametric
  form for the redshift distribution is above, with $(z_0,\alpha)=(0.26,0.94)$ for Y1 (see \appref{subsec:dndzphot}).  Based on
  the typical luminosity-bias relation, we use\footnote{The text states the intended bias-redshift
  relation, but see previous footnote for numbers that can be used when comparing data vectors and
  covariances with those from the DESC~SRD v1.} $b(z)=1.05/G(z)$ and use the same photo-$z$ scatter,
  $\sigma_z=0.03(1+z)$.
\end{itemize}

\subsubsection{Anticipated improvements}

\begin{figure}[!htbp]
  \begin{center}
    \includegraphics[width=0.8\textwidth]{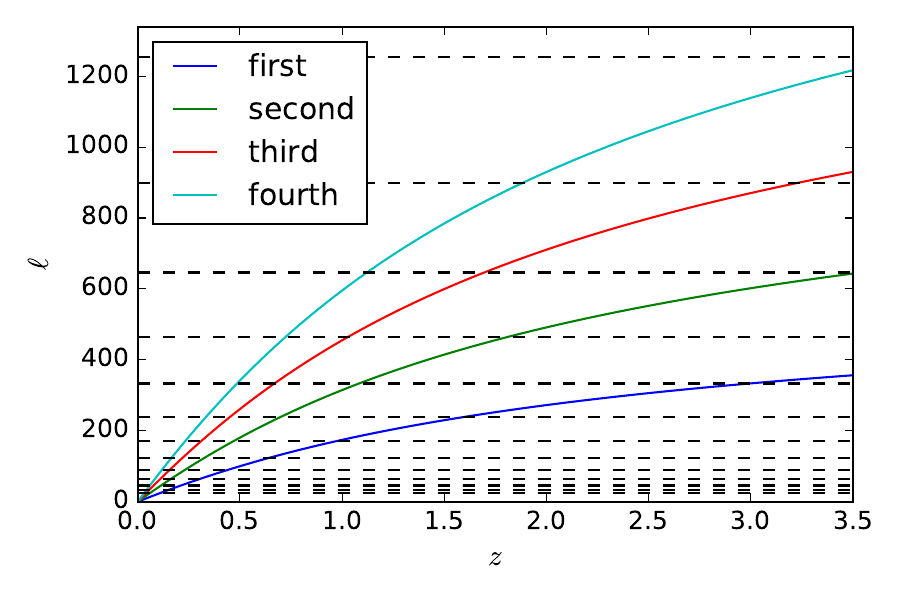}
  \end{center}
  \caption{The positions of the first four BAO peaks as a function of redshift are shown in color,
    while the horizontal lines show the centers of our $\ell$ bins in this DESC~SRD version.
In order to Nyquist sample, we
    need two $\ell$ bins (i.e., horizontal lines) between each colored line.  A finer binning, and
    possibly linear rather than logarithmic $\ell$ binning, would yield a more optimal measurement
    of the BAO feature from the galaxy angular power spectrum.
\label{fig:bao-binning}}
\end{figure}

Ideally the LSS analysis would make use of a red sample with better photo-$z$ in a multitracer
analysis using both the gold sample and the red sample.  Future DESC~SRD versions should use
multitracer analyses if possible, and also enable use of the ``lens'' galaxies beyond $z=1.2$ in the
clustering analysis.  While most information about cosmology is included in the auto-spectra, there
is some additional information in the cross-spectra between bins that it would be useful to include
in the future (especially once our analysis has a more complicated treatment of photo-$z$ errors, which
tend to increase the cross-bin correlations).  Modified $\ell$ binning will be important to more
optimally include the BAO feature; the current broad $\ell$ binning scheme is likely quite
suboptimal. \autoref{fig:bao-binning} may be used to guide future efforts to derive a better binning
scheme. Finally, incorporating the impact of survey inhomogeneity in Y1 would be
valuable, if a simple parametrization for its effect on the observables can be derived.

\subsubsection{Systematic uncertainties}

For LSS, we consider the following classes of systematic uncertainties in our two categories:
\begin{itemize}
\item Self-calibrated systematics: galaxy bias, magnification, baryonic effects on the matter power
  spectrum.
\item Calibratable systematics on which we place requirements: any photo-z issue,
  sky contaminants.
\end{itemize}

The models for the self-calibrated
systematic uncertainties are summarized in \autoref{tab:lsssys-selfcal}.  For each source of
systematic uncertainty, we describe how
it is included in this DESC~SRD version, as well as aspirations for more complex models to be
included in future.  Currently, our data vectors are produced assuming linear galaxy biasing due to
lack of an adopted and validated $b(k)$ model, and
marginalization will use $b$ defined in redshift bins without assuming a parametrized $b(z)$ model
-- hence $N$ bins leads to $N$ nuisance parameters.  For a given $z$ bin, the fact that we do not
have to marginalize over nonlinear bias implies more statistical power left for cosmology
constraints than would be present in reality (since we will have to marginalize over nonlinear
bias); but the fact that we need a per-bin bias (rather than having a parametrized model) does
result in many bias parameters.  Incorporating a realistically complex nonlinear bias model with an
appropriate number of nuisance parameters is the highest-priority
update to make for the LSS analysis in the next DESC~SRD version.

\begin{table}[!htbp]
\begin{center}
\begin{tabular}{p{0.8in}p{2.4in}p{2.8in}} \hline
Self-calibrated systematic uncertainty& Current model & Future plans \\ \hline
Galaxy bias & Linear galaxy bias, one value per tomographic bin (Gaussian prior, mean$=1.9$ and $\sigma=0.9$) & Nonlinear galaxy bias with a
redshift-dependent parametrization of the linear bias vs.\ redshift, and at least one nonlinear bias
parameter \\
Magnification & None & Self-consistent convergence field and luminosity function
as what goes into the shear and intrinsic alignments in 3x2pt analysis, following e.g.\
\protect\cite{Joachimi10}; marginalize over uncertainty in slope of number counts \\
Baryonic effects & None & Sufficiently complex nonlinear galaxy bias model that it can absorb
modifications to the matter power spectrum due to baryonic effects \\
\hline
\end{tabular}
\caption{Self-calibrated systematic uncertainties for LSS.}
\label{tab:lsssys-selfcal}
\end{center}
\end{table}

\begin{figure}[!htbp]
  \begin{center}
    \includegraphics[width=0.8\textwidth]{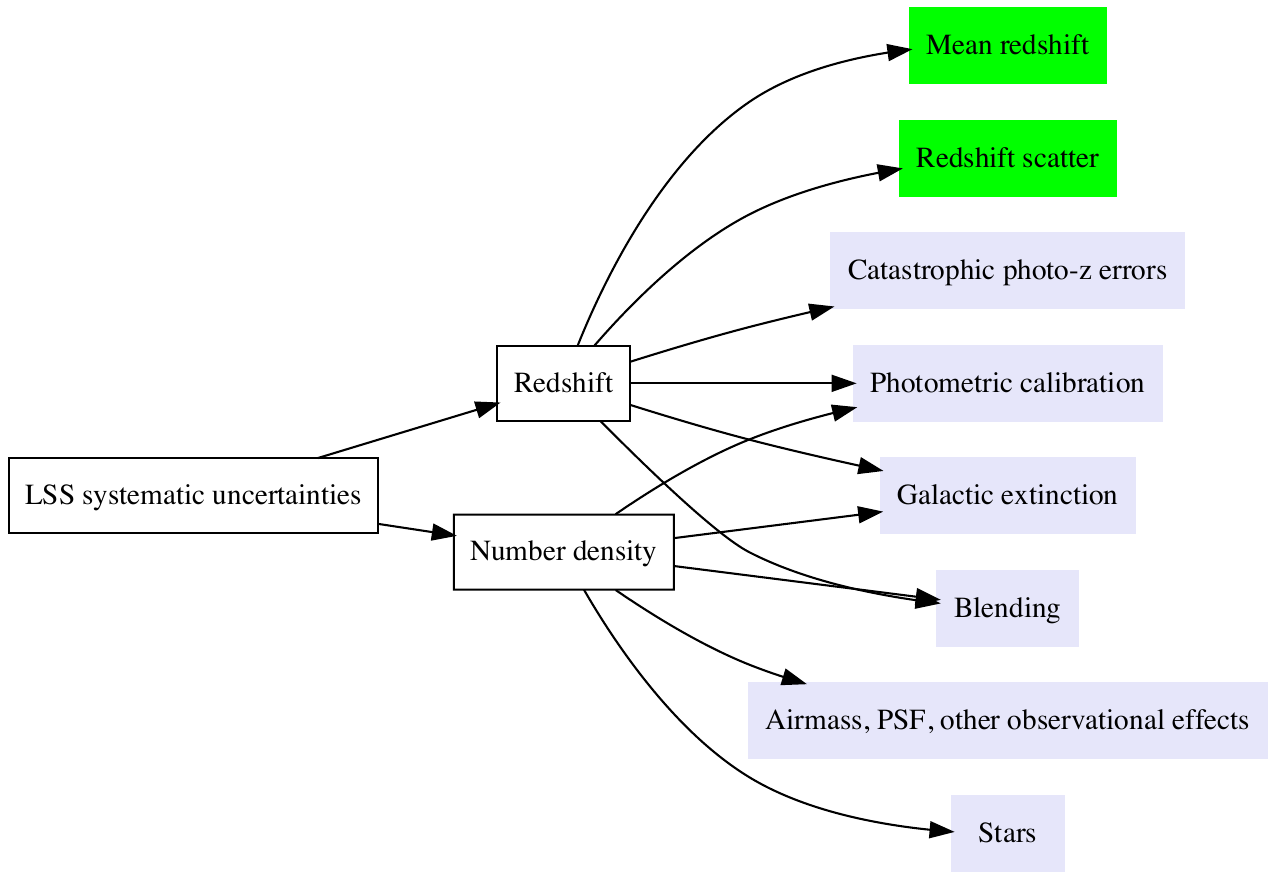}
  \end{center}
  \caption{Diagram indicating sources of systematic uncertainty for the LSS analysis on which we would like to place
    requirements in the DESC~SRD.   The direction of the arrows indicates the flow from overall
    systematic uncertainty to broad systematics categories to the specific physical effects on which
    we place requirements. As shown, there are several issues that
contribute to both redshift and number density uncertainty.  The green / lavender boxes indicate
sources of uncertainty on which we do / do not place requirements in this DESC~SRD version,
respectively.
\label{fig:lsssys-cal}}
\end{figure}

The residual calibratable systematic uncertainties on which we will place requirements can be divided into
two categories: those that cause uncertainties in the galaxy number density as a function of
position on the sky (and hence in the galaxy power spectrum), and those that cause redshift
uncertainties. Both classes of uncertainty are coupled by systematics such as dust extinction or
photometric calibration. A diagram of these calibratable systematics is shown in \autoref{fig:lsssys-cal},
while the current models and future plans for how to represent them can be found in
\autoref{tab:lsssys-cal}.  The DESC's DC2 analysis effort will provide useful guidance on how to
incorporate our need for knowledge of systematic uncertainty due to observational effects like
airmass and PSF effects and how to connect them to cosmological parameter estimates, so some models
for inclusion of these effects and how to place requirements on our knowledge of them (or associated
scale cuts) are still to be defined in future DESC~SRD versions.  Additional updates to our modeling
might result from DC2- or DC3-era decisions about whether the analysis will proceed in bins (which
requires the calibration of photo-$z$'s in each bin) or will rely on photo-$z$ PDFs (in which case
misspecification of PDFs may also be an issue, similar but not identical to catastrophic photo-$z$
outliers in impact).  In either case, a more sophisticated model for the redshift-dependent
photo-$z$ bias, scatter, and outlier rate is needed.

\begin{table}[!htbp]
\begin{center}
\begin{tabular}{p{1.3in}p{2.2in}p{2.5in}} \hline
Calibratable systematic uncertainty & Current model & Future plans \\
\hline
\multicolumn{3}{c}{Redshift uncertainties}\\
\hline
Mean redshift & Uncertainty in $\langle z\rangle$ for each tomographic bin & Investigate bins separately \\
Redshift width & A redshift bin width that is the same for each bin modulo $1+z$ factors &
Account for inflation of $\sigma_z$ at higher redshift compared to the $1+z$ model; use DC2 guidance
on $\sigma_z$
\\
Catastrophic photo-$z$ errors & None & To be decided based on DC2 \\
Galactic extinction & None & To be decided \\
Photometric calibration & None & To be decided \\
Blending & None & To be decided \\
\hline
\multicolumn{3}{c}{Number density uncertainties}\\
\hline
Galactic extinction & None & To be decided \\
Photometric calibration & None & To be decided \\
Stars & None & Templates for incomplete detection near bright stars, impact of bright stars on
background estimates, stellar
contamination of galaxy sample, \dots \\
Airmass, PSF, other observational effects & None  & To be decided based on DC2 \\
Blending & None & To be decided based on DC2 \\
\hline
\end{tabular}
\caption{Calibratable systematic uncertainties for LSS.}
\label{tab:lsssys-cal}
\end{center}
\end{table}

%% file: inc/wl.tex
The default weak lensing analysis is a tomographic analysis of shear-shear, galaxy-shear, and
galaxy-galaxy correlations (or ``3$\times$2-point'' analysis), which has become the standard in the
field of weak lensing due to the way it enables marginalization of both astrophysical and
observational systematic uncertainties in the shear signal.  Currently, this analysis is implemented
such that the LSS analysis described in the previous subsection is a strict subset of the $3\times
2$-point analysis, but that might not always be the case in future DESC~SRD versions.  In other
words, the lens sample for the 3$\times$2-point analysis was described in \appref{app:lss},
while the source sample is described here.

\subsubsection{Analysis choices}

Here we describe the essential points of the WL analysis setup in this version of the DESC~SRD:

\begin{itemize}
\item For the ``lens'' sample, we assume 10 tomographic bins spaced by $0.1$ in
  photo-$z$ between $0.2\le z\le 1.2$ for Y10, and 5 bins spaced by $0.2$ in photo-$z$ in that same
  redshift range for Y1.  This is the same as for the LSS analysis in the previous subsection.
\item For the ``source'' sample, we assume 5 redshift bins defined with equal numbers of source
  galaxies per bin for both Y1 and Y10. This is done using the true redshift distribution, and then the
  bins are convolved with the photo-$z$ error distribution to make the photo-$z$ distributions.
  While this technically does not mimic what is done in reality, it is quite a close approximation
  and easier to implement.  Unlike for the lens sample described in \appref{app:lss}, there is
  no upper redshift cutoff at $z=1.2$ for the source sample.
While this use of only 5 redshift bins is pessimistic compared
  to previous LSST and other Stage-IV survey forecasts, there are a few reasons to do this here:
  First, there were relatively larger numerical convergence issues seen with CosmoLike forecasts
  with 10 bins, and stronger discrepancies between CosmoLike and GoFish than for the 5-bin case
  (where agreement was excellent as described in \appref{app:software}).  Second, the
  requirements on calibration of the mean redshift of each tomographic bin were extremely tight in
  the 10-bin case,
  possibly unachievable even with acquisition of substantial additional data; hence opting for fewer
  tomographic bins and less constraining power constitutes a more realistic analysis choice.  Third,
  there were concerns that the overly simplistic photometric redshift error model may be
  particularly limiting in the 10-bin case, where each bin is relatively narrower and the impact of
  outliers (not yet modeled) is more important.
\item The data vector consists of the angular power spectra, $C_\ell$. Both auto- and cross-power
  spectra are included in the analysis for shear-shear and galaxy-shear correlations, but as in the
  previous subsection, only auto-power spectra are included for galaxy-galaxy correlations.
\item We use the same set of globally defined $\ell$ bins as described for the LSS analysis.  The
  actual choice of bins to include in the forecasting is made as follows:
\begin{itemize}
\item For galaxy-shear and galaxy-galaxy correlations, the $k_\text{max}$ value is chosen in the
  same way as for the LSS analysis (based on our ability to model galaxy bias), and $\ell$ bins are eliminated for each tomographic redshift bin
  based on the relationship between $k_\text{max}$ and $\ell_\text{max}$ as described in the
  previous subsection.
\item For shear-shear correlations, since the $C_\ell$ are determined based on an integral that goes
  from redshift $z=0$, it does not make sense to define a physical $k_\text{max}$.  Instead, we
  adopt $\ell_{\text{max,shear}}=3000$ in all tomographic bins.  This is based on assuming some
  improvements in our ability to model the impact of baryons on the matter power spectrum compared
  to the current state-of-the-art, though we have not factored in the fact that baryonic physics
  mitigation schemes will prevent our use of the full constraining power of the data on those scales.
\item For galaxy-shear tomographic cross-correlations, for bins $i$ and $j$ where $i\ne j$, we use
  the value of the $\ell_\text{max}$ for the lens sample (since the restriction is based on our
  ability to model nonlinear bias in the lens sample), and in all cases require $\ell_\text{max} \le
  \ell_{\text{max,shear}}$.
\end{itemize}
\item We use the same prescription for the nonlinear matter power spectrum as described in the
  previous subsection.
\item The covariance matrix estimation follows the same numerical prescription as in \citet{2017MNRAS.470.2100K}, with
  the only changes being those required by our new baseline analysis definition, tomographic
  binning, etc.
\item The number density and redshift distribution for the Y1 and Y10 lens samples are as described
  in the previous subsection.
\item We use a process similar to that of \citet{2013MNRAS.434.2121C} to estimate
  $n_\text{eff}$ for lensing source galaxies.  Following the calculations in \appref{subsec:neff} of
  this document, we use 10 and 27~arcmin$^{-2}$ as the lensing $n_\text{eff}$ in Y1 and Y10 as given
  in \autoref{fig:neffzy1y10}. We
  also use $\sigma_e=0.26$ per component, and $\sigma_z=0.05(1+z)$.  This $\sigma_z$ value is
  somewhat optimistic for $z\gtrsim 1.2$; future DESC~SRD versions should incorporate a stronger
  redshift-dependence of the scatter for greater realism.
\item For $n_\text{eff}(z)$, we use the same parametric form as \autoref{eq:nz}, with
  $(z_0,\alpha)=(0.13, 0.78)$ for Y1 and $(0.11,0.68)$ for Y10 (see legends of
  \autoref{fig:neffzy1y10} and discussion in \appref{subsec:dndzeff}
  for details).  These give mean effective source
  redshifts of $\sim 0.85$ and $\sim 1.05$, respectively.
\end{itemize}

\subsubsection{Anticipated improvements}

In future DESC~SRD versions, we may consider versions of the baseline WL analysis that allow a
different choice of ``lens'' galaxies from what goes into the LSS analysis, in order to optimize the
two science cases.  If multitracer analysis is deemed useful for WL, it may be pursued in future.
Other potential improvements include: use of ``lens'' galaxies beyond $z=1.2$ in the clustering
analysis; use of galaxy-galaxy cross-power spectra; more optimal tomographic binning; and incorporating the impact of survey inhomogeneity
in Y1, if a simple parametrization for its effect on the observables can be
derived.

\subsubsection{Systematic uncertainties}\label{subsubsec:wl-sysuncert}

For the WL analysis, we consider the following classes of systematic uncertainties in our two categories:
\begin{itemize}
\item Self-calibrated systematics: intrinsic alignments, baryonic effects, galaxy bias, magnification
\item Calibratable systematics on which we place requirements: any shear, photo-z, detector, or
  image processing issue
\end{itemize}

The models for the self-calibrated
systematic uncertainties are summarized in \autoref{tab:wlsys-selfcal}.  For each source of
systematic uncertainty, we describe how
it is included in this DESC~SRD version, as well as aspirations for more complex models to be
included in future.  Currently, our data vectors are produced assuming linear galaxy biasing, and
marginalization will use $b$ defined in redshift bins without assuming a parametrized $b(z)$ model
-- hence $N$ bins leads to $N$ nuisance parameters.  For a given $z$ bin, the fact that we do not
have to marginalize over nonlinear bias implies more statistical power left for cosmology
constraints than would be present in reality (since we will have to marginalize over nonlinear
bias); but the fact that we need a per-bin bias (rather than having a parametrized model) does
result in many bias parameters.  In one sense we are overestimating the number of needed bias
parameters, and in another we are underestimating compared to reality.

For intrinsic alignments (IA), we currently use the nonlinear alignment model as in section 4.4 of
\citet{2017MNRAS.470.2100K} for red galaxies, assuming that blue galaxies have no IA.  The red
galaxy fraction is computed from equation 25 and subsequent text in \citet{2016MNRAS.456..207K}; the
red galaxy luminosity function is from GAMA.  The definition of IA amplitude and luminosity
dependence is in the text around equations 7-8 and section 4.1 of that paper.  For the DESC SRD, the luminosity
function parameters were fixed and the four IA parameters allowed to vary within the ranges defined by
the following Gaussian priors:
%
\begin{itemize}
\item Overall intrinsic alignment amplitude $A_\text{IA}$: mean$=5$, $\sigma=3.9$
\item Power-law luminosity scaling $\propto L^{\beta_\text{IA}}$: $\beta_\text{IA}$ with mean$=1$, $\sigma=1.6$
\item Redshift scaling $\propto (1+z)^{\eta_\text{IA}}$: $\eta_\text{IA}$ with mean$=0$, $\sigma=2.3$
\item Additional high-redshift scaling parameter $\eta_\text{high-z}$: mean$=0$, $\sigma=0.8$
\end{itemize}

\begin{table}[!htbp]
\begin{center}
\begin{tabular}{p{0.8in}p{2.4in}p{2.8in}} \hline
Self-calibrated systematic uncertainty& Current model & Future plans \\ \hline
Galaxy bias & Linear galaxy bias, one value per tomographic bin  (Gaussian prior, mean$=1.9$ and $\sigma=0.9$) & Nonlinear galaxy bias with a
redshift-dependent parametrization of the linear bias vs.\ redshift, and at least one nonlinear bias
parameter \\
Magnification & None & Self-consistent convergence field and luminosity function
as what goes into the shear and intrinsic alignments in WL analysis, following e.g.\
\protect\cite{Joachimi10}; marginalize over uncertainty in slope of number counts \\
Intrinsic alignments & Nonlinear alignment model as in section 4.4 of \citet{2017MNRAS.470.2100K},
but with different priors as described in \appref{subsubsec:wl-sysuncert}
 & More complex model such as
\citet{2015JCAP...08..015B}, with IA and luminosity function parameters
marginalized \\
Baryonic effects & None & Hydrodynamic simulation-motivated emulator for baryonic effects in WL
\citep[e.g.,][]{2015MNRAS.450.1212H} \\
\hline
\end{tabular}
\caption{Self-calibrated systematic uncertainties for WL.}
\label{tab:wlsys-selfcal}
\end{center}
\end{table}

\begin{figure}[!htbp]
  \begin{center}
    \includegraphics[width=\textwidth]{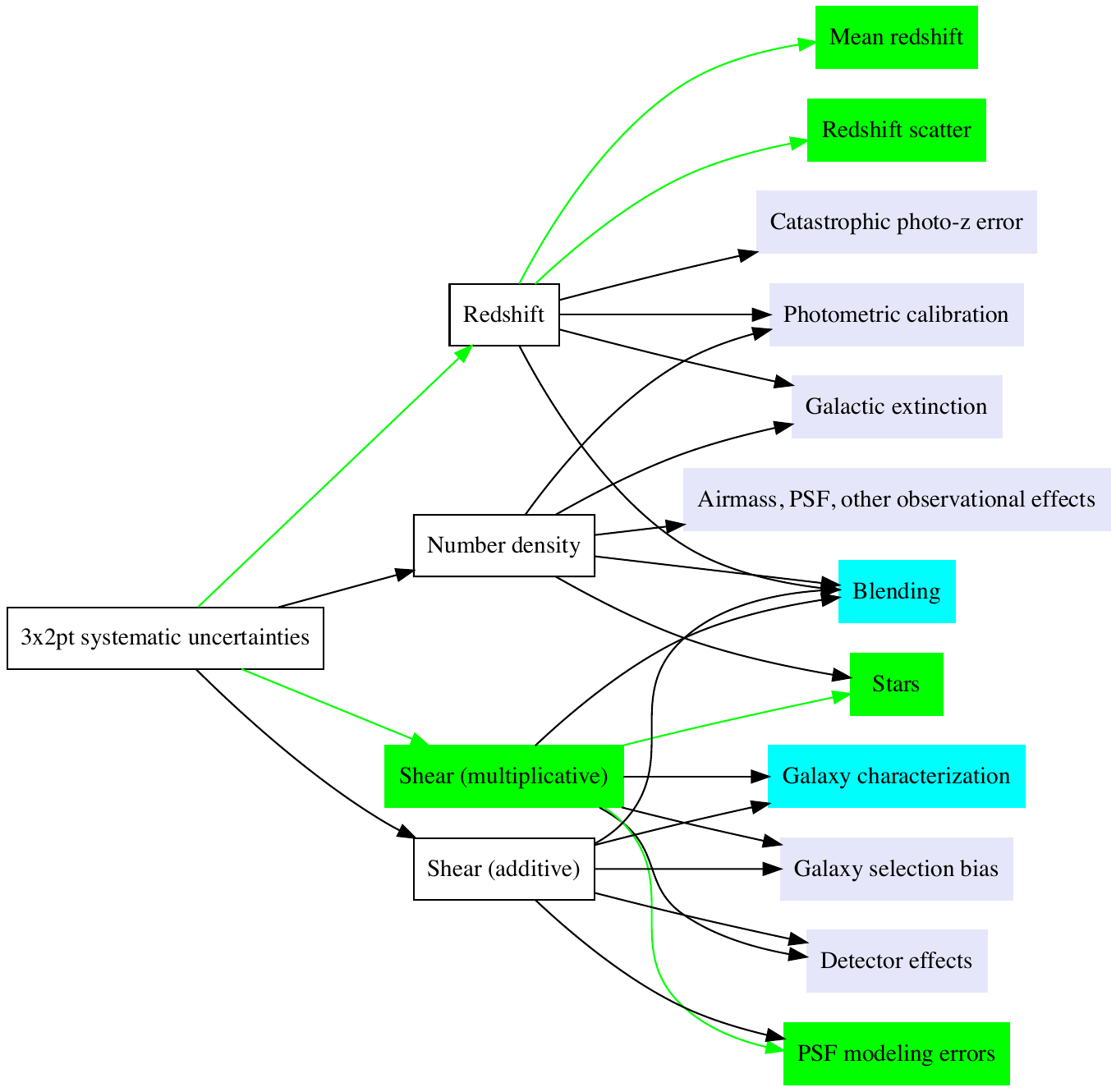}
  \end{center}
  \caption{Diagram indicating sources of systematic uncertainty for the WL analysis on which we would like to place
    requirements in the DESC~SRD.   The direction of the arrows indicates the flow from overall
    systematic uncertainty to broad systematics categories to the specific physical effects on which
    we place requirements.  As shown, there are several low-level issues in the right-hand
    column that
contribute to multiple categories of uncertainty in the middle column.  The green / lavender boxes indicate
sources of uncertainty on which we do / do not place requirements in this DESC~SRD version,
respectively.  The cyan boxes indicate those for which more R\&D beyond the DESC's DC2 may be needed
in order to place requirements.  For some of the green boxes, we currently only have software infrastructure to place
requirements through their impact on one class of uncertainty; such connections are shown as green
arrows.}
\label{fig:wlsys-cal}
\end{figure}

The residual calibratable systematic uncertainties on which we will place requirements can be divided into
four categories: redshift, number density, multiplicative shear, and additive shear uncertainties.
A diagram of these calibratable systematics is shown in \autoref{fig:wlsys-cal}, while the
current models and future plans for how to represent them is in \autoref{tab:wlsys-cal}. We note
that the residual multiplicative shear calibration bias $\Delta m$ is in principle a function of redshift, and
indeed it is redshift evolution of $\Delta m$ that can mimic changes in dark energy models (because it
implies a change in structure growth).  As a result, our model for $\Delta m(z)$ is not simply a constant
$m_0$; we confirmed that this gives extremely weak requirements on $m_0$ because, as noted
previously by \citet{2013MNRAS.429..661M}, $\partial C_\ell/\partial m_0$ differs significantly from
$\partial C_\ell / \partial w_0$ (or $w_a$) when considering the full data vector of $C_\ell$ across
all redshift bins.  In order to identify sensitivity of cosmological parameter constraints to
multiplicative shear calibration uncertainties, it is important to use a redshift-dependent model; we adopted
\begin{equation}
\Delta m(z) = m_0 \left(\frac{2 z - z_\text{max}}{z_\text{max}}\right)
\end{equation}
with $z_\text{max}$ set to the middle of the highest tomographic redshift bin.  Given this parametrization, the total variation in shear calibration across the redshift range used for the weak
lensing analysis is $2m_0$, and requirements on shear calibration should be interpreted as a
requirement on our
knowledge of redshift-dependent shear calibration trends across the sample.
In principle, there is some higher-order dependence on the
redshift-dependent function adopted, but do not explore this further in this version of the DESC~SRD.

 The DESC's DC2 analysis effort
will provide useful guidance on how to incorporate our need for knowledge of systematic uncertainty
due to observational effects like airmass and PSF effects and how to connect them to cosmological
parameter estimates, so some models for inclusion of these effects and how to place requirements on
our knowledge of them (or associated scale cuts) are still to be defined in future DESC~SRD
versions.  Additional updates to our modeling might result from DC2- or DC3-era decisions about
whether the analysis will proceed in bins (which requires the calibration of photo-$z$'s in each
bin) or will rely on photo-$z$ PDFs (in which case misspecification of PDFs may also be an issue,
similar but not identical to catastrophic photo-$z$ outliers in impact).   In either case, a more sophisticated model for the redshift-dependent
photo-$z$ bias, scatter, and outlier rate is needed.

It is worth noting that
several of the effects listed in the right-most column of \autoref{fig:wlsys-cal} contribute to
systematic uncertainties for WL in several ways (number density, redshift, and shear-related
uncertainties); it will be important to develop a self-consistent approach for how systematic
uncertainties due to blending, photometric calibration, galactic extinction, stars, galaxy
characterization, selection bias, detector effects, and PSF modeling errors propagate into all of
the observables.

\begin{longtable}{p{1.5in}p{1.8in}p{2.7in}}
\hline
Calibratable systematic uncertainty & Current model & Future plans \\
\hline
\multicolumn{3}{c}{Redshift uncertainties}\\
\hline
Mean redshift & Uncertainty in $\langle z\rangle$ for each tomographic bin & Investigate bins separately \\
Redshift width & A redshift bin width that is the same for each bin modulo $1+z$ factors &
Account for inflation of $\sigma_z$ at higher redshift compared to the $1+z$ model; use DC2 guidance
on $\sigma_z$
\\
Catastrophic photo-$z$ errors & None & To be decided based on DC2 \\
Galactic extinction & None & To be decided \\
Photometric calibration & None & To be decided \\
Blending & None & To be decided \\
\hline
\multicolumn{3}{c}{Number density uncertainties}\\
\hline
Galactic extinction & None & To be decided \\
Photometric calibration & None & To be decided \\
Stars & None & Templates for incomplete detection near bright stars, impact of bright stars on
background estimates, stellar
contamination of galaxy sample, \dots \\
Airmass, PSF, other observational effects & None  & To be decided based on DC2 \\
Blending & None & To be decided based on DC2 \\
\hline
\multicolumn{3}{c}{Shear (multiplicative) uncertainties}\\
\hline
Blending & None & To be decided based on DC2 \\
Stars & Fractional contamination of galaxy sample by stars & To be decided based on DC2 \\
Galaxy characterization & None & To be decided \\
Galaxy selection bias & None & To be decided \\
Detector effects & None & To be decided based on DC2 \\
PSF modeling errors & PSF model size requirement based on second moments & To be decided based on DC2 \\
\hline
\multicolumn{3}{c}{Shear (additive) uncertainties}\\
\hline
Blending & None & To be decided based on DC2 \\
Galaxy characterization & None & To be decided \\
Galaxy selection bias & None & To be decided \\
Detector effects & None & To be decided based on DC2 \\
PSF modeling errors & $\rho$ statistics & To be decided based on DC2 \\
Member galaxy contamination & None & To be decided \\
\hline
\caption{Calibratable systematic uncertainties for WL.}
\label{tab:wlsys-cal}
\end{longtable}

Finally, several boxes in the right-most column of \autoref{fig:wlsys-cal}
implicitly include
multiple effects.  For completeness, we note the primary contributors to these,
which  will eventually be important for modeling their impact on the observable quantities:
\begin{itemize}
\item Detector effects: brighter-fatter, glowing edges, tree rings, and others.
\item PSF modeling errors: differential chromatic refraction, chromatic seeing, other chromatic
  effects in the optics and sensors, color gradients, relative astrometry between exposures, model
  bias, PSF interpolation, contamination of the PSF star sample by binaries.
\item Galaxy characterization: insufficient PSF correction method, pixel-noise bias, model bias
\item Blending: effects on detection, astrometry, photometry, and shapes due to undetected blended
  objects, selection effects, increased model and/or noise bias
\end{itemize}
These may not be explicitly modeled independently in the end, but understanding the nature of these
contributing factors may be important for building up templates for systematics.

%% file: inc/cl.tex
For the purpose of this first version of the DESC~SRD, the default galaxy cluster abundance analysis
incorporates the cluster counts as a function of richness and redshift, along with stacked cluster
weak lensing in the 1-halo regime to constrain parameters of the cluster mass-observable relation (MOR).

\subsubsection{Analysis choices}

Here we describe the essential points of the CL analysis setup in this version of the DESC~SRD.  It
largely follows the treatment in \cite{2017MNRAS.470.2100K}, with a few variations:

\begin{itemize}
\item For the source sample, we use the same sample as defined for WL in
  Y1 and Y10. See \appref{app:wllss} for number densities, redshift distributions, tomographic
  bin definitions, and other
  relevant parameters.
\item The cluster-shear data vector consists of the angular power spectra, $C_\ell$, defined in
  redshift and richness bins, along with the cluster counts on those bins $N(\lambda, z)$.
\item We use the same set of globally defined $\ell$ bins as described for the LSS analysis.  Since
  the analysis for now uses only the 1-halo regime, we restrict to $3~382 \le \ell \le 15~000$.  The
  choice of lower $\ell$ limit is due to technical issues in covariance estimation; lower $\ell$
  values may be used in future DESC~SRD versions.
\item The cluster-shear data vector is a stack of the Fourier transform of NFW profiles given the redshift and mass distribution of the clusters and a concentration-mass relation from \cite{2013ApJ...766...32B}. At high redshift, the lowest $\ell$ bins contain a contribution
  from the 2-halo term.  This term is constructed using the nonlinear matter
  power spectrum along with the halo bias vs.\ mass relation from \citet{2010ApJ...724..878T}.
  Currently the bias is fixed and is not fit for as a free parameter.
\item Covariances are calculated following \cite{2017MNRAS.470.2100K}.
\item The cluster sample binning is defined as follows:
\begin{itemize}
\item $\lambda=[20,30],[30,45],[45,70],[70, 120], [120,220]$
\item $z=[0.2,0.4],[0.4,0.6],[0.6,0.8],[0.8,1.0]$ for Y10, with the highest redshift bin omitted in
  Y1 due to the shallower depth of the Y1 imaging data.
\end{itemize}
\item For the first version of the DESC SRD, we assume perfect knowledge of cluster redshift, i.e., there is no photo-$z$ error.
\end{itemize}

\subsubsection{Anticipated improvements}

The above baseline analysis does not correspond to either cluster cosmology analysis approach that
is under consideration within the DESC.  The clusters working group is considering two approaches, at least one of
which should be included in a future DESC~SRD version.  The two approaches are as follows:
\begin{itemize}
\item An extension of the approach described above to include cluster-shear correlations in the
  2-halo regime, along with cluster-cluster correlations.
\item An approach that involves using individual cluster shear profiles, rather than a stacked
  analysis.
\end{itemize}
In addition, external X-ray and SZ data will be used to place priors on certain astrophysical
nuisance parameters (MOR, optical cluster miscentering).  A more direct connection to that data,
given some reasonable assumptions about what will be available when the LSST survey starts, would be
desirable -- i.e., using sample sizes to connect to the size of the priors on the MOR parameters.

Additional extensions for consideration include stacking in real space using physical coordinates
for better reconstruction of cluster profiles, going to lower $\ell$, and a comparison of stacking shear vs.\ surface densities.

\subsubsection{Systematic uncertainties}\label{subsubsec:cluncert}

For the CL analysis, we consider the following classes of systematic uncertainties in our two categories:
\begin{itemize}
\item Self-calibrated systematics: mass-observable relation, intrinsic
  alignments, other theory uncertainties (e.g., mass function), cluster miscentering, cluster
  large-scale bias, baryonic effects on cluster halo profiles and mass function, projection effects
\item Calibratable systematics on which we place requirements: any shear, photo-z, detector, or
  image processing issue
\end{itemize}

The models for the self-calibrated
systematic uncertainties are summarized in \autoref{tab:clsys-selfcal}. For cluster abundance
measurements, the MOR tends to be the dominant self-calibrated systematic uncertainty (and dominates
over purely statistical uncertainties).
For each source of
systematic uncertainty, we describe how
it is included in this DESC~SRD version, as well as aspirations for more complex models to be
included in future.

For the first version of the DESC~SRD, we use the MOR from \cite{2018ApJ...854..120M} with an extension of redshift dependence. Specifically, the mean relation is defined as
\begin{equation}\label{eq:cl:mor}
\ln \lambda(M,z|A, B, C) = A + B\ln\left(\frac{M}{M_{\rm pivot}}\right)+C\ln\left(1+z\right),
\end{equation}
and the mass-dependent mass-observable scatter is defined as
\begin{equation}\label{eq:cl:morscat}
\sigma_{\ln \lambda|M}(M, z| \sigma_0, q_M, q_z)=\sigma_0+q_M\ln\left(\frac{M}{M_{\rm pivot}}\right)+q_z\ln\left(1+z\right).
\end{equation}
This is a slight update from \cite{2017MNRAS.470.2100K} where they used a constant mass-observable
scatter. We assume fiducial values for $A$, $B$, and $\sigma_0$ used in \cite{2018ApJ...854..120M},
fiducial values 0 for $C$, $q_m$, and $q_z$. The pivot mass is $M_{\rm  pivot} =3\times 10^{14}h^{-1}M_\odot$. We use Gaussian priors for $A$ and $B$ with a width
corresponding to 80\% of the probability within the flat prior ranges for $A$ and $B$ used in
\cite{2018ApJ...854..120M} and a Gaussian prior with $\sigma=1.2$ for $C$. For nuisance parameters in the
mass-observable scatter, we use priors which are not directly on these parameters but on scatters at
three different richnesses and redshifts, $P(\sigma_{\ln \lambda|M}(M_{\rm fid}(\lambda_i, z_i),z_i| \sigma_0, q_M, q_z))={\rm Gauss}[\sigma_{\ln \lambda|M}(M_{\rm fid}(\lambda_i, z_i),z_i|\sigma_{0,\rm{fid}}, q_{M,\rm{fid}}, q_{z,\rm{fid}}), \sigma_i]$, where $\sigma_i = 0.1$ and
$M_{\rm fid}(\lambda_i, z_i)$ is the inverse of the mean relation with the fiducial nuisance
parameters; there are different sets of
$(\lambda_i, z_i)$ and different assumptions about how well the scatter can be constrained for Y1 and Y10.  The priors for Y1 and Y10 are based on reasonable
assumptions of ancillary datasets that will be available at those times:
\begin{itemize}
\item Y1: The constraints are $\sqrt{N}$ extrapolations from current constraints of $\sim$10\% on
  the scatter in the mass-observable relation for
  $N=30$ clusters.  Based on this assumption, we use $(\lambda_i, z_i) = (90, 0.2)$,
  $(30, 0.1)$, and $(100, 0.8)$, assuming the scatter in the MOR is known to $0.06$,
  0.05, and 0.04 respectively, corresponding to 100, 150, and 200 clusters.  The 100 cluster assumption is based on expectations for Chandra$+$SPT follow-up,
  the 150 cluster assumption is based on an already-completed observing program with Swift, and the
  200 cluster assumption is from combined SPT$+$ACT.
\item Y10: We use $(\lambda_i, z_i) = (80, 0.2)$,
  $(30, 0.1)$, and $(90, 0.8)$, assuming the scatter in the MOR is known to $0.03$,
  0.05, and 0.03 respectively, corresponding to 500, 150, and 500 clusters,
  respectively.  The 500 cluster assumption is based on expectations for eROSITA and CMB-S4, and is
  conservative for those surveys (but we assume there will be some systematics floor that prevents
  further constraint on scatter in the MOR even for higher $N$).  The 150 cluster assumption for low redshift is
  based on an already-completed observing program with Swift.
\end{itemize}

\begin{table}[!htbp]
\begin{center}
\begin{tabular}{p{1.5in}p{2.3in}p{2.3in}} \hline
Self-calibrated systematic uncertainty& Current model & Future plans \\ \hline
\hline
MOR & See discussion surrounding \autoref{eq:cl:mor} and~\autoref{eq:cl:morscat} & Further
exploration of parametrizations with more complexity and possibly modified priors \\
Intrinsic alignments & None & Self-consistent model with WL analysis \\
Mass function uncertainty & None & At least one parameter overall rescaling, but possibly with
mass-dependence as well \\
Baryonic effects & None & Self-consistent inclusion of baryonic effects on mass function, cluster
shear profiles \\
Cluster large-scale bias & None & Once 2-halo regime and/or cluster clustering is included, will need a model
that includes nonlinear bias. \\
Other theoretical uncertainty (e.g., halo definition biases) & None & Further exploration needed to
determine approach \\
Cluster miscentering & None & Approach to be decided after DC2 \\
Projection effects & None & Approach to be decided after DC2 \\
\end{tabular}
\caption{Self-calibrated systematic uncertainties for CL.}
\label{tab:clsys-selfcal}
\end{center}
\end{table}

\begin{figure}[!htbp]
  \begin{center}
    \includegraphics[width=0.65\textwidth]{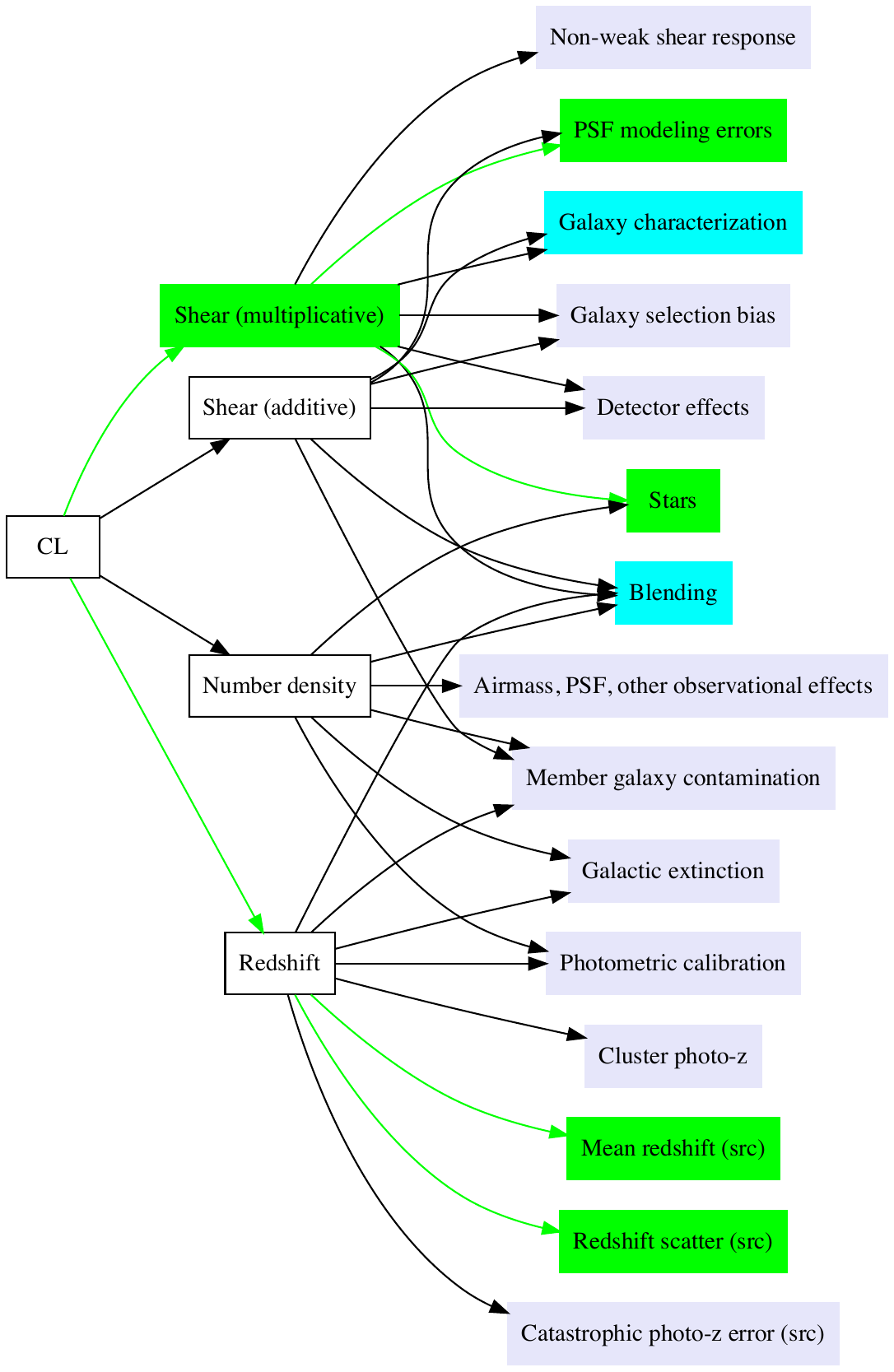}
  \end{center}
  \caption{Diagram indicating sources of systematic uncertainty for the CL analysis on which we would like to place
    requirements in the DESC~SRD.   The direction of the arrows indicates the flow from overall
    systematic uncertainty to broad systematics categories to the specific physical effects on which
    we place requirements.  As shown, there are several low-level issues in the right-hand
    column that
contribute to multiple categories of uncertainty in the middle column.  The green / lavender boxes indicate
sources of uncertainty on which we do / do not place requirements in this DESC~SRD version,
respectively.  The cyan boxes indicate those for which more R\&D beyond the DESC's DC2 may be needed
in order to place requirements.  For some of the green boxes, we currently only have software infrastructure to place
requirements through their impact on one class of uncertainty; such connections are shown as green
arrows.}
\label{fig:clsys-cal}
\end{figure}

The residual calibratable systematic uncertainties on which we will place requirements can be divided into
four categories: redshift, number density, multiplicative shear, and additive shear uncertainties.
A diagram of these
calibratable systematics is shown in \autoref{fig:clsys-cal}, while the current models and future
plans for how to represent them is in \autoref{tab:clsys-cal}. The DESC's DC2 analysis effort
will provide useful guidance on how to incorporate our need for knowledge of systematic uncertainty
due to observational effects like airmass and PSF effects and how to connect them to cosmological
parameter estimates, so some models for inclusion of these effects and how to place requirements on
our knowledge of them (or associated scale cuts) are still to be defined in future DESC~SRD
versions.   It is worth noting that
several of the effects listed in the right-most column of \autoref{fig:clsys-cal} contribute to
systematic uncertainties for CL in several ways (number density, redshift, and shear-related
uncertainties); it will be important to develop a self-consistent approach for how systematic
uncertainties due to blending, photometric calibration, galactic extinction, stars, galaxy
characterization, selection bias, detector effects, and PSF modeling errors propagate into all of
the observables.

The parameters related to shear calibration, shear sample tomographic bins, and source photo-$z$ are shared with the WL analysis
(\appref{app:wllss}).

\begin{longtable}{p{1.5in}p{1.8in}p{2.7in}}
\hline
Calibratable systematic uncertainty & Current model & Future plans \\ \hline
\multicolumn{3}{c}{Redshift uncertainties}\\
\hline
Mean redshift & Uncertainty in $\langle z\rangle$ for each tomographic bin & Investigate bins separately \\
Redshift width & A redshift bin width that is the same for each bin modulo $1+z$ factors &
Account for inflation of $\sigma_z$ at higher redshift compared to the $1+z$ model; use DC2 guidance
on $\sigma_z$
\\
Catastrophic photo-$z$ errors & None & To be decided based on DC2 \\
Cluster photo-$z$ errors & None & To be decided \\
Galactic extinction & None & To be decided \\
Photometric calibration & None & To be decided \\
Member galaxy contamination & None & To be decided \\
\hline
\multicolumn{3}{c}{Number density uncertainties}\\
\hline
Galactic extinction & None & To be decided \\
Photometric calibration & None & To be decided \\
Blending & None & To be decided based on DC2 \\
Stars & None & Templates for incomplete detection near bright stars, impact of bright stars on
background estimates, stellar
contamination of galaxy sample, \dots \\
Airmass, PSF, other observational effects & None  & To be decided based on DC2 \\
Member galaxy contamination & None & To be decided \\
\hline
\multicolumn{3}{c}{Shear (multiplicative) uncertainties}\\
\hline
Blending & None & To be decided based on DC2 \\
Stars & Fractional contamination of galaxy sample by stars & To be decided based on DC2 \\
Galaxy characterization & None & To be decided \\
Galaxy selection bias & None & To be decided \\
Detector effects & None & To be decided based on DC2 \\
PSF modeling errors & PSF model size requirement based on second moments & To be decided based on DC2 \\
Non-weak shear response and flexion & None & To be decided \\
\hline
\multicolumn{3}{c}{Shear (additive) uncertainties}\\
\hline
Blending & None & To be decided based on DC2 \\
Galaxy characterization & None & To be decided \\
Galaxy selection bias & None & To be decided \\
Detector effects & None & To be decided based on DC2 \\
PSF modeling errors & $\rho$ statistics & To be decided based on DC2 \\
Member galaxy contamination & None & To be decided \\
\hline
\caption{Calibratable systematic uncertainties for CL.}
\label{tab:clsys-cal}
\end{longtable}

Finally, we note that several boxes in the right-most column of \autoref{fig:clsys-cal}
implicitly include multiple effects.  The primary contributors to these are listed in
\appref{subsubsec:wl-sysuncert}.  Where possible, adopting a common approach to these
systematics and their impact on LSS, WL, and CL would be desirable.

%% file: inc/sn.tex
\subsubsection{Analysis choices}

Here we describe the essential points of the SN analysis setup in this version of the DESC~SRD:

The surveys are simulated in four separate parts: a one-year DDF survey, a ten-year DDF survey, a
ten-year WFD survey and an external low-redshift sample. For all simulations we
use simulation libraries from \citet{biswas_rahul_2017_1006719} to take the
OpSim \verb minion_1016 ~run and provide inputs to the SuperNova
ANAlysis \citep[SNANA][]{2009PASP..121.1028K}
software to generate supernova light curves.  Then we fit those light curves with the SALT2 model \citep{2007A&A...466...11G, betoule}, and compute distance moduli using the \cite{2011ApJ...740...72M} approach. The analysis follows the prescription of \cite{2018ApJ...859..101S}, which was most recently used for the Pantheon sample of supernovae.

These statements assume that a commissioning mini-survey will reach `template depth' of 5-10 times
the single night coadd depth of the DDF (around 3000s exposures) in the DDF regions to produce deep templates in order to
detect SN from difference images in the first year of operation. In the absence of such a mini-survey,
the first year survey results should be considered to have significantly fewer SN\footnote{This is
not just because of the time to make the observations that will be used to build the templates, but because of
the need for a time lag to avoid the supernovae we want to measure being in the templates.} -- or,
effectively, the first year forecasts actually describe our capabilities after 2-3 years
rather than one year. Similarly, the first few years of observation of the wide field will be used to generate templates, hence we do not assume a WFD year-one survey, and restrict ourselves to those obtained after 10 years.

The current software framework for forecasting cosmological constraining power with SNe and
the impact of systematics involves a sample of simulated Type Ia supernovae, with selection cuts,
that get passed through light curve fitters.
The simulated SN analyses currently assume that the supernova redshifts are being determined through
spectroscopic follow-up of the host galaxies. Such spectroscopic follow-up of hosts is considerably
simpler than spectroscopic followup of active supernovae, as it does not need to be done during the
few-week window when a supernova is bright.  Also, several supernova hosts within a small patch of sky
aggregated over time may be simultaneously followed up using suitable multi-object
spectrographs\footnote{Such follow-up, which is used in current photometric supernova surveys like DES
  and Pan-STARRS, might be easiest to combine with needs of other working groups like
  photometric redshifts.}.

250,000 fiber hours have been allocated on 4MOST for the 4MOST-TiDES observations of transients, SN
hosts, and AGN reverberation mapping -- with LSST being the primary source of transients. The usage of that
allocated time is being decided, with a view to target as many viable hosts as possible (including
galaxies where the SN candidate is not obvious). 4MOST can reach a depth of $r\sim 22$ with 1-hour exposures, which
corresponds to roughly $z\sim 0.4- 0.6$ (in OzDES). This will cover the bulk of the wide-survey
candidates.  In future, it will be valuable to estimate how many of the WFD supernovae may have
redshifts from the DESI survey, but this is likely to be subdominant to the number that will be
acquired by 4MOST.  For the purpose of this forecast we assume the redshifts in the DDF will come from
Subaru (see below).

\begin{figure}
  \begin{center}
    \includegraphics[width=0.8\textwidth]{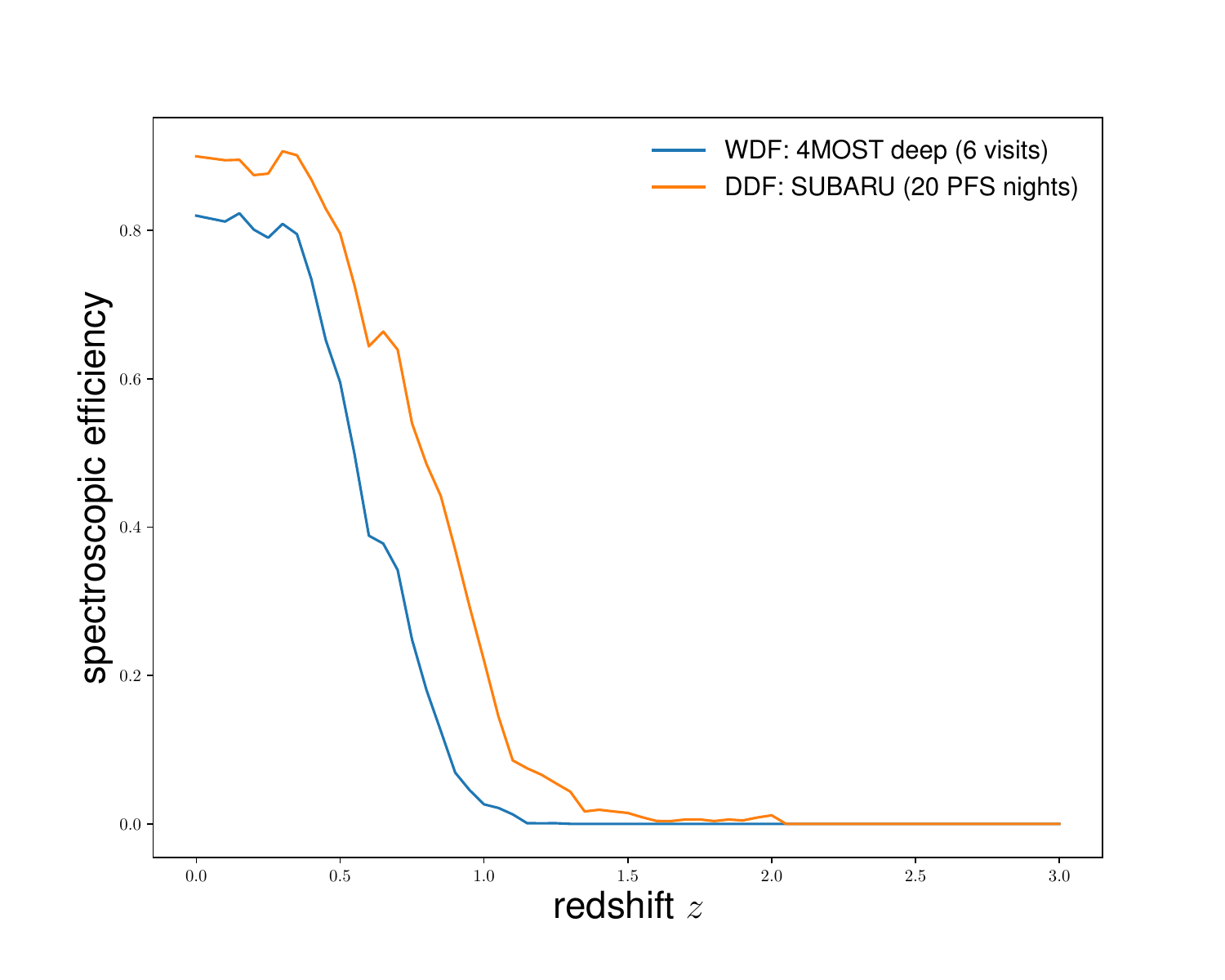}
  \end{center}
  \caption{Host spectroscopic redshift efficiencies for the WFD and DDF fields, assuming follow-up using a
combination of maximum-depth observations with the 4MOST survey (for WFD) and 20 nights of observing
time with the Prime Focus Spectrograph on the Subaru telescope (for DDF). The curves correspond to
the fraction of Type Ia supernovae at a given redshift which have hosts bright enough to be targeted
for spectroscopy (corresponding to a limit of $i=22$ for the maximum possible 6 visits with 4MOST,
or $i=22.86$ for this Subaru survey). We place additional restrictions on the total number of SNe
that can be followed up with 4MOST, given the total allocation of 250k fiber-hours. In addition to
the losses from hosts too faint to target that are plotted here, we assume that 20\% of targets will
fail to yield highly-secure redshifts based on past experience with data of the expected S/N.
\label{fig:snsys-eff}}
\end{figure}

We reduce the full simulated sample to a subset which would have spectroscopic host information
assuming 20 nights of observing time with the Prime Focus Spectrograph on the Subaru telescope in
the Deep Drilling Fields, which can reach $i=22.86$ hosts, plus 250k fiber-hours with the 4MOST
survey spread over the WFD area, which can reach $i=22$ with the maximum possible 6 visits per
fiber. The fraction of Type Ia supernovae at a given redshift which are expected to have hosts bright
enough for spectroscopy based on the stellar mass/SFR/photometric redshift catalogs of
\cite{2018MNRAS.474.5437L} are shown in \autoref{fig:snsys-eff}.  We note that a similar (same
order of magnitude) amount
of observing time on DESI in the DDFs could garner a comparable number of host redshifts, given that
DESI has a larger FoV, compensating for its smaller collecting area.  Hence there are two potential
pathways to obtain host redshifts for the DDF supernovae.

The restriction on fiber hours leads to a total restriction of 100k supernovae which we obtain by scaling the overall efficiency curves until we obtain the 100k objects in the final sample (see the note on fitting light curves below). For both surveys we also assume an 80\% secure redshift rate given the S/N for the assumed exposure times, and assuming that some time is spent obtaining redshifts for non-Ias and other objects (converting an initial estimate of 125k supernova candidates to 100k).  In this forecast, we assume the supernovae without redshifts will not be
used for cosmological constraints.  Work is ongoing within the DESC to study both the impact of host
photo-$z$ contamination on the overall science FoM, including mis-identification of hosts, and the best practice for mitigating redshift
error (\citealt{2017JCAP...10..036R}; Malz, Peters, Hlozek \etal in
preparation\footnote{\url{https://github.com/aimalz/scippr}}). It is worth noting that the dark energy task force (DETF) assumed a much larger sample of 400k photometric supernovae with an `optimistic photo-$z$' error. We restrict ourselves to the more conservative case described above.

The Foundation survey \citep{2018MNRAS.475..193F} will observe $\sim 800$ low redshift SNe by the time it completes the survey
in 2020. For the Y10 survey, we triple this number of low-redshift SNe, given the number of
low-redshift surveys (e.g. ZTF) that will be online and will yield impressive numbers of supernovae
by the end of the LSST Y10 survey. We include this sample by decreasing the error on the distance
modulus in our final Y10 sample by $\sqrt{3}$ for $z<0.05-0.1$. This low-$z$ sample is an essential
anchor to the Hubble diagram. For the low-$z$ sample, we do not include here any individual
calibration systematics for the different samples and filters.  Instead, we include the SALT2
calibration systematic and uncertainty in the HST calibration which affects both low-$z$ and LSST
supernovae.  Future versions of the DESC~SRD should factor in a reasonable level of calibration
uncertainties for the external low-$z$ sample, which will mildly change its constraining power.

The supernovae are then fit in two stages: the data are fit with light curve templates based on the
SALT2 model \citep{2007A&A...466...11G, betoule}. At this stage, detection quality cuts are applied,
and the number of supernovae reduces slightly compared to the original simulated light curve
points. There is a quality cut to restrict the number of objects that we would put fibers on (for
host-$z$ determination): we require objects to have three epochs that have signal-to-noise ratios
in the nightly coadds of greater than five.  An epoch here is based on a detection where the likelihood of detection follows an SNR-dependence as modelled by the Dark Energy Survey. The quality cut at this stage is to ensure that any fibers used to obtain host galaxy spectra are placed on `high quality' candidates.

The separate subsamples described above are then combined to form the overall Y1 and Y10 samples as
follow: DDF Y1 and an 800-SN low-$z$ sample to form the overall Y1 sample; DDF Y10, a 2400-SN
low-$z$ sample, and WFD Y10 sample to form the overall Y10 sample.

We apply the following quality cuts for inclusion of the SNe in the final cosmological sample:
$|c|<0.3, |x_1|<3, \sigma(x_1) < 1, \sigma(t_0) < 2,$
where $c, x_1, t_0$ are the color, stretch and explosion time respectively.
Following the cuts on light curve fit quality, we arrive at a sample of $\sim 2400$ SNe for the Y1 sample, and $\sim 104000$ SNe for the Y10 sample. The distance moduli are included as a likelihood in
CosmoSIS\footnote{\url{https://bitbucket.org/joezuntz/cosmosis/wiki/Home}} \citep{2015A&C....12...45Z},
which makes use of algorithms for efficient sampling of cosmological parameter
space \citep{2000ApJ...538..473L, 2012JCAP...04..027H}.

The number of SNe obtained and their light curve quality depend strongly on the survey cadence. Initial studies show that the total number of high-quality SNe in the 10-year survey can change by a factor of $\sim 1.5$ with cadences that are optimized for their detection, which will not only improve the statistical uncertainty, but also allow for new studies of supernova systematics\footnote{Exploration of the impact of cadence on cosmology with SNe is a key ongoing DESC activity.}.

For the WFD survey, the corresponding numbers depend even more strongly on the cadence strategy
employed.  With that caveat, we obtain a final sample combining the WFD and DDF and low-$z$ data of
$\sim$112~000 SNIa with sufficiently well-sampled light curves, host galaxy redshifts, with a
distribution peaking around $z\sim 0.4$.
The WFD and low-$z$ sample significantly improve the SN science case for the following reasons:
\begin{itemize}
\item The resulting low-redshift sample is complete up to $z\sim 0.4$, which adds significantly to the quality of the Hubble diagram constraints.
\item $z<0.5$ is roughly the redshift range where BAO is limited by cosmic variance. An all-sky $z \lesssim 0.5$
    SN sample over ten years is unique as a distance indicator in this region. Further, we note that
    there is no other SN survey capable of producing a sample of around 100 thousand supernovae.
\item A rolling cadence or other more optimal cadence may substantially increase the number of SNe with
  well-sampled light curves, allowing us to reach the nominal statistics in a fraction of the duration of the
  survey (modulo uncertainties in the host redshift), and providing a larger sample with which to
  study systematic effects. It may also enable the construction of a deeper supernova sample,
  extending the redshift lever arm for this probe.
\item The low-redshift sample will also be useful both in constraining non-standard cosmological
  models and improving our understanding of the SN population and underlying correlations with
  environments.  Moreover, it allows for the opportunity of probing structure growth with
  supernovae.
\end{itemize}

In contrast, the deep drilling fields will generate a superb `gold' sample of supernovae out to
$z\sim 1.2,$ which not only allow for cosmological constraints, but will enable us to study the
redshift evolution of supernova systematics (e.g., evolution of SN populations).

\subsubsection{Anticipated improvements}

In future DESC~SRD versions,
we believe we can improve both the precision and accuracy of supernova cosmology with several
changes to the baseline analysis presented here. These new approaches are still a subject under
development, and the tradeoff in terms of quantitative success versus additional computational cost
will determine which approaches are adopted in the end.
We present a brief outline of such newer analysis methods below:
\begin{enumerate}
    \item Standardization of supernova light curves: Historically, supernova cosmology has used a
      two-step standardization approach of (a) compressing SNIa light curves into a universal
      function of a few light curve model parameters thereby dimensionally reducing the multi-band
      time series dataset, and then (b) using a correlation between these light curve model
      parameters to an intrinsic brightness or distance of the supernova. Currently, surveys use
      models~\citep{2007A&A...466...11G} which include a scatter $\sim 0.1 $ mags in the resultant
      redshift-distance relationship,  part of which correlates with the environment of the
      supernova, potentially leading to biases in cosmological analysis. Following current standard
      approaches, the baseline analysis tries to account for this via post-facto correctional terms
      based on host galaxy stellar mass, and an attempt to distribute the scatter in the different
      model parameters. A better approach that minimizes the intrinsic dispersion and accounts for
      such environmental dependence in training the model itself requires the development of newer
      SN light curve models~\citep{2013ApJ...766...84K,2014ApJ...784...51K,2018arXiv180206125H},
      which is a key project in the SRM. These more sophisticated models will also make it possible to incorporate redshift dependencies of the components of the models, which presumably are primarily related to the environmental correlations with redshift (known and unknown).

    \item Supernova selection: The selection criteria for the supernova sample in the baseline analysis is designed to yield a sample which has a small impurity of moving objects. While this might have a small impact on surveys designed for SN cosmology, it is possible that this choice has a large impact on the supernova sample for a multipurpose survey like LSST. Therefore, we would like to investigate the possibility of less aggressive selection combined with a more computationally-intensive filtration step to a sample purity similar to the current strategy. For certain observing strategies, this could result in an increase in good quality supernovae in the final cosmological sample.
    \item Cosmology Inference: The current baseline analysis uses a method of estimating the
      distance modulus~\citep{2011ApJ...740...72M} and intrinsic dispersion which has been validated
      on SDSS data, and compares binned theoretical and `observed' values of distance moduli for
      supernovae binned in redshift. In the future, we would like to better model the relationships
      between different parameters used in the cosmological inference (e.g.\ Hinton \etal in
      preparation\footnote{\url{http://dessn.github.io/sn-doc/doc/out/html/index.html}}), including
      the systematic parameters, leading to a likelihood of a large number of parameters. An
      important feature of these methods is that they enable better treatment of systematics,
      leading to better knowledge of them after the analysis. A problem with these methods is the
      integration with other probes in the joint likelihood analysis (e.g.\ with TJP tools due to software and parallelization issues).
    \item Photometric Classification and Cosmological Inference: The current baseline analysis
      assumes that (binary) photometric classification (using PSNID;
      \citealt{2011ApJ...738..162S,2013ApJ...763...88C}) is performed accurately. In practice, any
      photometric classification algorithm will mis-classify objects, causing biases. This can be
      accounted for using algorithms like BEAMS~\citep{2012ApJ...752...79H,2017ApJ...843....6J} in
      the future. Further, we plan to study the extension of cosmological inferences with
      photometric classification to include supernovae whose host redshifts are not
      known~(\citealt{2017JCAP...10..036R}; Malz, Peters, Hlozek \etal\ in
      preparation\footnote{\url{https://github.com/aimalz/scippr}}).
\item Cross-correlations between probes: Some of the systematic effects that affect the supernova
  science case can be mitigated by cross correlating supernova with other probes, and indeed can
  mitigate the systematics of other probes. For example, measurements of the LSST SN magnifications
  could be used to calibrate multiplicative errors in the LSST cosmic shear measurements \citep{2015ApJ...806...45Z}, while measurements of the skewness and kurtosis of the SN magnitude distribution (see
  \citealt{2017MNRAS.467..259M}) will help constrain $\sigma_8$. Further study of the impact of systematics
  on these cross-correlations will be an important area of research and development with DC2 data and
  beyond.
\item Retraining the SALT2 model: \ref{reqd:snlightcurve} shows that it is essential to retrain the
  SALT2 models as part of the analysis.  Also, doing this retraining simultaneously with the
  cosmology inference has the potential to add constraints on filter wavelength calibration.
\end{enumerate}

On a more technical note, non-Gaussian contours in cosmological parameter space may need to be
more fully addressed in future DESC~SRD versions; they were neglected in this one.

The issue of redshift uncertainty and its impact on the supernova science case is an active area of
study within the DESC~SNWG. In this analysis we assume the existence of host spectroscopic redshifts for all
supernovae that form part of the final sample. In the next iteration of the DESC~SRD, we will deviate from this default, by also assuming a simple photo-$z$ model for the majority of objects. This subsample will again be selected with an efficiency that roughly matches expected spectroscopic redshift yield. This source of systematic uncertainty is one of the most important ones to model in the future, as this affects both axes on the Hubble diagram ($\mu$ fits in addition to $z$).

In this analysis, we have also made a large simplifying assumption that the cosmology sample
consists of only Type Ia supernovae that have been selected with perfect efficiency and purity. The
potential bias from misclassification (so-named classification uncertainty), will be introduced in the
future by including some percentage (a few percent, given reasonable estimates of classification
purity, see e.g.\ \citealt{2013ApJ...763...88C}) of non-Ias for which the light curves get fit as if they were Ias, and checking the induced bias.  This results in a conservative requirement that ignores improvements that can be gained from probabilistic inference methods (e.g., \citealt{2017JCAP...10..036R}), and would represent our `worst case' scenario for classification bias.

\subsubsection{Systematic uncertainties}

For the SN analysis, we consider the following classes of systematic uncertainties in our two categories:
\begin{itemize}
\item Self-calibrated systematics: astrophysical systematics (errors in the modelling of SN and
  their standardization based on the supernova light curve, the environment, and any other redshift dependencies,including those induced by weak lensing magnification and peculiar velocities)
\item Calibratable systematics on which we place requirements: calibration (uncertainty on filter
  zero points, the transmission function wavelength, wavelength-dependent flux calibration, uncertainties in Galactic
  extinction corrections); redshift uncertainty; detection, classification
\end{itemize}

The models for the self-calibrated
systematic uncertainties are summarized in \autoref{tab:snsys-selfcal}.  For each source of
systematic uncertainty, we describe how
it is included in this DESC~SRD version, as well as aspirations for more complex models to be
included in the future.
Generally speaking, the systematic uncertainties are estimated by simulating the supernova sample with the given systematic effect included and comparing the distance modulus to the case where no systematic is included. This difference in the distance modulus redshift-by-redshift produces a systematics covariance matrix for each systematic, which is added to the intrinsic scatter. In these cases, the systematic uncertainty is naturally estimated as a data covariance ($\mu$ estimates) and then propagated into cosmological parameter space.

We assume that the error in the intrinsic dispersion modelling will be reduced by two times its current value given the large sample of SNe in the full survey that can be used to study the scatter, in combination with spectra from the ground-based collaboration. Similarly, for the host
mass-SN luminosity correlations, a bias in the high-mass sample following the model from
\cite{2017arXiv170201747H} is introduced into the simulations and ignored in the fitting, and this
is used to estimate the resulting systematic uncertainty.

We introduce a SN-color population drift (again following the prescription in \citealt{2017arXiv170201747H}) that gets ignored in the fitting to determine the systematic uncertainty.
Finally, we include a systematic due to the potential from evolution of the $\beta$ SALT2 parameter which describes the colour-luminosity correlation, which we include as half of the measurement uncertainty of $d\beta/dz$ as determined from our simulated LSST sample, because we can take the average of the observed and expected values.

\begin{table}[!htbp]
\begin{center}
\begin{tabular}{p{1.5in}p{2.2in}p{2.2in}} \hline
Self-calibrated systematic uncertainty& Current model baseline & Future plans \\ \hline
Standardization of color-luminosity law and its $z$-dependence & Marginalization over SALT-II
parameter $\beta$ \cite{2014ApJ...795...45S} at half the predicted measurement uncertainty & Same as current model \\ \hline
Intrinsic scatter & Modelling differences between {\cite{2010A&A...523A...7G}} and {\cite{2011A&A...529L...4C}} at half model differences & To be decided \\
Host mass-SN luminosity correlations & High-mass sample bias as in \cite{2017arXiv170201747H} &
  To be decided \\
Magnification-induced covariances (especially for DDFs) & Not modeled & To be decided \\
Individual low-$z$ calibration uncertainty & Not modeled & To be decided \\
 \hline\hline
\end{tabular}
\caption{Self-calibrated systematic uncertainties for SN.}
\label{tab:snsys-selfcal}
\end{center}
\end{table}

\begin{figure}[!htbp]
  \begin{center}
    \includegraphics[width=\textwidth]{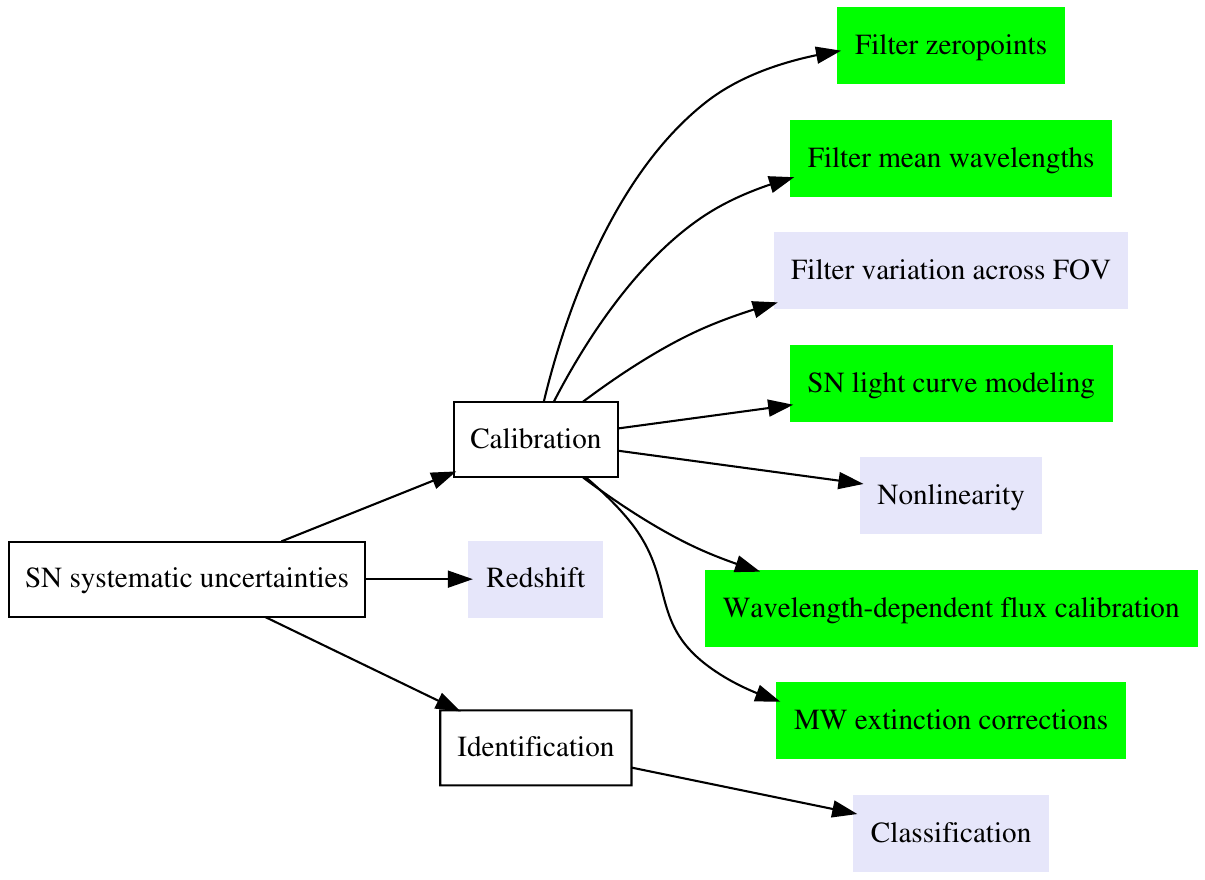}
  \end{center}
  \caption{Diagram indicating sources of systematic uncertainty for the SN analysis on which we would like to place
    requirements in the DESC~SRD.   The direction of the arrows indicates the flow from overall
    systematic uncertainty to broad systematics categories to the specific physical effects on which
    we place requirements. The green boxes indicate
sources of uncertainty on which we place requirements in this DESC~SRD version,
respectively.  In some cases, as discussed in \autoref{tab:snsys-cal}, the systematics models will
have to be updated.}
\label{fig:snsys-cal}
\end{figure}

The residual calibratable systematic uncertainties on which we will place requirements can be divided into 3
categories: calibration, redshift, and identification-related systematic uncertainties.  A diagram
of these calibratable systematics is shown in \autoref{fig:snsys-cal}, while the current models
and future plans for how to represent them is in \autoref{tab:snsys-cal}.

The DESC's DC2 analysis effort will provide more sophisticated models for the impact of several of these effects than the
ones used in these simulations.

\begin{table}[!htbp]
\begin{center}
\begin{tabular}{p{1.5in}p{1.8in}p{2.7in}} \hline
Calibratable systematic uncertainty & Current model baseline & Future plans \\
\hline
\multicolumn{3}{c}{Calibration}\\
\hline
Filter zero points & $\mathcal{N}(0,1~mmag)$ offset per band & Same as current model \\
Filter mean wavelength & $\mathcal{N}(0,1~\mathrm{\AA})$ & Same as current model \\
Filter variation across FOV & None & 10 measurements across FoV with 5 \AA~mean wavelength uncertainty \\
SN light curve modeling &  One-third scaling of \cite{betoule} SALT2 parameter covariance matrix & Same as current model\\
Nonlinearity & None & 3 mmag over 5 mag \\
Wavelength-dependent flux calibration & 5 mmag slope over 7000 \AA \ & Same as current model \\
MW extinction corrections & 5\% scaling of \cite{2014ApJ...789...15S} model & Same as current model \\
\hline
\multicolumn{3}{c}{Redshift}\\
\hline
Redshift & None & To be included based on DC2 \\
\hline
\multicolumn{3}{c}{Identification}\\
\hline
Classification & None & To be included based on DC2 or DC3 \\
\hline
\end{tabular}
\caption{Calibratable systematic uncertainties for SN. The numbers given in the table represent a
  base value for the level of uncertainty; the requirements given in \autoref{subsec:sn} were placed using contaminated data vectors
  with various multiples of this best guess.  Note that the `base' value of the systematics was chosen as larger than our intended baseline, in order to fully investigate the systematics requirements.}
\label{tab:snsys-cal}
\end{center}
\end{table}

One might expect systematic errors to be correlated (for example the filter mean wavelengths and
cutoff positions). Here we have assumed that all systematics can be varied individually, and
estimated the impact of them separately. For the band-dependent systematics we then include their
combined contribution through a Monte Carlo simulation of the individual bias vectors. This allows
for the separate uncertainties for different bands.

Finally, we note that the extinction requirement is placed without allowing for uncertainty in the
Milky Way dust law, which is yet another layer of complexity with the potential to introduce
apparent differences between low- and high-redshift supernova populations.  We defer consideration
of this effect to future versions of the DESC~SRD.

%% file: inc/sl.tex
\subsubsection{Analysis choices}

Here we describe the essential points of the SL analysis setup in this version of the DESC~SRD:

The Y10 strong lensing sample for our baseline analysis consists of time delay systems and compound
lenses.  The time delay cosmography sample is defined as 400 lensed quasar systems monitored with
LSST through 2032 and followed-up with TMT/GMT/JWST/E-ELT, so as to provide a 7\% time delay
distance measurement after marginalization over lens model parameters (including lens galaxy mass
distribution, environment and line of sight structure).  This marginalization is implicit, not
explicit.  The sample size was estimated using OM10
predictions\footnote{\url{https://github.com/drphilmarshall/OM10}} \citep{2010MNRAS.405.2579O} and
results of the DESC's Time Delay Challenge, with requirements on length of the time delay, image
separation, and ability to obtain follow-up data.

The Y10 compound lens sample is assumed to include 87 double source plane lenses. This calculation
uses LensPop\footnote{\url{https://github.com/tcollett/LensPop}} \citep{2015ApJ...811...20C}
modified to include compound lenses using best seeing single epoch imaging alone. Thus all systems
are bright enough that follow up is practicable.  We assume that each is followed up with TMT/GMT/JWST/E-ELT so as to provide fractional precision on $\beta$ of
  $\left[(0.01/R_{\text{ein},1})^2 + (0.01/R_{\text{ein},2})^2 + 0.01^2\right]^{1/2}$ after marginalization over lens model parameters.

  For Y1, lensed quasars are challenging due to the need to generate high-quality template imaging
  that can be used as a reference for identifying time-varying objects 
  in the first $\sim 2$ years of the survey. As for supernovae, we assume that a commissioning
  mini-survey may result in templates being available slightly earlier than would otherwise be
  possible.  We assume 20 of the brightest and most variable lenses will get measured time delays
  early.  We can, however, reliably assume there will be of order ten
  compound lenses in Y1.

In both Y1 and Y10, the likelihood function is taken as gaussian in $D_{dt}$ for time delay lenses
(no requirement on $D_A$ from lens velocities). For compound lenses the likelihood is taken as
Gaussian.

These forecasts assume 1 high resolution image per lens (space or AO), and spectroscopic redshifts
for lens and source in each system. In \citet{2016arXiv161001661N}, it was estimated that adaptive
optics IFU spectroscopy for characterizing lens galaxy velocity dispersions and image positions for
100 strong lens systems would require roughly 100 hours on GSMTs in total (or greater amounts of
time on 8-10m telescopes).  Given the relatively modest requirements when spread out over a number
of years, we therefore assume that the needed spectroscopy will be done. As a backup plan, WFIRST or
Euclid could provide some of the high-resolution imaging.

\subsubsection{Anticipated improvements}

This version of the DESC~SRD does not include lensed supernovae in the forecasts, because of the
additional need for high-resolution cadenced imaging.  Further developments with respect to
follow-up telescope resources may lead to their inclusion in the future.  We anticipate that the Y10
sample could include 50-500 lensed SNe systems (strongly dependent on follow-up resources), each
providing a 5\% time delay distance measurement after marginalization over lens model parameters.

There are also $\sim$3500 lensed quasars that could yield precise time delays with supplementary
data points on their light curves. There are up to 2000 fainter compound lenses that are
discoverable with LSST coadds. Exploiting these fainter systems will require significantly more
high-resolution imaging time, and hence we defer consideration of these additional systems for the
future.

\subsubsection{Systematic uncertainties}

For the SL analysis, we consider the following classes of systematic uncertainties in our two categories:
\begin{itemize}
\item Self-calibrated systematics: lens model assumptions (lens galaxy mass distribution),
  environment effects (including halo vs.\ stellar mass relation), line-of-sight structure, compound
  lensing in double source plane
\item Calibratable systematics on which we will eventually place requirements: time delay measurement systematics,
  selection bias, photometry issues including blending, photo-z and $M_*$ errors in environment
  analysis
\end{itemize}

Currently the self-calibrated systematics are all included implicitly in the forecasts through
adoption of a larger uncertainty in distance estimates than are obtained from statistical error in
the time delay measurements.  Future work within the DESC will include a more physical
forward-modeling of these effects and their impact on dark energy observables.

Developing models for how the calibratable systematics enter the observable quantities for LSST
(specifically through challenges in time delay estimation or sample characterization) will be an
important task for the DESC SL working group during DC2 and DC3.  In this DESC~SRD version, in the
absence of such models, no requirements will be placed for the SL analysis.

%% file: inc/synthesis.tex
\subsection{Systematic uncertainties in this DESC~SRD version}\label{subset:synth:thisversion}

In this section, we synthesize the list of systematic uncertainties on which requirements are placed
in this DESC~SRD version (see \autoref{tab:synth-req}).  As in the appendices on individual
probes, we divide them into categories (e.g., uncertainties in redshifts, number densities, and so
on).  The
top two categories are considered only for probes of structure growth, while the last is considered
only for SN.

\begin{table}[hp]
\begin{center}
\begin{tabular}{p{2.5in}p{1.5in}p{2in}} \hline
Systematic uncertainty & Probes & Note \\
\hline
\multicolumn{3}{c}{Redshift uncertainties}\\
\hline
Mean redshift & LSS, WL, CL & For CL, only source redshift uncertainties were considered \\
Redshift scatter & LSS, WL, CL & \\
\hline
\multicolumn{3}{c}{Shear (multiplicative) uncertainties}\\
\hline
Overall multiplicative bias & WL, CL & This requirement encompasses all sources of
multiplicative bias  \\
Stellar contamination & WL, CL & Subset of ``Overall multiplicative bias'' \\
PSF modeling errors & WL, CL & Subset of ``Overall multiplicative bias'' \\
\hline
\multicolumn{3}{c}{Photometric calibration uncertainties}\\
\hline
Filter zeropoints & SN & \\
Filter mean wavelengths & SN & \\
Wavelength-dependent flux calibration & SN & \\
Light curve modeling & SN & \\
MW extinction corrections & SN & \\
\hline
\end{tabular}
\caption{List of all sources of systematic uncertainty for which we place requirements in this
  version of the DESC~SRD.}
\label{tab:synth-req}
\end{center}
\end{table}

\subsection{Systematic uncertainties deferred for future work}

In this DESC~SRD version, we have focused on a limited list of systematic uncertainties for which
there is a clear prescription for describing how they modify the observable quantities for dark
energy analysis.  More complete wish-lists for calibratable systematic uncertainties on which
requirements could be placed in future may be found for each probe in \appref{app:baselines}.
More generally, as the DESC's analysis pipelines are constructed and models for systematics
characterization and mitigation are built, interconnections between classes of uncertainties should be
noted.  For example, if a given effect could result in systematic uncertainty in both redshift and
number densities, a self-consistent model for those uncertainties should be built.  Similarly,
self-consistent models should be built for how a given systematic uncertainty affects the
observables in different cosmological probes.

As described in \autoref{sec:definitions}, the choice not to place requirements on
model-sufficiency for either type of  systematic uncertainty may also be revisited in future
DESC~SRD versions.

While not strictly a systematic uncertainty, one issue affecting all probes that is important for
fulfilling our high-level objectives (\ref{obj:notsysdom}) and requirements (\ref{high:blinding}) is
the development of blinding methods that work at the level needed for single and joint probe
analysis.

%% file: inc/dndmag.tex
For the WL, LSS, and CL analyses, we need to define Y1 and Y10 source galaxy number densities and
redshift distributions.  LSS and WL also require number densities and redshift distributions for the
photometric lens sample.  Older estimates, e.g., in the LSST Science Book \citep{2009arXiv0912.0201L}, used the best deep
ground-based data and spectroscopic redshift samples available at the time.  We would like to confirm these
estimates for Y10 using more recent datasets, and produce a Y1 estimate.

Below we derive the photometric sample number density (\appref{subsec:overalln}), the source
sample number density (\appref{subsec:neff}), the photometric sample redshift distribution
(\appref{subsec:dndzphot}), and the source sample redshift distribution
(\appref{subsec:dndzeff}).  These numbers are somewhat different than common assumption
on LSST densities found in the LSST science book based on updates in our understanding both in
observations and simulations, as will be described in detail in each subsection.

\subsection{Photometric sample number density}\label{subsec:overalln}

To estimate the overall galaxy number density for a photometric sample with a given flux limit, we
use the HSC\footnote{\url{http://hsc.mtk.nao.ac.jp/ssp/survey/}} Deep field $i$-band catalogs from the HSC Survey Public Data
Release\footnote{\url{https://hsc-release.mtk.nao.ac.jp/doc/}} 1 \citep{2018PASJ...70S...8A}.
This sample is useful because it requires relatively minimal extrapolation to LSST depths, is a
large enough field that cosmic variance is not too significant, and the bandpass is similar to the
expected LSST $i$-band.

\begin{figure}[!htbp]
  \begin{center}
    \includegraphics[width=0.65\textwidth]{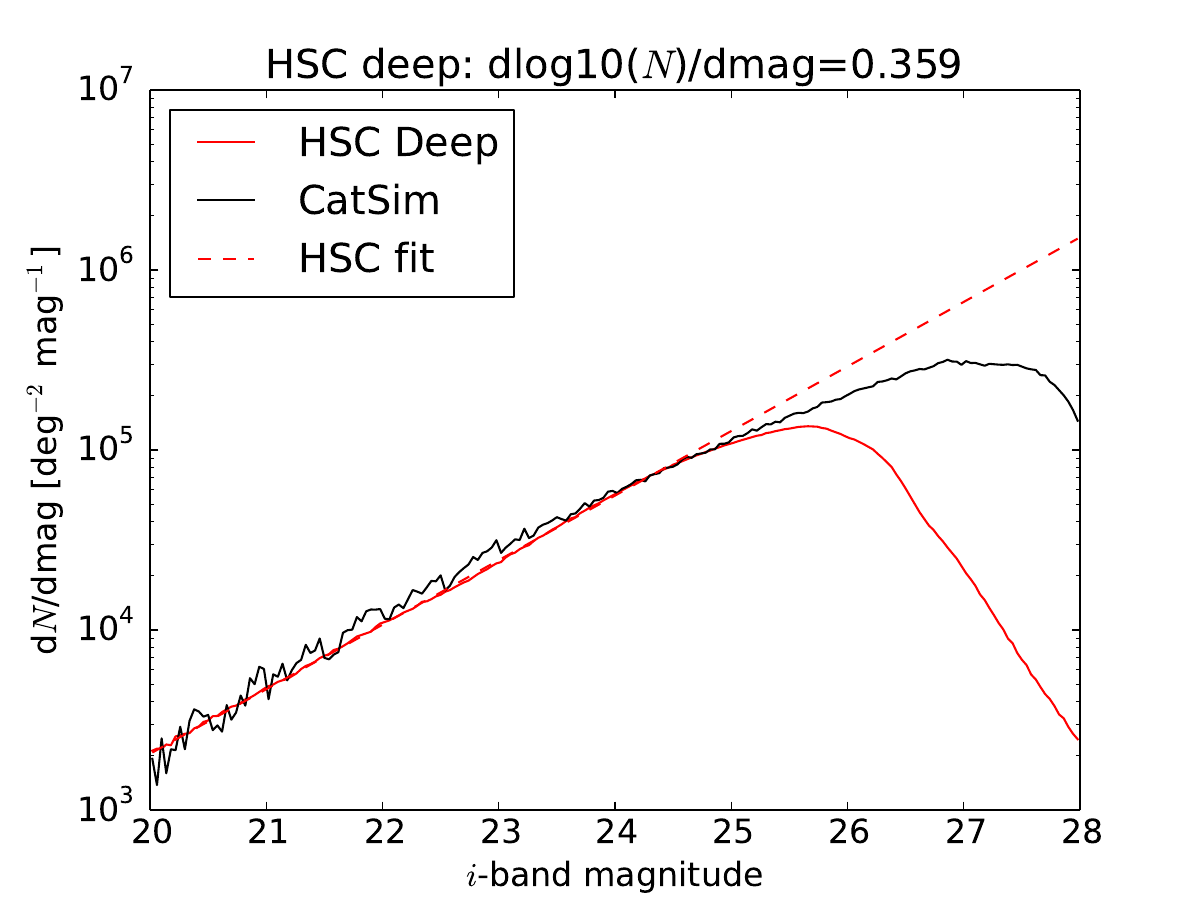}
    \includegraphics[width=0.65\textwidth]{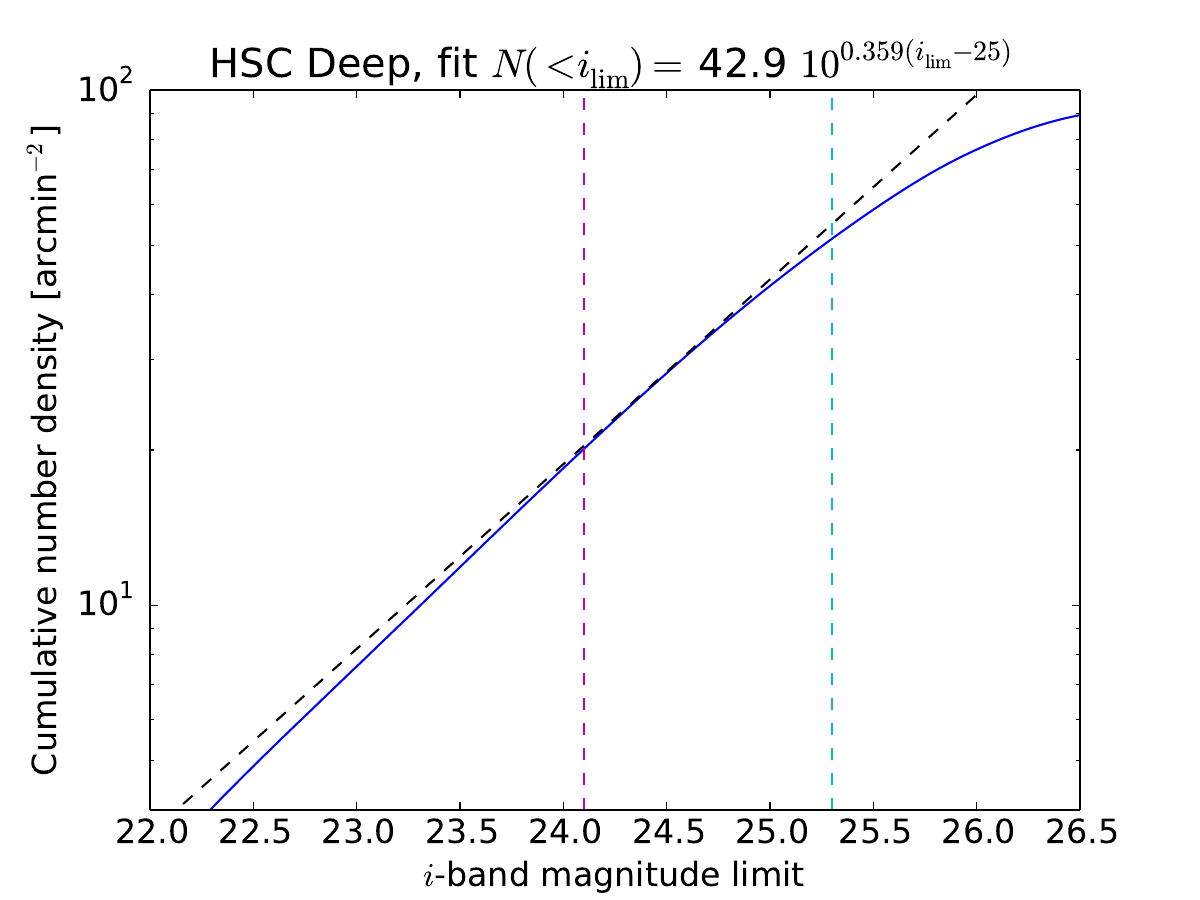}
  \end{center}
  \caption{{\em Top:} The differential distribution of $i$-band galaxy magnitudes in the HSC Deep
    survey. Before the turnover due to incompleteness, the
    data (red solid curve) were fit to a power-law, shown as a red dotted line, with a slope indicated in
    the plot title.  Finally, the \dndmag\ in the CatSim catalog used for WeakLensingDeblending
    simulations (described in more detail in \appref{subsec:neff}) is shown in
    black.  {\em Bottom:} The cumulative number density of galaxies as a function of $i$-band limiting magnitude
    $i_\text{lim}$. The power-law fit to the differential counts from
    the left panel was used to get an extrapolated version of the cumulative counts, shown as
    a dashed line.  Vertical lines show the Y1 and Y10 limits of the photometric sample used for
    clustering analysis, as described in \appref{subsec:overalln}.  The power-law equation is
    shown in the plot title.
    \label{fig:dndmag}}
\end{figure}

The HSC data were downloaded with an SQL query that was designed to get a complete galaxy sample
that goes as deep as possible, with only minimal flag cuts (e.g., star/galaxy classification, only
primary detections, and nothing with saturated/interpolated pixels near the center of the galaxy).
A bright star mask was imposed.  Random points were also downloaded from the HSC deep database while
imposing the same flag cuts, in order to properly calculate the area\footnote{The real and random
  SQL queries can be found in the Requirements repository, `number\_density/hsc.sql' and
  `number\_density/hsc\_rand.sql'.  The script that used the resulting HSC catalogs to carry out the
  calculations described below and make \autoref{fig:dndmag} is
  `number\_density/dndmag\_hsc.py'.}.

Using the random points, we calculated the area as $26.1$~deg$^2$.  Note that this includes a
$\sim$12\% reduction factor for masks due to various image defects, bright stars, and so on.  The
number densities we calculate do {\em not} include this reduction at the outset.  Since our adopted
survey area does not account for masking, we need to include it by reducing our number densities at
the end by a factor we will call $1-f_\text{mask}$.  Assuming that $f_\text{mask}=0.12$ would
amount to an assumption that HSC and LSST will have similar levels of masking.

The differential \dndmag\ (using the $i$-band forced \texttt{cmodel} magnitudes) is shown in
the top panel of \autoref{fig:dndmag}, along with a power-law fit that is extrapolated to faint magnitudes.
Analysis of deep pencil-beam HST surveys suggests that extrapolating a power-law \dndmag\ is a
reasonable approximation.

We can use this power-law and the area of the survey to get the cumulative counts as a function of
limiting $i$-band magnitude $i_\text{lim}$.  The result is shown in the bottom panel of
\autoref{fig:dndmag}.  Including the factor of $f_\text{mask}$, our adopted number density as
a function of $i_\text{lim}$ is
\begin{equation}\label{eq:nltilim}
N(<i_\text{lim}) = 42.9~(1-f_\text{mask})~10^{0.359(i_\text{lim}-25)}~\text{arcmin}^{-2}
\end{equation}
For the calculations in this version of the DESC~SRD, we define the Y1 and Y10 gold samples using $i_\text{lim}=24.1$
and $25.3$, respectively.  These limiting magnitudes come from defining the gold sample magnitude
limit one magnitude brighter than the median survey depth at any given time (see
\appref{subsec:assump-strategy} for assumptions about survey depth).  We also adopt
$f_\text{mask}=0.12$.  This results in Y1 and Y10 photometric sample number densities of 18 and
48~arcmin$^{-2}$, respectively.

\subsection{Source sample number density}\label{subsec:neff}

For weak lensing forecasts, we need the source effective number density \neff\ (accounting for the
necessary downweighting for low signal-to-noise shear estimates; e.g.,
\citealt{2013MNRAS.434.2121C}) for Y1 and Y10.  To estimate this quantity, we use a method similar
to that of \citet{2013MNRAS.434.2121C}. Specifically, we use the WeakLensingDeblending
(WLD\footnote{\url{https://github.com/LSSTDESC/WeakLensingDeblending}}) package to simulate the LSST
CatSim galaxy catalog in the $i$ and $r$ bands, and use the following values for each simulated
galaxy:
\begin{itemize}
    \item True redshift $z$,
    \item Estimated measurement error $\sigma_m$ for each shape component calculated using the parameterization $\sigma_m(\nu, R)$ of \citet{2013MNRAS.434.2121C}, and
    \item Estimated purity $\rho$ (a measure of blendedness in the range 0--1, with 1 for a perfectly isolated source).
\end{itemize}
The code used for this analysis is in a jupyter notebook\footnote{{\tt
  notebooks/RedshiftDistributions.ipynb}} in the Requirements repository. We have performed detailed
comparisons of WLD results against \citet{2013MNRAS.434.2121C} and discovered some issues that were
later determined to be a problem in that work rather than in this notebook, but the overall
agreement is good.

We are interested in different sample definitions using:
\begin{itemize}
    \item Y10 or Y1, where Y1 is defined as 0.1 of the total exposure time in each band for the purposes of WLD simulations.
    \item The combined $i + r$ sample or else $i$ and $r$ individually. For the combination, we use only galaxies detected in both bands and define the combined $\sigma_{m,i+r} ^{-2} = \sigma_{m,i}^{-2} + \sigma_{m,r}^{-2}$.
    \item A weak-lensing sample selected using $\sigma_m < k\cdot \sigma_{SN}$ with $k = 1$ (nominal), $0.5$ (conservative) or $2.0$ (optimistic).  We fix $\sigma_{SN} = 0.26$ for this cut.
    \item With or without blending corrections.  Blending reduces $\nu$, which in turn reduces $\sigma_m(\nu, R)$, and thereby the selection fraction and the $n_\text{eff}$ weight. We also remove galaxies with $\rho < \rho_{min}$ from the sample, assuming that they are too blended for reliable photo-z and shear estimation.
\end{itemize}
\autoref{table:neff} summarizes the results with $k=1$.  Note that our treatment of blending effects is not particularly conservative: we assume that all galaxies below a certain overlap fraction ($\rho_{min} = 0.85$) are unusable and all other galaxies are optimally deblended.

\begin{table}[!thbp]
\begin{center}
    \begin{tabular}[htb]{rcccrr}
        Epoch & Bands & Blended & $n$ & $n_\text{eff}$ \\
        \hline
        Y10 & $i+r$ & Y & 34.091 & \textcolor{red}{27.737} \\
        Y10 & $i+r$ & N & 41.765 & 34.581 \\
        Y10 & $i$   & Y & 25.658 & 20.732 \\
        Y10 & $i$   & N & 32.557 & 26.562 \\
        Y10 & $r$   & Y & 26.594 & 21.402 \\
        Y10 & $r$   & N & 34.319 & 27.922 \\
        Y1  & $i+r$ & Y & 13.969 & \textcolor{red}{11.112} \\
        Y1  & $i+r$ & N & 16.174 & 13.052 \\
        Y1  & $i$   & Y & 10.230 &  8.051 \\
        Y1  & $i$   & N & 12.317 &  9.744 \\
        Y1  & $r$   & Y & 10.402 &  8.170 \\
        Y1  & $r$   & N & 12.622 & 10.024 \\
        \hline
    \end{tabular}
\end{center}
\caption{Summary of results with $k = 1$.  The last two columns give
the integrated densities $n(z)$ and $\neff (z)$, respectively, in units of galaxies per square
arcminute. These densities are not corrected for any masking effects. The scenarios that we use for
forecasts are shown in red.\label{table:neff}}
\end{table}

\subsection{Photometric sample redshift distribution}\label{subsec:dndzphot}

While the photometric sample number density as a function of limiting magnitude can be obtained from
HSC, this is not possible for the redshift distributions, due to the lack of redshift information in
HSC.  Hence we derive redshift distributions from the CatSim input catalog used for
WLD, with a strict $i$-band limit as for the photometric samples discussed in
\appref{subsec:overalln}.
As a sanity check, they were compared against parametric fits to spectroscopic redshift
distributions estimated using DEEP2\footnote{\url{http://deep.ps.uci.edu/}} data,
correcting for the impact of selection/color cuts using
the targeting weighting factors. The data only include redshifts up to $z\sim 1.4$, so the redshift
distribution beyond that is an extrapolation.  In addition, DEEP2 is limited to $R<24.1$, so the
distribution is extrapolated to fainter magnitudes as well.

\begin{figure}[!htbp]
  \begin{center}
    \includegraphics[width=0.7\textwidth]{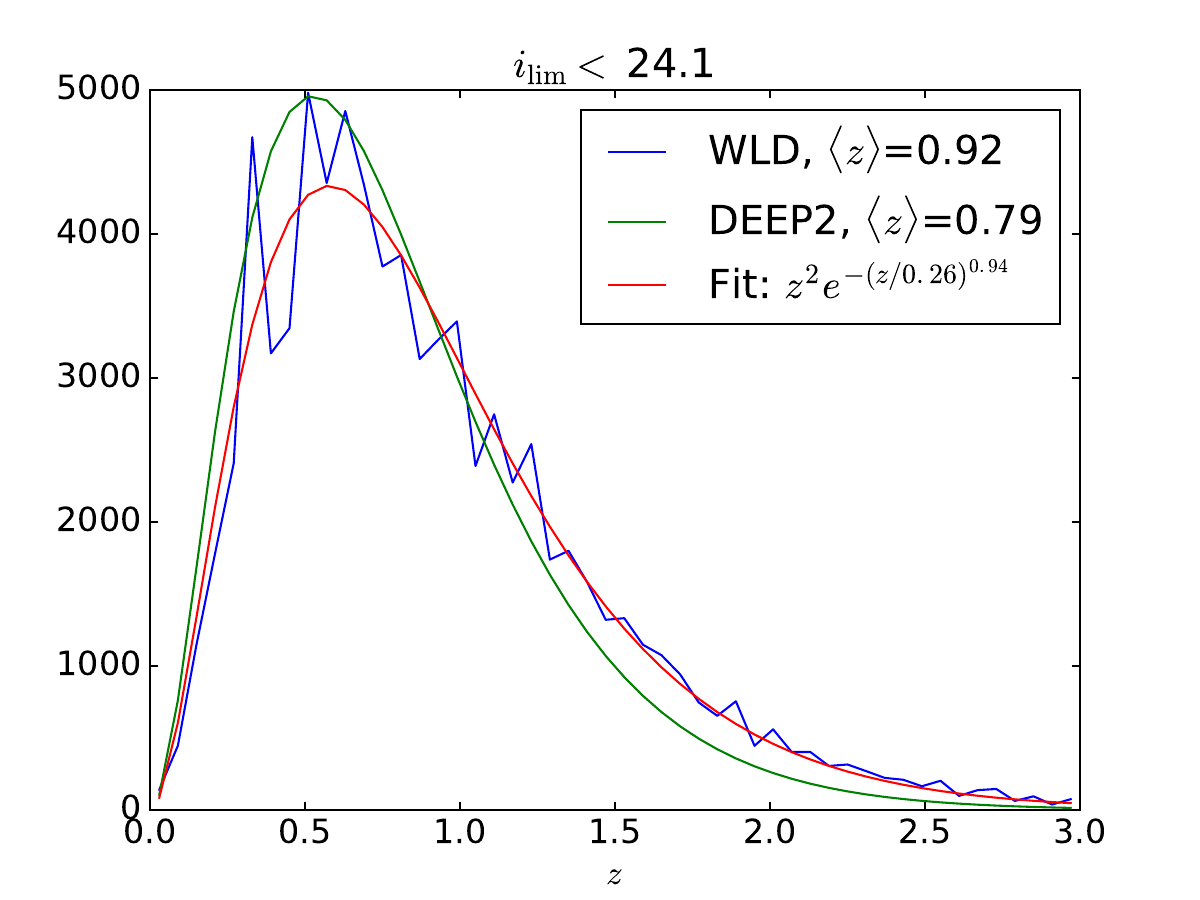}
    \includegraphics[width=0.7\textwidth]{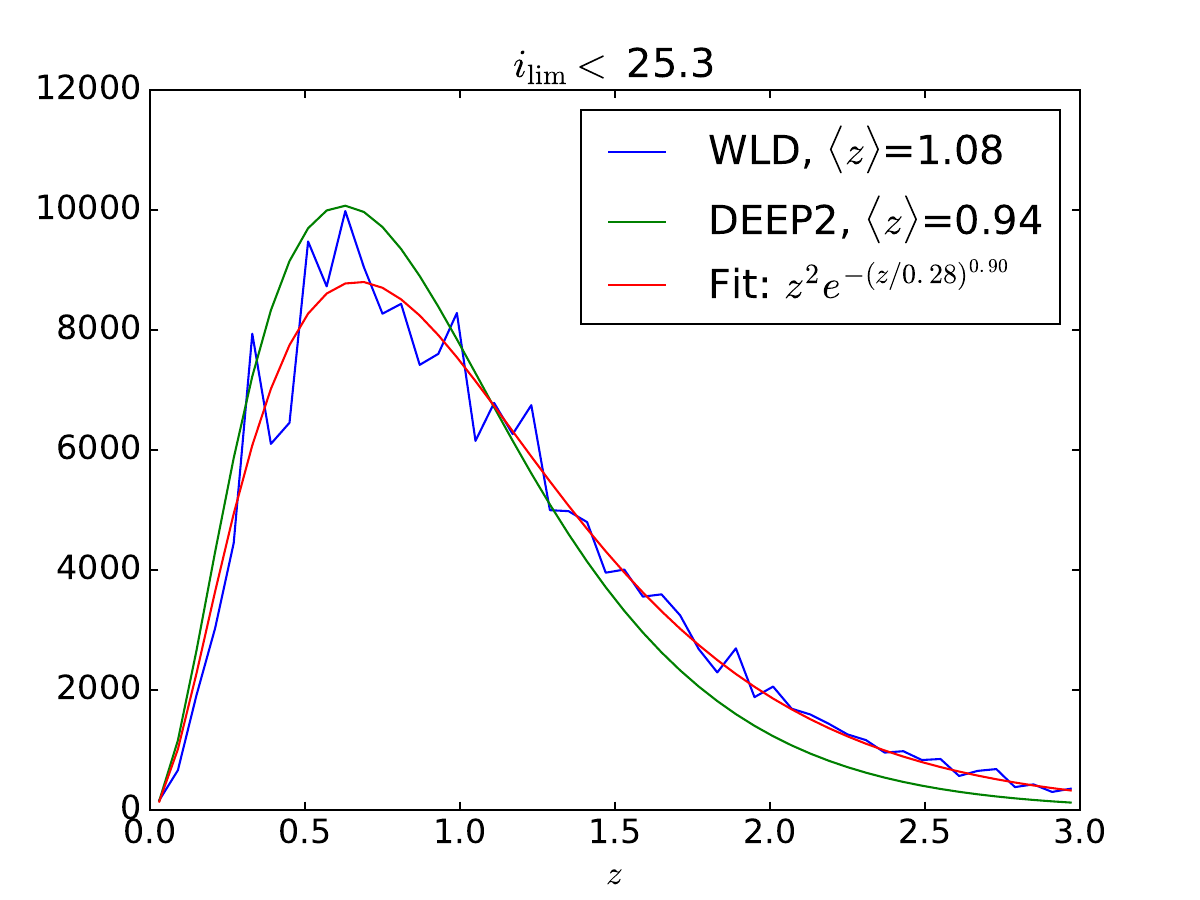}
  \end{center}
  \caption{The redshift distribution for the Y1 (top) and Y10 (bottom) photometric sample used for clustering, as predicted based
    on the CatSim input catalog used by WLD, a parametric fit to that data used
    in \appref{app:wllss}, and DEEP2 (parametric fit based on data
    to $z=1.4$ only). \label{fig:nzy1y10}}
\end{figure}

The comparison between the input catalog to WLD, our best-fitting parametric
distribution in \autoref{eq:nz}, and the best-fitting DEEP2 distribution\footnote{See
  `number\_density/dndz.py' in the Requirements repository.} is shown for Y1 and Y10 in
\autoref{fig:nzy1y10}.  As shown, the results for $z\lesssim 1.4$
(where less extrapolation is required) agree quite well between CatSim and DEEP2.

\subsection{Source sample redshift distribution}\label{subsec:dndzeff}

\begin{figure}[!htbp]
  \begin{center}
    \includegraphics[width=0.6\textwidth]{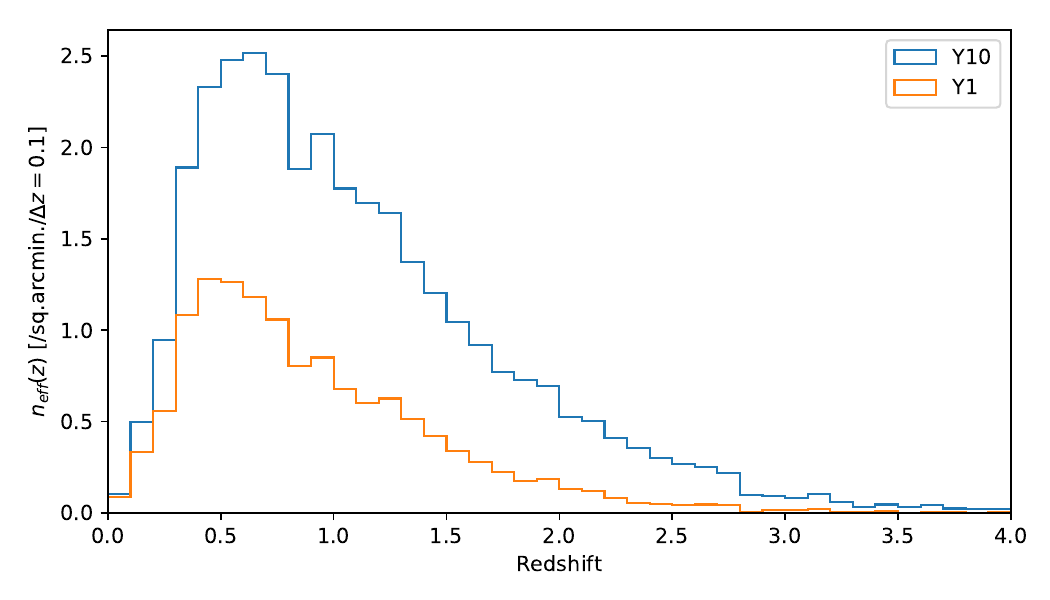}
    \includegraphics[width=0.6\textwidth]{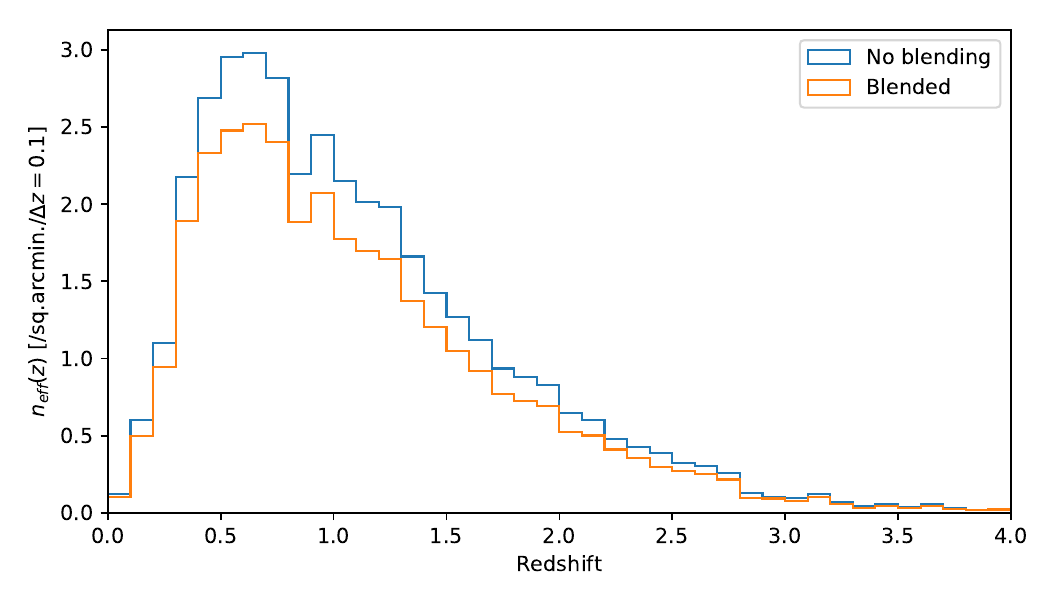}
\includegraphics[width=0.6\textwidth]{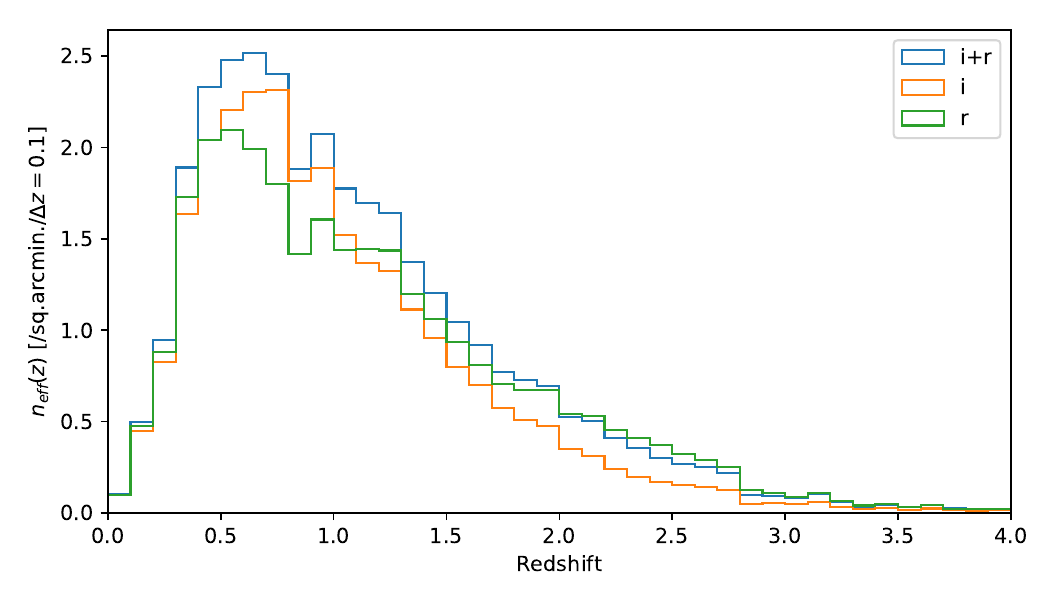}
  \end{center}
  \caption{{\em Top:} The effective $k=1$ redshift distribution for the Y10 and Y1 samples of $i+r$,
    including the effects of blending. {\em Middle:} The effective $k=1$ redshift distribution for
    the Y10 $i+r$ sample, with and without including the effects of blending. {\em Bottom:} The effective $k=1$ redshift distribution for the Y10 samples in $i+r$, $i$ and $r$, including the effects of blending. \label{fig:neff_choices}}
\end{figure}

The jupyter notebook mentioned in \appref{subsec:neff} saves the $n(z)$ and $\neff (z)$
distributions for each row of \autoref{table:neff} to a subdirectory {\tt notebooks/neff/} using
plain text format and a file name based on the first three columns.  \autoref{fig:neff_choices}
has several panels comparing the resulting \neff\ distributions to show
the impact of different choices (Y1 vs.\ Y10, blending, single band vs.\ both).

\begin{figure}[!htbp]
  \begin{center}
    \includegraphics[width=0.7\textwidth]{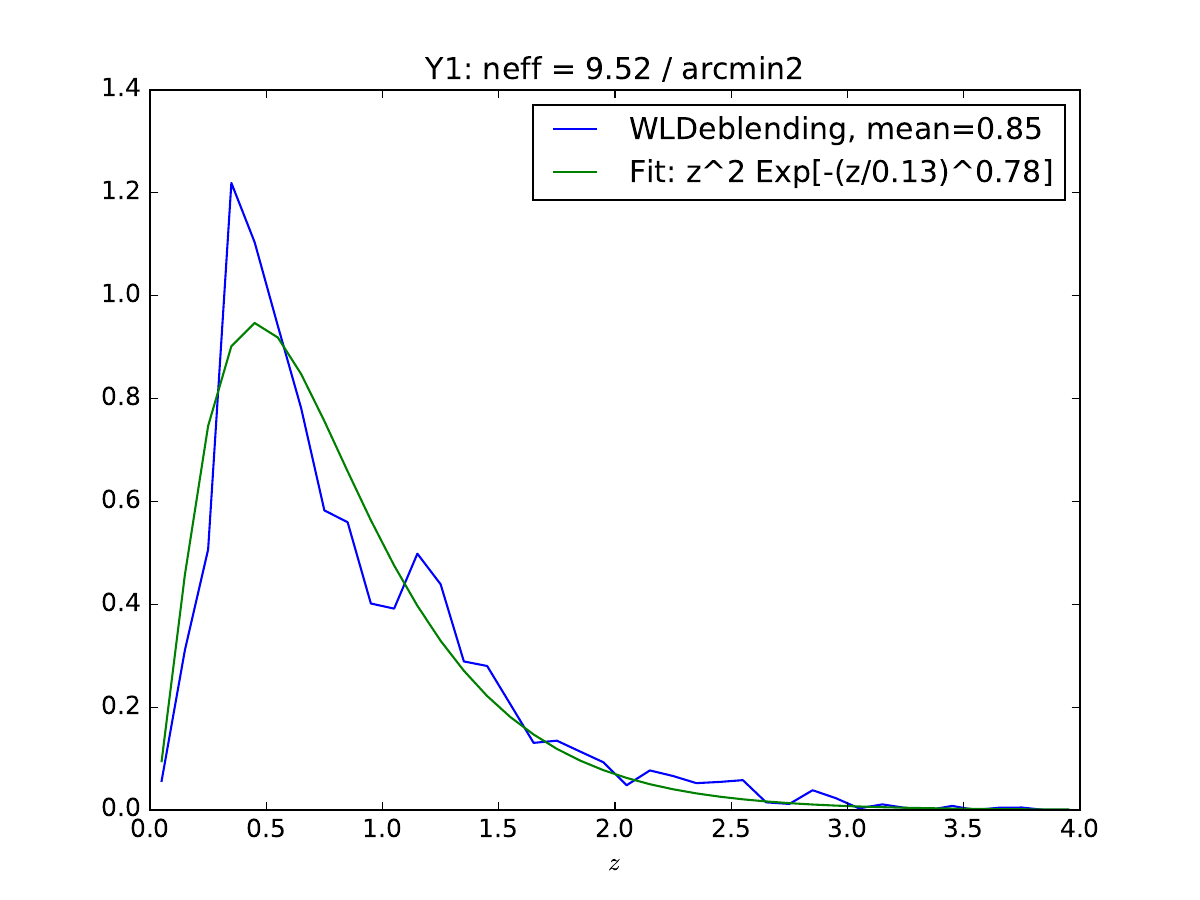}
    \includegraphics[width=0.7\textwidth]{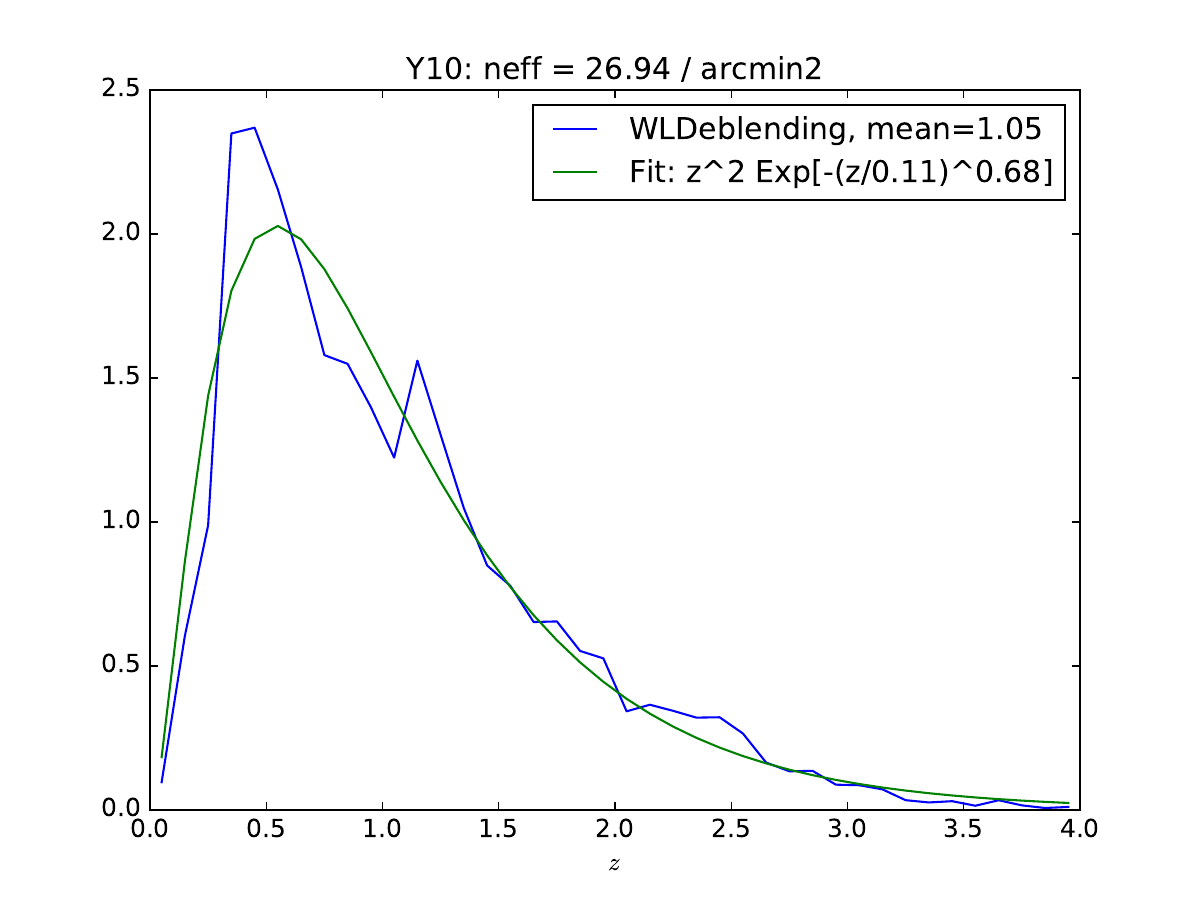}
  \end{center}
  \caption{The effective redshift distribution $\neff (z)$ for the Y1 (top) and Y10 (bottom) source
    sample from WLD,
    and a parametric fit to that data used
    in \appref{app:wllss}. \label{fig:neffzy1y10}}
\end{figure}

The resulting $n_\text{eff}(z)$ (histograms and parametric fits) are shown for Y1 and Y10 in
\autoref{fig:neffzy1y10}.
